\documentclass[aip,pop,amsmath,amssymb, reprint,twocolumn,showkeys,floatfix]{revtex4-1} 
\usepackage[T1]{fontenc}
\usepackage[english]{babel}
\usepackage{natbib}
\usepackage[colorlinks=true,linkcolor=blue]{hyperref}
\usepackage{graphicx}
\usepackage{amsmath}
\usepackage{txfonts}
\usepackage{verbatim}
\usepackage{comment}
\usepackage{setspace}
\usepackage{color}

\newcommand{\Fref}[1]{Figure~\ref{#1}}

\newcommand{\fref}[1]{figure~\ref{#1}}
\newcommand{\eref}[1]{equation~(\ref{#1})}

\begin{document}

\title{Sub-Grid-Scale Description of Turbulent Magnetic Reconnection in Magnetohydrodynamics}


\author{F.~Widmer}
\email{widmer@mps.mpg.de}
\author{J.~B\"uchner}
\affiliation{Max Planck Institute for Solar System Research, G\"ottingen, Germany}
\affiliation{Georg-August-Universit\"at G\"ottingen, Germany}
\author{N.~Yokoi}
\affiliation{Institute of Industrial Science, University of Tokyo}

\begin{abstract}
     Magnetic reconnection requires, at least locally, a non-ideal plasma response. In collisionless
space and astrophysical plasmas, turbulence could permit this instead of the too rare binary collisions. We investigated the possible influence of turbulence on the reconnection rate in the framework of a single fluid compressible MHD approach through simulations of a double Harris and force free current sheets, with finite guide magnetic fields. The goal is to find out, whether unresolved, sub-grid for MHD simulations, turbulence can enhance the reconnection process in high Reynolds number astrophysical plasma including force free and guide magnetic field. For this sake we solve, simultaneously with the grid-scale MHD equations, evolution equations for the sub-grid turbulent energy and cross helicity according to Yokoi's model\cite{Yokoi4} where turbulence is self-generated and -sustained through the inhomogeneities of the mean fields. This way we consider the feedback of the turbulence into the MHD reconnection process. The MHD equations together with evolution equation for the sub-grid turbulence are solved with a second order accurate in time and space MacCormack scheme. Harris and force free sheets are simulated in a box using double periodic boundary conditions. A small perturbation of the magnetic field, satisfying $\nabla\cdot\boldsymbol B=0$, is used to trigger reconnection. It has been shown that the turbulence timescale controls the regimes of reconnection
.\cite{Yokoi3} New results are obtained about the dependence on resistivity for large Reynolds number for Harris-type as well as force free current sheets with guide field. We interpret our results about obtaining the limit of fast magnetic reconnection and obtained important relation between the molecular and turbulent resistivity. The turbulence timescale $\tau_t$, parametrising the sub-grid model controls the regime of reconnection rate in both
Harris and force free current sheets. It decides whether reconnection takes place or if the system is just turbulent. The amount of energy transferred from large to the small scales is enhanced in case of fast turbulent reconnection and energy spectra are used to interpret the obtained regime of reconnection. The overall process is even faster for larger Reynolds numbers controlled by the background molecular resistivity $\eta$, as long as the initial level of turbulence is not too large. This implies that turbulence plays an important role on fast reconnection at situation of large Reynolds number while the amplitude of turbulence can still be small.
\end{abstract}

\maketitle

\section{Introduction}
\label{intro}
Magnetic reconnection is the most efficient process to convert stored magnetic energy into plasma kinetic energy and heat. For the Sun,
reconnection is at the origin of many dynamical processes in the corona such as CMEs or solar flares. Based on Parker's model of reconnection through a current sheet,\cite{JGR:JGR677} the rate of energy conversion during reconnection is as small as $\sim S^{-1/2}$ where $S$ is the Lundquist number (Reynolds number for the Alfv\'en velocity $v_{A}$). A typical Lundquist number for the solar corona
is of the order of $10^{12}$, giving an associated time scale $t\approx \sqrt{S}L/v_A\approx 0.3$ years.\cite{RevModPhys.82.603} On the other hand, the duration of the impulsive phase of a flare is of the order of $100s$.
%
Different plasma processes has been proposed for enhancing reconnection such as Petchek's model. For collisionless plasmas kinetic models were developed,\cite{Schindler} anomalous
resistivity have been used,\cite{yokoyama1994condition,ugai1977magnetic} Hall effects were taken into account \cite{terasawa1983hall,JGRA:JGRA15381} as well as plasmoid instabilities \cite{Loureiro:2007gv} in MHD.
On the other hand in astrophysical plasma turbulence is ubiquitous and \textit{in-situ} observation in the solar wind
show a close relation between magnetic reconnection and turbulence.\cite{Carbone1,Matthaus1} It was proposed that turbulence can lead to fast reconnection \cite{Matthaeus_Lamkin_1985,lazarian1999reconnection,kowal2009numerical} with a reconnection rate $M_A\cong0.1$. Since reconnection is multiple scale process where large scale magnetic field supplies the small scales where diffusion takes place, an important open question is how to bridge the gap of this multiple scale interaction. One possible way to increase reconnection in the Sweet-Parker scaling is to increase the thickness of the diffusion layer such that it is comparable to its length. Turbulence might provide such situations. For instance, the theory of turbulent reconnection by Lazarian and Vishiniac proposes that the effective thickness of the diffusion layer is largely increased by the stochastic movement of the magnetic field lines.\cite{lazarian1999reconnection} However, the dynamics of mean fields is largely affected by turbulence, through enhanced transport effects while at the same time, large scale fields determine the properties of turbulence. In this sense, treating turbulence and mean-field simultaneously instead of imposing turbulence as an external component of any (astro-)physical process is necessary. In situation of high Lundquist number, mean field approach is well suited since it can easily handle turbulence which, in such situation, reduces to statistical quantities. However in most astrophysical plasmas the extreme space- and astro-physical parameters do no facilitate numerical simulations of turbulent reconnection with realistic parameters. In such situation, the magnetic Reynolds number is $\sim 10^{12}$ and it is therefore impossible to resolve all scales of turbulence even numerically. In fact, the number of grid points necessary for direct numerical simulation (DNS) would be $O(R_e^{9/4})$ which is far too much for existing computing system. For that reason, modelling of MHD turbulence is necessary. In particular, e.g., two approaches to turbulence are used:
large eddy simulations (LES) with sub-grid scale (SGS) terms and Reynolds averaged Navier-Stokes (RANS) equation solvers. The LES based models allow that the resolved fields vary in space while only some part of the fluctuations are
included (for example non-linear Smagorinsky-type model \cite{GreteDimitar}). RANS models, on the other hand, model the effect of the sub-grid scale turbulence by solving evolution equations only for the mean field. These kinds of modelling provide a powerful and effective way to incorporate turbulence in numerical simulations. Such description of RANS MHD modelling where turbulence is generated and sustained by the inhomogeneities of the mean fields, has been presented to increase reconnection.\cite{Yokoi1} The generation of turbulent energy, localised around the diffusion region of reconnection by a turbulent cross-helicity (cross-correlation between magnetic-field and velocity-field fluctuations) plays the role of a turbulent diffusivity enhancing reconnection. By means of this model, reconnection can be made fast, bridging the timescale discrepancy between theory and observation. Of course a confirmation of mean field models is required even though mean field approach of turbulence as been able to successfully reproduce the Alfv\'enicity of the solar wind\cite{2007PhPl...14k2904Y} and that the expression of turbulence for the electromotive force (\eref{eq:Emf}) has been confirmed by direct numerical simulation (DNS) for a Kolmogorov flow.\cite{YokBal} Here in this paper, we
apply a RANS model \cite{Yokoi1} to compressible MHD.\linebreak
We utilize a simplified approach that contains a free parameter, the turbulence timescale.\cite{Yokoi3} The effect of this parameter on the
reconnection process for Harris equilibrium has previously been shown to control the regime of magnetic energy release, as \textit{laminar reconnection},
\textit{turbulent reconnection}, enhancing the reconnection rate strongly, and \textit{turbulent diffusion} regime. We investigate the three regimes of reconnection for a Harris equilibrium with guide field and dependence on the resistivity by carrying out large Reynolds number simulations. We also investigate the consequence of an out-of-plane guide magnetic field considered to be co-aligned or anti-aligned with the mean current density.
\linebreak In addition, the influence of the initial turbulence in the system is tested by varying the initial (constant) background resistivity as well as $\beta_0$ related in the model to the initial amount of turbulent energy in the system. A spectral analysis approach of the kinetic and magnetic energy is presented in order to understand the physical process controlling the different regimes found for different initial amount of turbulence in the system.
Finally, to study these effects in a more realistic equilibrium for solar corona, current force free sheets are considered. Kinetic simulations have shown an effect of strong guide fields on non-linear phase, through a Harris or force free equilibrium, in delaying the reconnection process as well as to reduce the maximum reconnection rate.\cite{Pato2015,Ricci:2003yc} This was also seen in MHD experiments.\cite{RevModPhys.82.603} We show that a RANS turbulence model reproduces a decreased reconnection rate in the presence of a strong guide field. We conjecture that this reduction, even in two dimensions, is due to the turbulent magnetic helicity ($\alpha$ effect) on the reconnection rate. Remembering, however, that in three dimensions the situation could be different.
\section{Turbulence Modelling}
\label{Turbulence}
The turbulence model we are going to use is based on a Reynolds decomposition where instantaneous physical quantities are divided into mean and turbulent ones: $f=F+f'$ where the capital letter stands for mean field $F=\left<f\right>$ and $\left<...\right>$ is the ensemble average for which $\left<f'\right>\equiv 0$. 
Applying the decomposition to the induction equation, an electromotive force (EMF) $\left<\boldsymbol{v}'\times \boldsymbol{b}'\right>$ appears in the mean induction equation which can be written as a function of the mean quantities
\begin{equation}
	\langle{{\bf{u}}'\times{\bf{b}'}}\rangle = -\beta\mu_0\boldsymbol{J} +\gamma_t \sqrt{\mu_0\rho}\boldsymbol{\Omega}+\alpha\boldsymbol{B}
\label{eq:Emf}
\end{equation}
Here $\boldsymbol{\Omega}=\nabla\times\boldsymbol{U}$ is the curl of the mean velocity and the coefficients $\beta$, $ \gamma_t$ and $\alpha$ have to be determined. In turbulent reconnection, the modeling of the turbulent diffusivity $\beta$ is extremely important and must represent the reduced information of averaged turbulence. In the framework of the present model, statistical quantities are chosen to represent the coefficients in (\ref{eq:Emf}), such that the turbulent magnetic resistivity $\beta$ is related to the turbulent energy $K$, the $\gamma$ term to turbulent cross-helicity $W$ and the $\alpha$ term to the turbulent helicity $H$ which are statistically defined as:
\begin{eqnarray}
K&=&\frac{1}{2}\left<\boldsymbol{u}'^2 + \frac{\boldsymbol{b}'^2}{\mu_0\left<\rho\right>}\right>\label{Kmath}\\
W&=&\textcolor{white}{\frac{1}{2}}\left<\frac{\boldsymbol{u}'\cdot\boldsymbol{b}'}{\sqrt{\mu_0\left<\rho\right>}}\right>\label{Wmath}\\
	H &=& \left<-\boldsymbol{u}'\cdot\boldsymbol{\omega}' + \frac{\boldsymbol{b}'\cdot\boldsymbol{j}'}{\left<\rho\right>} \right>
\label{prime}
\end{eqnarray}
The modeling of the turbulent transport coefficient by these statistical quantities is
\begin{eqnarray}
\beta &=& C_\beta\tau_t K \label{ModK} \\
\gamma_t &=& C_{\gamma_t}\tau_t W \label{ModW} \\
\alpha &=& C_\alpha\tau_t H \label{ModH}
\end{eqnarray}
with $C_\beta$, $C_{\gamma_t}$ and $C_\alpha$ model constants and $\tau_t$ the timescale of turbulence. The derivation and physical origins of the model expression (\ref{eq:Emf}) have already been presented.\cite{Yokoi4} Note that in the following we will not consider $\alpha$ related term in the electromotive force, even though it may play an important role in two dimensions when an out-of-plane guide magnetic field is considered. In such a case $\boldsymbol{B}\cdot\boldsymbol{J}$ will be the main source for (magnetic) turbulent helicity. If no guide magnetic field is considered, the turbulent helicity $H$ will vanish identically due to mirror symmetry of the current sheet equilibrium used.
\clearpage
\section{Basic equations and numerical implementation}
\label{Basic}
\subsection{Resistive MHD with sub-grid model equations}
\label{equations}
Sub-grid turbulence, as defined under section \ref{Turbulence}, are added to the MHD
equations for a compressible, isotropic plasma and solved consistently with
GOEMHD3 code.\cite{Skala:2014cwa} In order to close the system of equations,
evolution equations for the turbulent energy $K$ and cross-helicity $W$, for which effects of mean-field inhomogeneities are taken into account, are used together with the full system of mean field MHD dimensionless equations (\ref{eq:density})-(\ref{eq:W})
\begin{eqnarray}
	\hspace{-0.5cm} \frac{\partial\rho}{\partial t}&=&-\nabla\cdot(\rho\boldsymbol{U})  \label{eq:density} \\ 
      \frac{\partial\rho \boldsymbol U}{\partial t}&=&-\nabla\cdot\left[ \rho\boldsymbol{U}\otimes\boldsymbol{U}+\frac{1}{2}(p+B^2)\boldsymbol{I} 
- \boldsymbol{B}\otimes\boldsymbol{B}\right]+{\textstyle{\chi}}\nabla^2 \rho\boldsymbol{U}\label{eq:momentum}\\
\frac{\partial\boldsymbol{B}}{\partial t}& =&\textcolor{white}+ \nabla\times\left(\boldsymbol{U}\times \boldsymbol{B}+\langle{{\bf{u}}'\times{\bf{b}}}'\rangle\right)+\eta\nabla^2\bf{B}\label{eq:induction}\\
\frac{\partial h}{\partial t}&=&-\nabla\cdot(h\boldsymbol U)+\frac{\gamma-1}{\gamma h^{\gamma-1}}(\eta\boldsymbol{J}^2+\frac{\rho K}{\tau_t})+\chi\nabla^2h\label{eq:entropy}\\ 
\frac{\partial K}{\partial t}&=&-\boldsymbol{U}\cdot \nabla K -\langle{{\bf{u}}'\times{\bf{b}}}'\rangle\cdot\bf{J}+ \frac{\boldsymbol{B}}{\sqrt{\rho}}\cdot\nabla W - \frac{K}{\tau_t}\label{eq:K}\\
&&\nonumber\\
\frac{\partial W}{\partial t}&=&-\boldsymbol{U}\cdot \nabla W -\langle{{\bf{u}}'\times{\bf{b}}}'\rangle\cdot\bf{\Omega}+\frac{\boldsymbol{B}}{\sqrt{\rho}}\cdot\nabla K - {\textstyle{c_w}}\frac{W}{\tau_t}\label{eq:W}
\end{eqnarray}
where the symbols $\rho$, $\boldsymbol U$,  and $\boldsymbol B$
denote the mean variables, density, velocity, and the magnetic field, respectively. Note that the mean
entropy $h$ is used instead of the internal energy in order to have
the equation in conservative form. The entropy is related to the thermal pressure
by the equation of state $p=2h^\gamma$ for the ratio of specific heats for adiabatic conditions $\gamma = 5/3$. The entropy is conserved
if no turbulence, Joule or viscous heating is present. 
The current density is calculated via the
Ampere law as $\mu_0\boldsymbol J = \nabla\times \boldsymbol B$.
The symbol $\boldsymbol{I}$ denotes
the three-dimensional identity matrix. The molecular resistivity of the
plasma is given by $\eta$.
The parameters $\chi$ are used for stability, they are switched on locally if
the derivative of the associated quantity (for example $\rho\boldsymbol{U}$ in the momentum equation) has a minimum (maximum).
The set of equations (\ref{eq:density})-(\ref{eq:W}) uses dimensionless variables obtained using a typical
length scale $L_0$, a normalizing density $\rho_0$ and magnetic
field strength $B_0$.
The normalization of the remaining variables and parameters is given by
$p_0=\frac{B_0^2}{2\mu_0}$ for the pressure,
$U_{\mathrm A}=\frac{B_0}{\sqrt{\mu_0\rho_0}}$ for velocities, and $\tau_{\mathrm A}$ for the Alfv\'en crossing time over a
distance $L_0$. The current density is normalized by
$J_0=\frac{B_0}{\mu_0L_0}$, the resistivity by
$\eta_0=\mu_0 L_0 U_{\mathrm A}$ and the energy by $E_0=B_0^2L_0^2/\mu_0$.
Finally turbulence quantities are normalised by $K_0=U_{\mathrm A}^2$, $W_0=U_{\mathrm A}^2$ for
the turbulent energy and cross-helicity and by the Alfv\'en crossing time $\tau_A$ for the
turbulence timescale $\tau_t$.
In equations (\ref{eq:K})-(\ref{eq:W}), the production term ($P_K$) of the turbulent energy $K$ and the production
of the turbulent cross-helicity ($P_W$) are given by
\begin{eqnarray}
P_K&=& \tau_t\left(C_\beta K\frac{\boldsymbol{J}^2}{\rho} - C_{\gamma_t}W \frac{\boldsymbol{\Omega}\cdot\boldsymbol{J}}{\sqrt{\rho}}\right) \label{eq:PK}\\
P_W &=& \tau_t\left(  C_\beta K\frac{\boldsymbol{\Omega}\cdot \boldsymbol{J}}{\sqrt{\rho}} - C_{\gamma_t} W\boldsymbol{\Omega}^2\right)\label{eq:PW}
\end{eqnarray}
which shows that the mean current density $\boldsymbol{J}$ and the mean vorticity $\boldsymbol{\Omega}$ play an important role in turbulence production. Since the cross-helicity always have the same sign as the mean vorticity (see \fref{fig:contours}), its generation will always act to reduce the production of turbulent energy.
\subsection{Numerical implementation}
\label{Numerical Implementation}
In order to deal with the non-linearity of the turbulence term in the evolution equations (\ref{eq:K})-(\ref{eq:W}),
a MacCormack scheme (a second-order spatial finite difference method and second-order in time) is used within the GOEMHD3D code instead of a staggered-grid leap-frog scheme.
MacCormack is an explicit two-step approach solving first a predicted value ($Q_{Pr}$) then a corrected one
($Q_{Co}$) and finally use
the mean of these values for the final result. In conservative form, for a variable $Q$
and a flux $F$, the scheme reads
\begin{eqnarray}
\left(Q_{Pr}\right)_{j}^{n+1}&=& Q_{j}^n-\frac{\Delta t}{\Delta y}\left(F^n_{j+1}-F^n_{j}\right)\label{eq:Qpr}\\
\left(Q_{Co}\right)_{j}^{n+1}&=& Q_{j}^n-\frac{\Delta t}{\Delta y}\left(F^n_{j}-F^n_{j-1}\right)\label{eq:Qco}\\
Q^{n+1}_{j}&=& \frac{1}{2}\left(\left(Q_{Pr}\right)_{j}^{n+1}+\left(Q_{Co}\right)_{j}^{n+1}\right)-\Delta t S^n\label{eq:Qend}
\end{eqnarray}
where the variable $j$ stands for spatial coordinates and $n$ for temporal ones. The $S$ term
represents source terms like $\nabla^2\boldsymbol B$, $\eta \boldsymbol J^2$, $\rho K/\tau_t$ and the
diffusive terms proportional to $\chi$.
This form of the scheme is applied to equations (\ref{eq:density})-(\ref{eq:entropy}) which are
written in conservative form plus the source terms. This is unfortunately not the case for equation (\ref{eq:K})-(\ref{eq:W}). There the variables on the right-hand side are treated like source terms without flux and
computed as in (\ref{eq:Qpr})-(\ref{eq:Qend}) but with the $F$ terms set to zero. The source terms are then computed by the predicted values ($S^n_{Pr}$) and corrected ones ($S^n_{Co}$) advanced similarly as (\ref{eq:Qend})
\begin{eqnarray}
\left(Q_{Pr}\right)_{i}^{n+1}&=& Q_{i,j}^n-2 \Delta t S_{Pr}^n\label{eq:QprK}\\
\left(Q_{Co}\right)_{i}^{n+1}&=& Q_{i,j}^n-2 \Delta t S_{Co}^n\label{eq:QcoK}\\
Q^{n+1}_{i,j}&=& \frac{1}{2}\left(\left(Q_{Pr}\right)_{i,j}^{n+1}+\left(Q_{Co}\right)_{i,j}^{n+1}\right)\label{eq:QendK}
\end{eqnarray}
\section{Simulation settings}
\label{Settings}
In the reconnection plane perpendicular to the initial current sheets, dimensions are given by the unit vectors $e_y$ (across the current sheets) and $e_z$ (along the
current sheets), $e_x$ is
the out-of plane direction. The typical length scale is the half width $L_0$ of the sheets.
The box size is $0.4\times 64\times64 L_0^3$ and the number of grid points is $4\times1026\times 1026$. Each current sheet halfwidth is initially resolved by 16 grid points.
In the presentation of the results of the simulations the half width $L_0$ and asymptotic magnetic field $B_0$ are chosen to be equal to 1. The initial conditions
for the turbulent energy $K$ and cross-helicity $W$ are the same in each equilibrium and are
\begin{eqnarray}
K &=& 0.001\label{eq:initK}\\
W &=& 0 \label{eq:initW}
\end{eqnarray}
 
\subsection{Double Harris current sheets}
\label{HarrisSetting}
The initial conditions for the Harris equilibrium are
\begin{eqnarray}
\rho&=& 1\\
\boldsymbol{U}&=&0\\
\boldsymbol{B} &=& B_{0}\left(\tanh{\left(\frac{y+d}{L_0}\right)} - \tanh{\left(\frac{y-d}{L_0}\right)}-1\right)\boldsymbol{e}_z \label{eq:BzHarris}\\
h &=& \frac{1}{2}\left(1+\beta_p-\boldsymbol{B}^2\right)^{\frac{1}{\gamma}}
\end{eqnarray}
where $L_0$ is the current sheet half width, $B_{0}$ is the asymptotic value of the magnetic field and $\beta_p$ is the plasma beta normalized to the asymptotic magnetic field $B_0$. The position of the sheets are given by $\pm d=\pm 16L_0$.  
In order to trigger reconnection, a perturbation is imposed in accordance with $\nabla\cdot \boldsymbol{B}=0$ as
\begin{eqnarray}
-\delta b_{per}\cos(2\pi z/L_z+\pi/2)\sin(2 \pi y/L_y)^2\boldsymbol{e}_y && \label{eq:ByPert}\\
\textcolor{white}-\delta b_{per}\frac{L_z}{L_y}\sin(2\pi z/L_z+\pi/2)\sin(4\pi y/L_y)\boldsymbol{e}_z&&  \label{eq:BzPert}
\end{eqnarray}
where $L_i=(L/L_0)\boldsymbol{e_i}$, for $i=x,y,z$ are the normalized lengths and $\delta b_{per}$ is the
magnitude of the magnetic field perturbation.
\subsection{Force free current sheets}
\label{ForceFreeSettings}
\begin{figure}[t]
  \centering
  \includegraphics[width=0.7\hsize,angle=0]{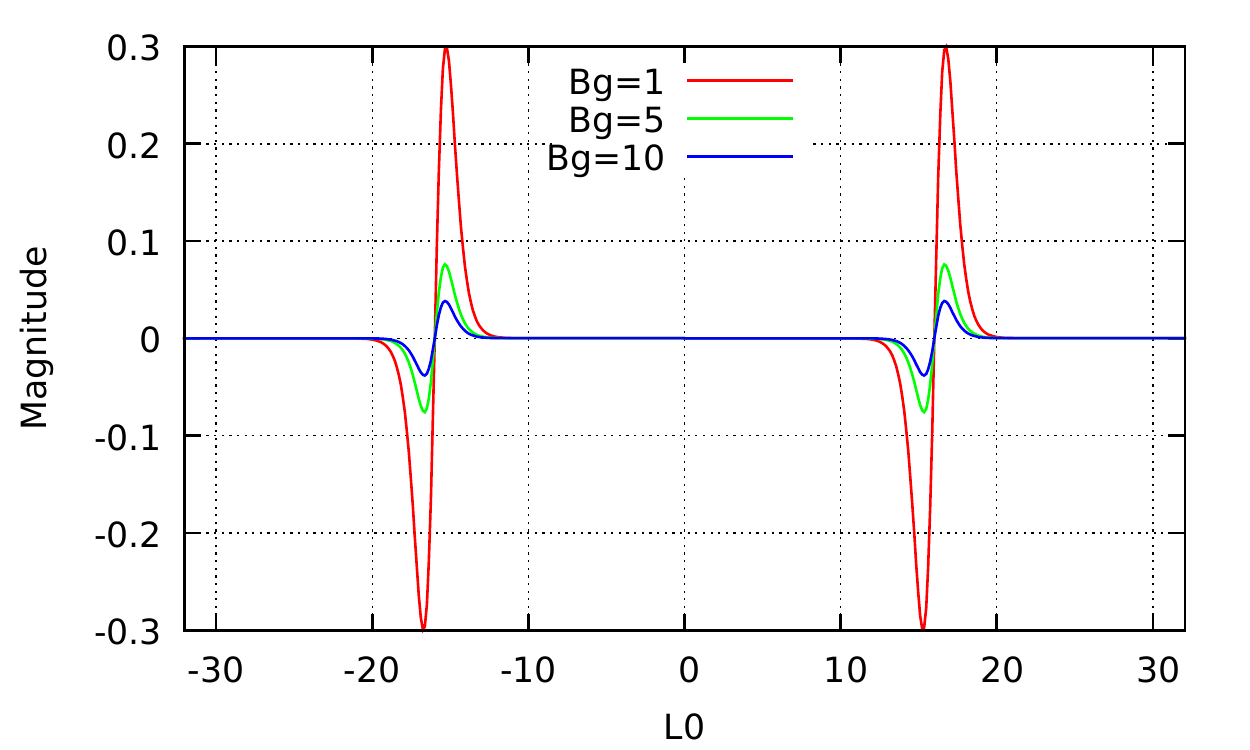}
  \caption{Initial in-plane current for different guide field $b_g$}
  \label{fig:CurrentBG}
\end{figure}

Since for most of the solar corona the plasma beta $\beta_p=2\mu_0p^2/B^2$ is smaller than unity, the field can be considered as force-free. Hence for solar applications it is appropriate to consider force-free current sheets with $\boldsymbol{J}\times\boldsymbol{B}=0$ initially. In particular, we assume the following force-free
current sheet \cite{Pato2015}
\begin{eqnarray}
\rho&=& 1 \label{eq:rhoFF}\\
\boldsymbol{U}&=&0\\
\boldsymbol{B} &=&\textcolor{white}+ B_0 \sqrt{b_g^2 +\cosh^2{\left(\frac{(y+d)}{L_0}\right)}+\cosh^2{\left(\frac{(y-d)}{L_0}\right)}}\boldsymbol{e_x}\label{eq:BxForceFree}\nonumber\\
&&+B_0\left(\tanh{\left(\frac{y+d}{L_0}\right)} - \tanh{\left(\frac{y-d}{L_0}\right)}-1\right)\boldsymbol{e}_z \label{eq:BzHarrisFF}\\
h &=& \frac{1}{2}\left(\beta_p\right)^{\frac{1}{\gamma}} \label{eq:hFF}
\end{eqnarray}
where $b_g$ is the guide field normalised by the asymptotic magnetic field, i.e., $b_g=B_g/B_0$.
To trigger the reconnection process, the same initial current sheets perturbation (\ref{eq:ByPert})-(\ref{eq:BzPert}) is applied as in our investigation of the Harris-type current sheets.
For equilibrium (\ref{eq:rhoFF})-(\ref{eq:hFF}), the initial current density is according to Ohm's law
\begin{eqnarray}
\boldsymbol J = \cosh^2{\left(\frac{y+d}{L_0}\right)}+\cosh^2{\left(\frac{y-d}{L_0}\right)}\boldsymbol{e_x}&&\label{eq:JxFF}\\
+\frac{\tanh\left(\frac{y+d}{L_0}\right)\cosh^{-2}{\left(\frac{y+d}{L_0}\right)}+ \tanh\left(\frac{y-d}{L_0}\right)\cosh^{-2}{\left(\frac{y-d}{L_0}\right)}}{\sqrt{b_g^2+\cosh^{-2}{\left(\frac{y+d}{L_0}\right)}+\cosh^{-2}{\left(\frac{y-d}{L_0}\right)}}}\boldsymbol{e_z}&&\label{eq:JzFF}
\end{eqnarray}
In this model the in-plane current $J\boldsymbol{e_z}$ decreases as the guide field $b_g$ increases (see \fref{fig:CurrentBG}). Note that that the current sheets are around at $\pm16L_0$
\section{Simulation results}
We characterize the evolution of the magnetic reconnection by calculating the reconnected magnetic flux 
\begin{equation}
\frac{\phi}{B_0L_0}=\int\limits_{z_O}^{z_X} \frac{B_y dz}{B_0L_0}\label{eq:Rflux}
\end{equation}
where $z_0$ is the 'O' point location in the center of the magnetic island and $z_X$ is the 'X' point location. The integration over $dz$ is carried out along the center of the current sheet. The reconnection rate is the time derivative of the magnetic reconnected flux (\ref{eq:Rflux}) as
\begin{equation}
M_A = \frac{\partial_t \phi}{(B_0L_0/\tau_a)}\label{eq:Rrate}
\end{equation}
\subsection{Case of a Harris-type current sheet without guide field}
\label{SimHarris}
\begin{figure}[h]
\centering
\begin{tabular}{cc}
	{\includegraphics[width=4.25cm, keepaspectratio]{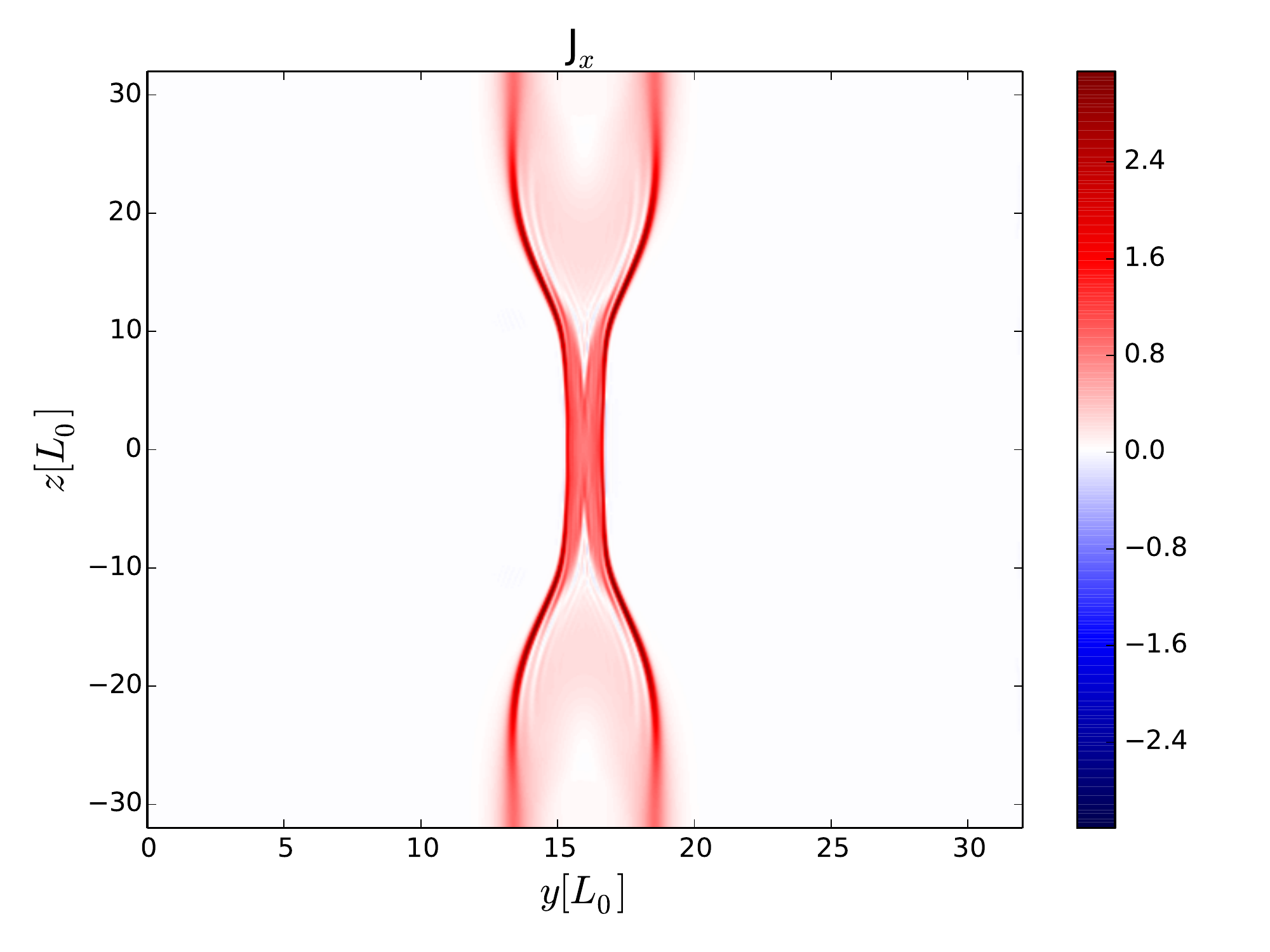}} & {\includegraphics[width=4.25cm, keepaspectratio]{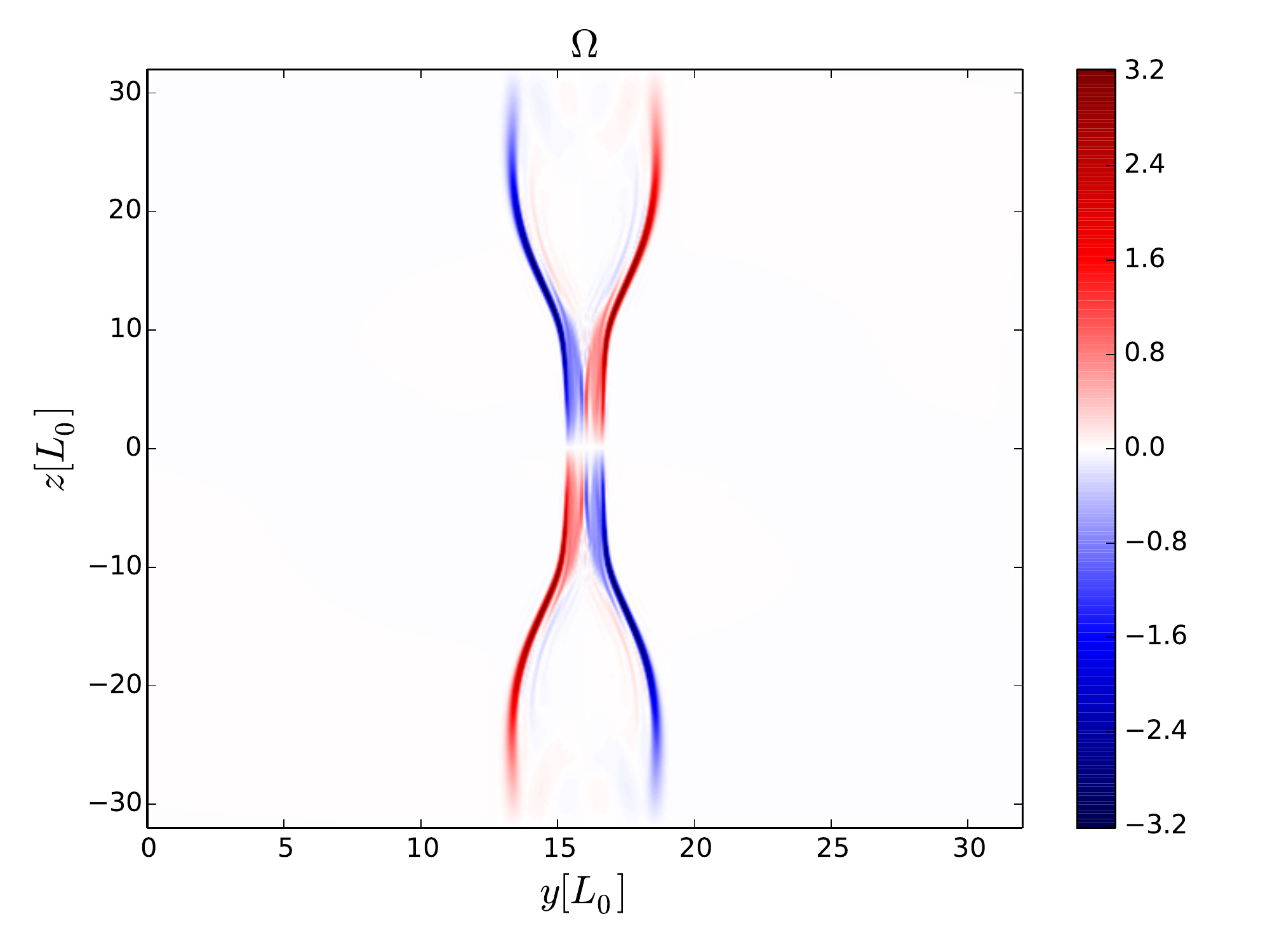}} \\
	(a) \small{Mean current density $J$} & (b) \small{Mean vorticity $\Omega$}\\
	{\includegraphics[width=4.25cm, keepaspectratio]{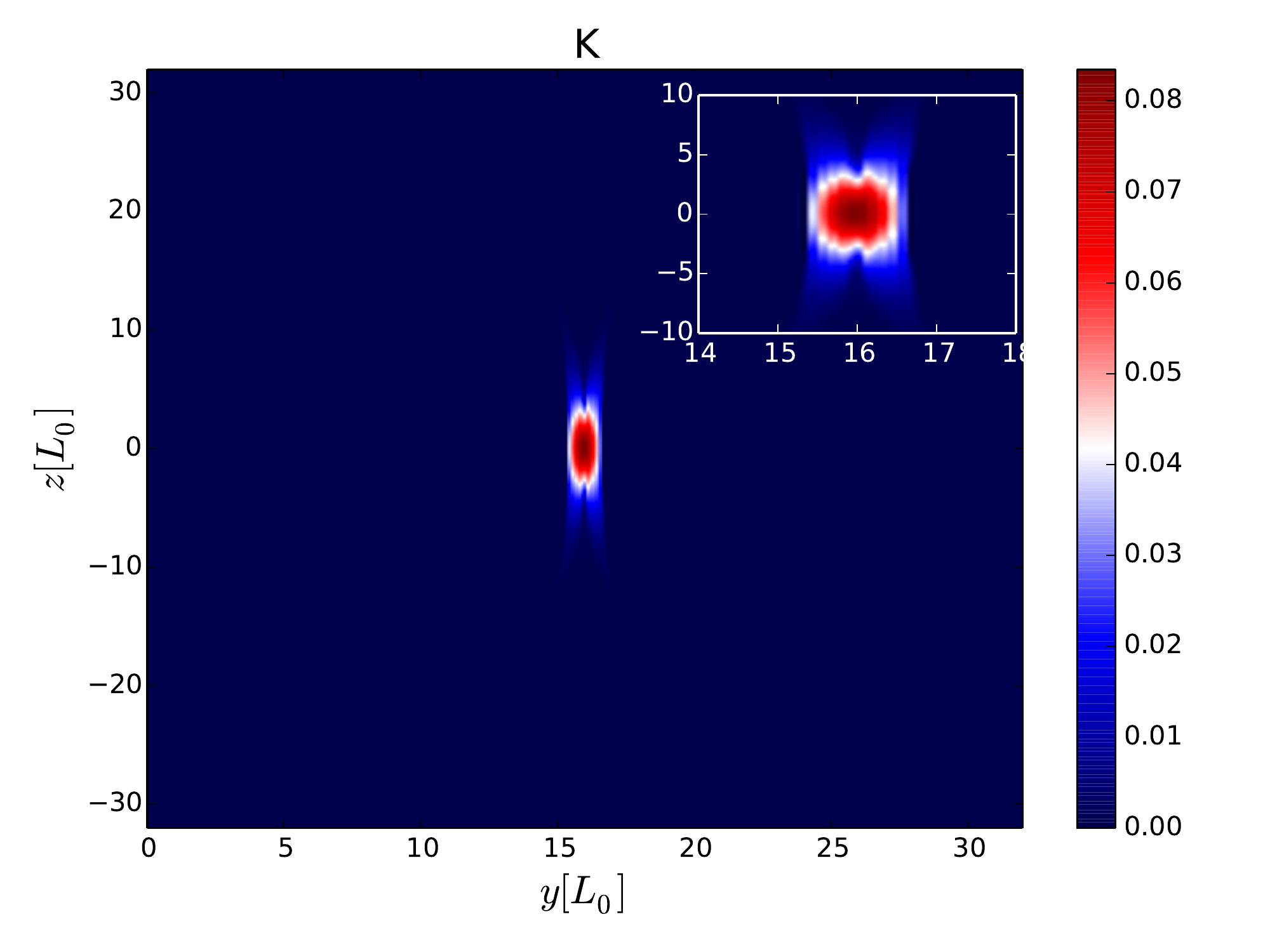}} & {\includegraphics[width=4.25cm, keepaspectratio]{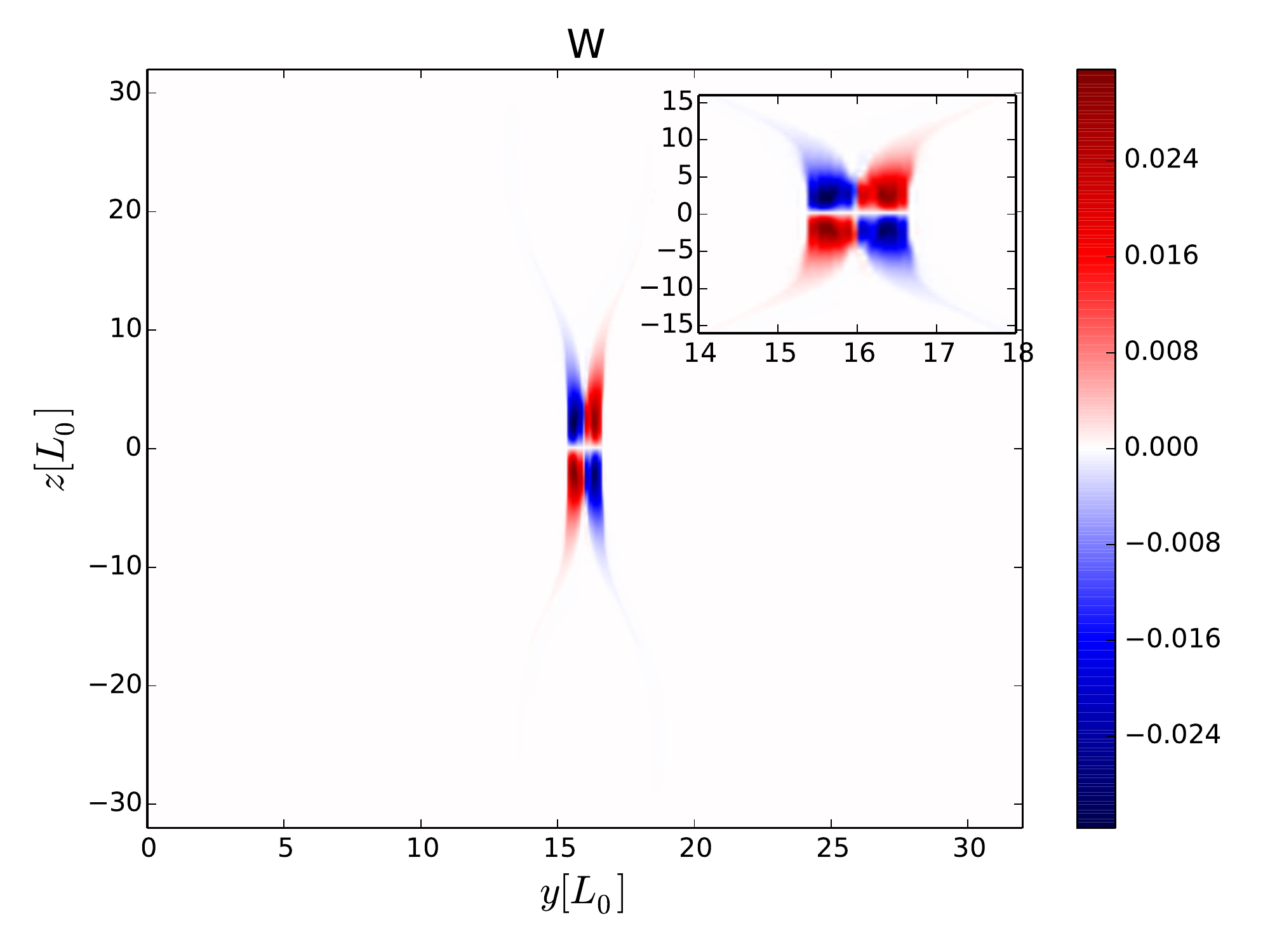}}\\
	(c) \small{Turbulent energy $K$} & \small{(d) Turbulent cross-helicity $W$}\\

    \end{tabular}
    \caption{Spatial distribution of (a) mean current density $J$, (b) mean vorticity $\Omega$, (c) turbulent energy $K$ and (d) cross-helicity $W$ in Harris equilibrium at $t=250\tau_A$ for $\eta=10^{-3}$ and $\tau=1.3$}
  \label{fig:contours}
\end{figure}

\begin{figure}
  \centering
\begin{tabular}{c}
	{\includegraphics[width=5.5cm, keepaspectratio]{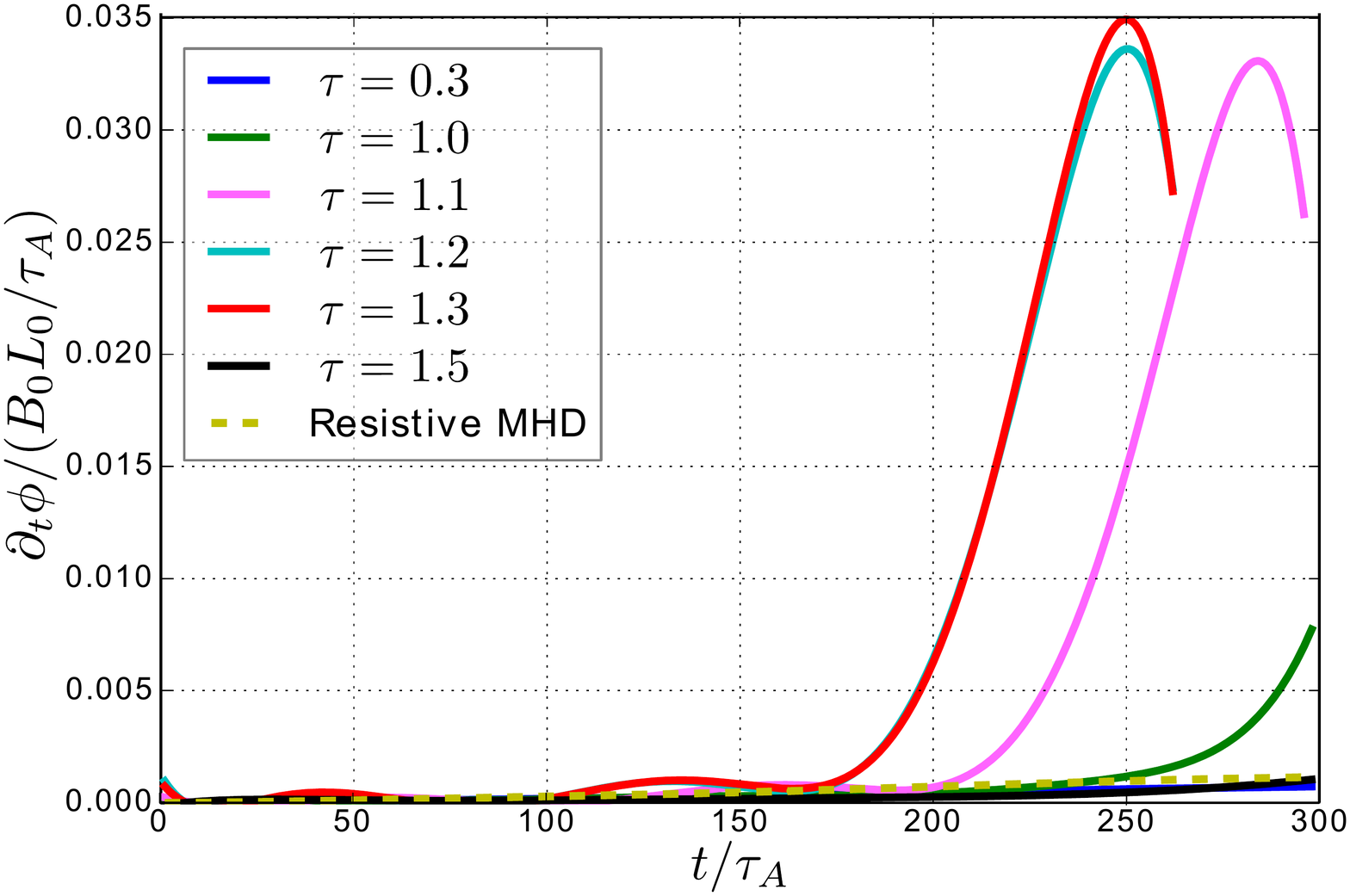}}\\
	(a) \small{Varying $\tau$, $\eta=10^{-3}$}\\
	{\includegraphics[width=5.5cm,keepaspectratio]{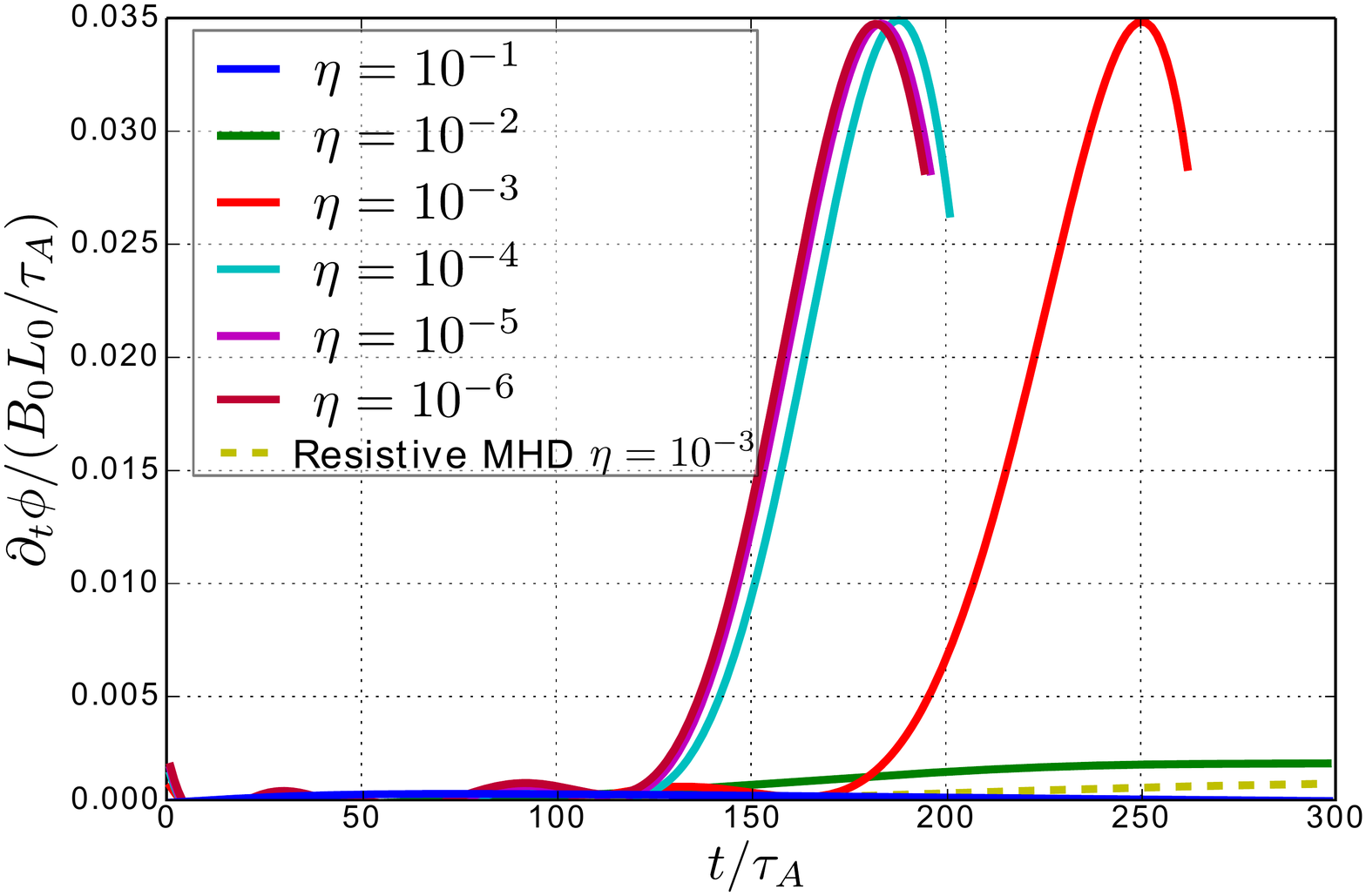}}\\
	(b) \small{Varying $\eta$, $\tau=1.3$} \\
\end{tabular}
\caption{Reconnection rates in time for different parameters}
\label{fig:RecHarrisTime}
\end{figure}

\begin{figure}
  \centering
\begin{tabular}{c}
	{\includegraphics[width=5.7cm, keepaspectratio]{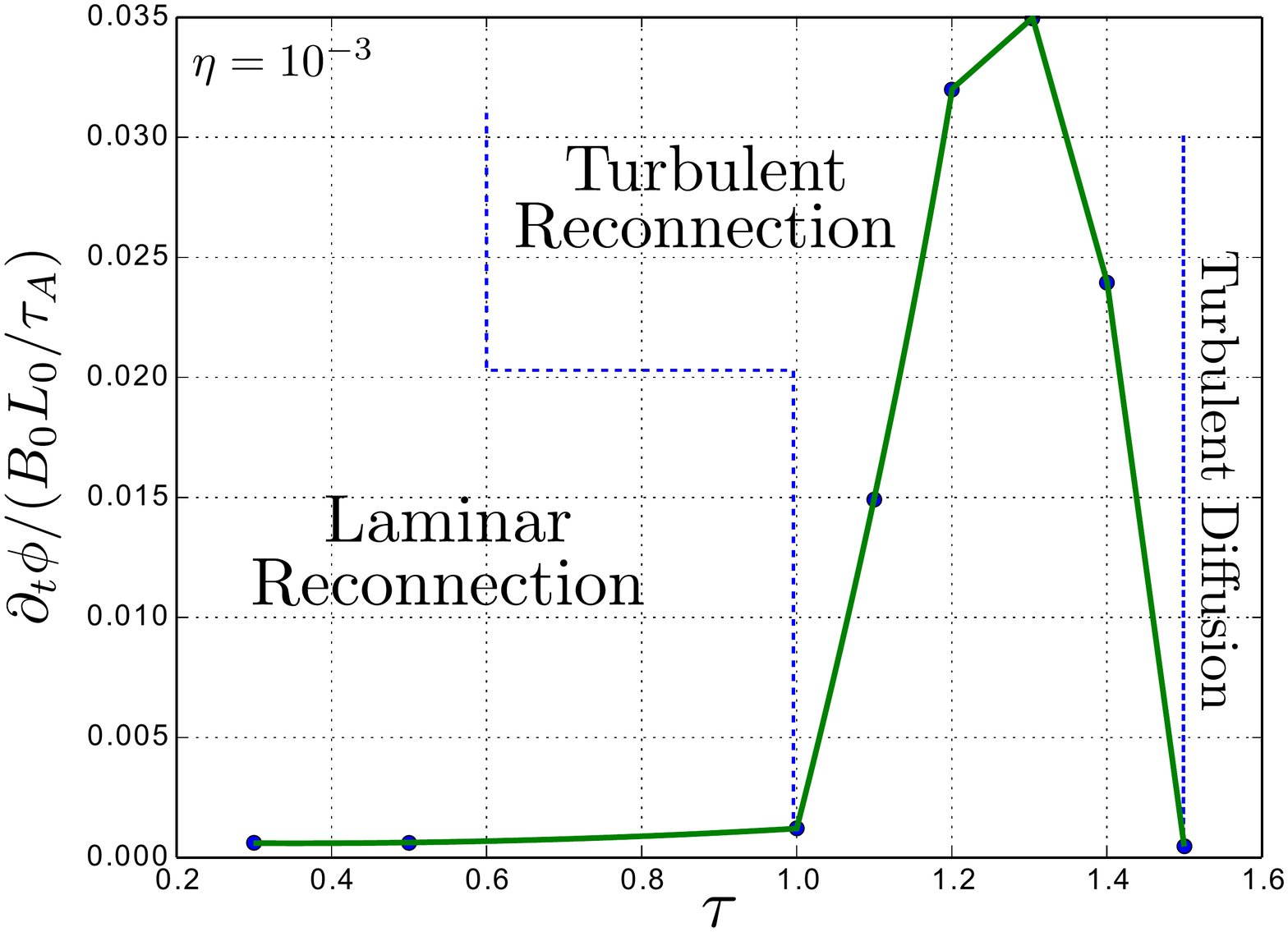}} \\
	(a) \small{Varying $\tau$, $\eta=10^{-3}$} \\
	{\includegraphics[width=5.5cm, keepaspectratio]{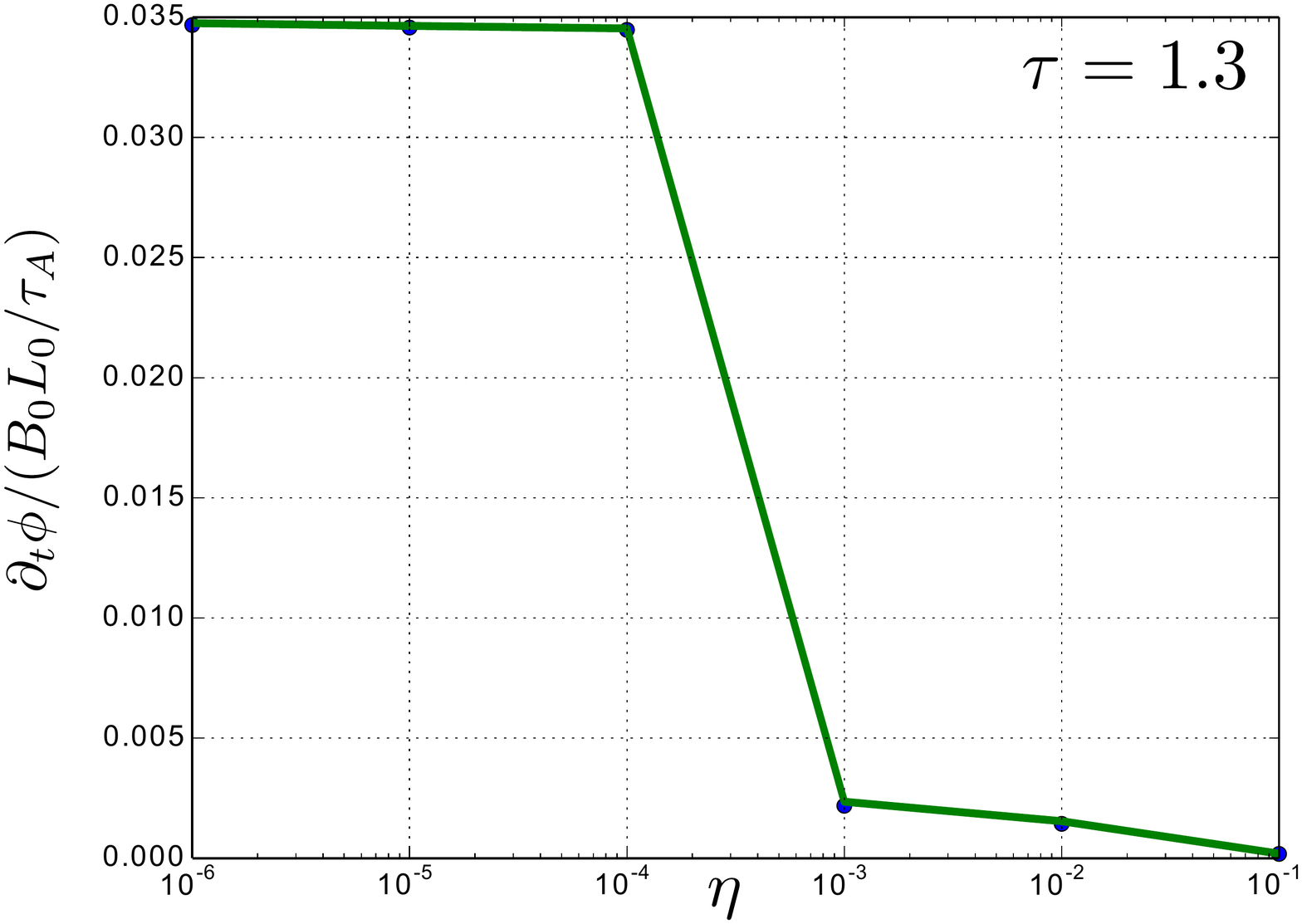}} \\
	(b) \small{Varying $\eta$, $\tau=1.3$} \\
\end{tabular}
\caption{Reconnection rates at $250\tau_A$ for different parameters}
\label{fig:RecHarrisFixed}
\end{figure}

The consequence of varying the turbulence timescale $\tau_t$ is investigated by varying $\tau\equiv\frac{\tau_t}{\tau_0}$, where $\tau_0=\frac{1}{\sqrt{C_\beta}}\frac{\sqrt{\rho}}{\sqrt{\mu_0}}\frac{1}{|\boldsymbol{J}|}$. Study already carried for an Harris equilibrium without guide magnetic field.\cite{Yokoi3} This is done solving equations (\ref{eq:density})-(\ref{eq:W}) with initial turbulence (\ref{eq:initK})-(\ref{eq:initW}). \Fref{fig:contours} is the spatial distribution of the mean current density $\boldsymbol{J}$, mean vorticity $\boldsymbol{\Omega}$
, turbulent energy $K$ and turbulent cross-helicity $W$ taken at the time of the reconnection peak. The turbulent energy $K$ is located around the diffusion
region where the current density accumulates and the cross-helicity as a quadrupole structure around the diffusion region, similarly to the mean vorticity, reflecting its pseudo-scalar nature. It has to be noted that the cross-helicity has the same sign as the mean vorticity. 
It can be seen on \fref{fig:RecHarrisTime} and \ref{fig:RecHarrisFixed} that the turbulence
timescale controls the regime of reconnection, e.g., for $\tau=1.3$ the reconnection process is the
fastest.
Since the turbulent energy acts as a turbulent resistivity localized in the diffusion region (\fref{fig:contours}), different values of the molecular (constant) resistivity are tested which correspond to a variation of the Reynolds number. In our normalisation, the Reynolds number is the inverse of the molecular resistivity.
According to dimensional analysis of the reconnection rate, the Alfv\'en Mach number $M_A$ in the inflow region, according to the Sweet-Parker (SP) model should be proportional to $\sqrt{\eta}$. Assuming a steady state, a similar derivation to the resistive MHD gives
\begin{equation}
M_A^2 = \eta+\beta\left(1+\frac{|\gamma|}{\beta}\eta^{3/2}\right)\label{eq:mach}
\end{equation}
This equation shows that for $\eta\gg\beta$ the reconnection should be similar to an SP current sheet while for $\eta\ll\beta$ turbulence should dominate the reconnection rate as long as the initial level of turbulence is not too large. With a growing turbulence level, the ratio $|\gamma|/\beta\cong 1$ and the reconnection rate is enhanced by both $\beta$ and $\gamma$. For MHD turbulence, it has been shown that by decreasing the molecular resistivity, the rate of reconnection is increased at the nonlinear stage.\cite{Matthaeus_Lamkin_1985} According to (\ref{eq:mach}), for a small molecular resistivity (large Reynolds number) the $\beta$ term should be the leading term for the reconnection rate. \Fref{fig:RecHarrisFixed} (b) shows that indeed a decreased molecular resistivity results in an enhancement of the reconnection rate, limited numerically to 
a minimum of $\eta=10^{-6}$ after which the reconnection rate does change anymore since numerical resistivity associated with the finite grid resolution is reached.
This implies that the reconnection rate is controlled by turbulence in a high Reynolds number plasma.
 
 \begin{figure*}
  \centering
  \begin{tabular}{ccc}
	  {\includegraphics[width=5.5cm, keepaspectratio]{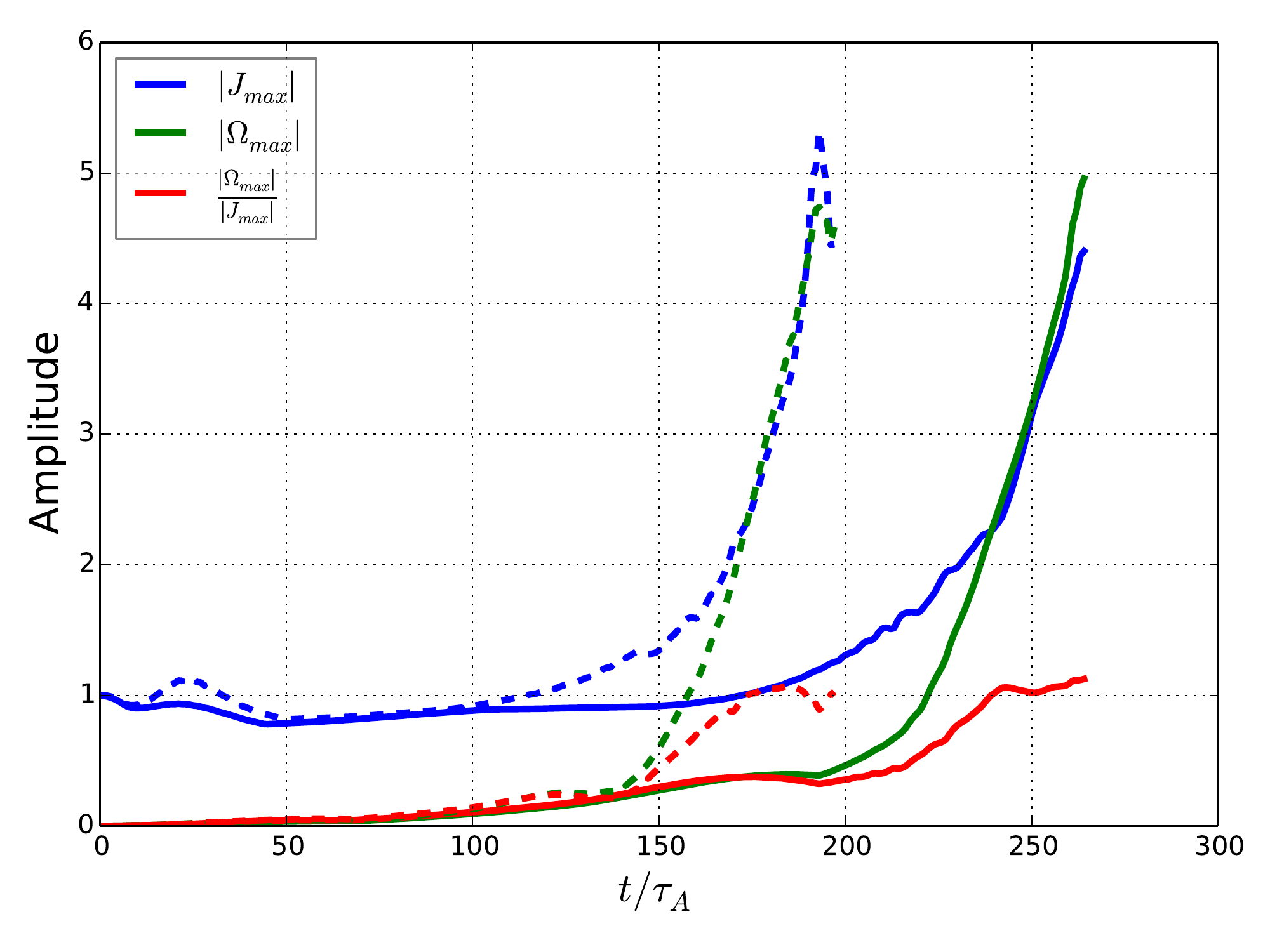}} & {\includegraphics[width=5.5cm, keepaspectratio]{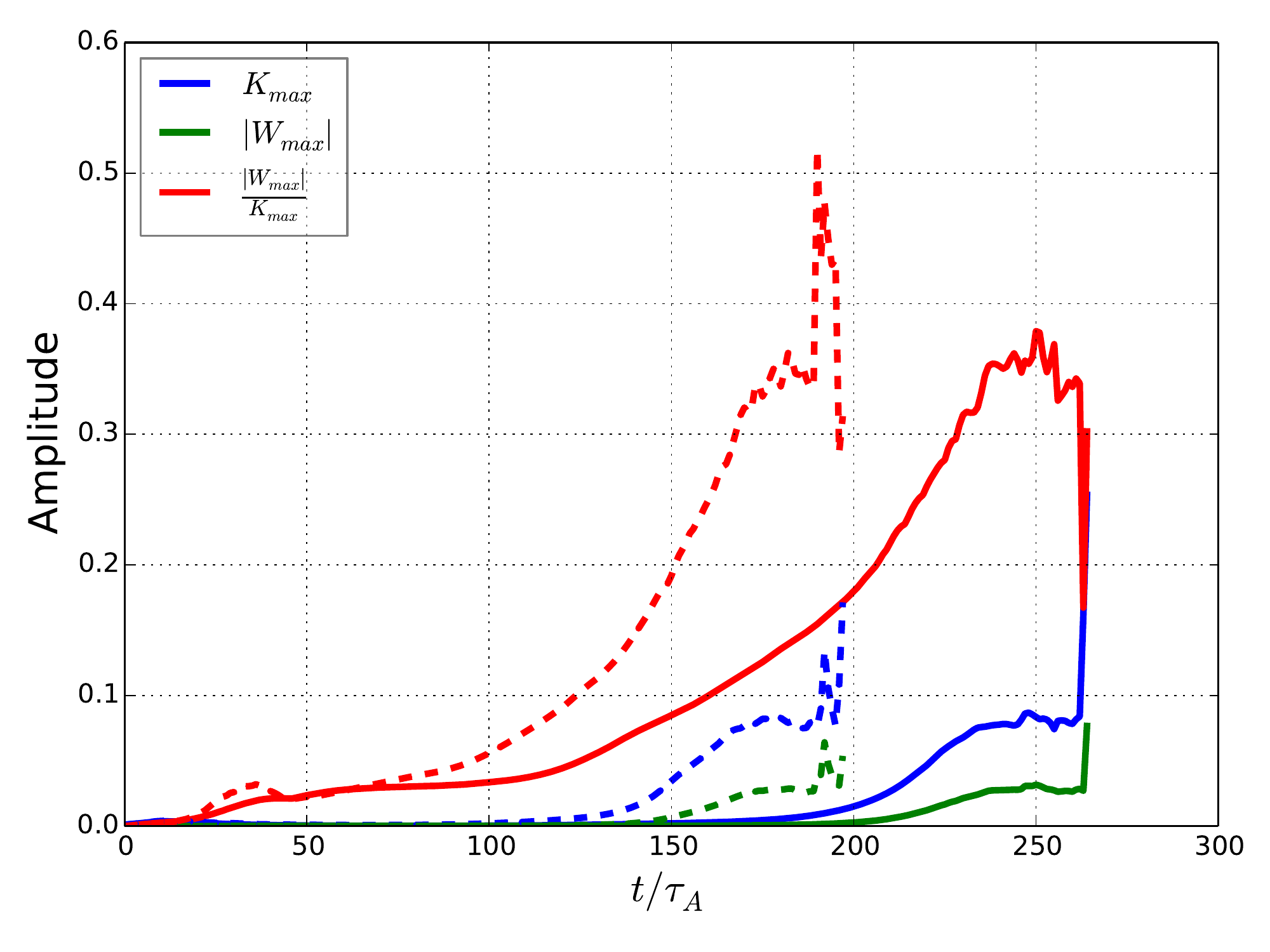}}&{\includegraphics[width=5.5cm, keepaspectratio]{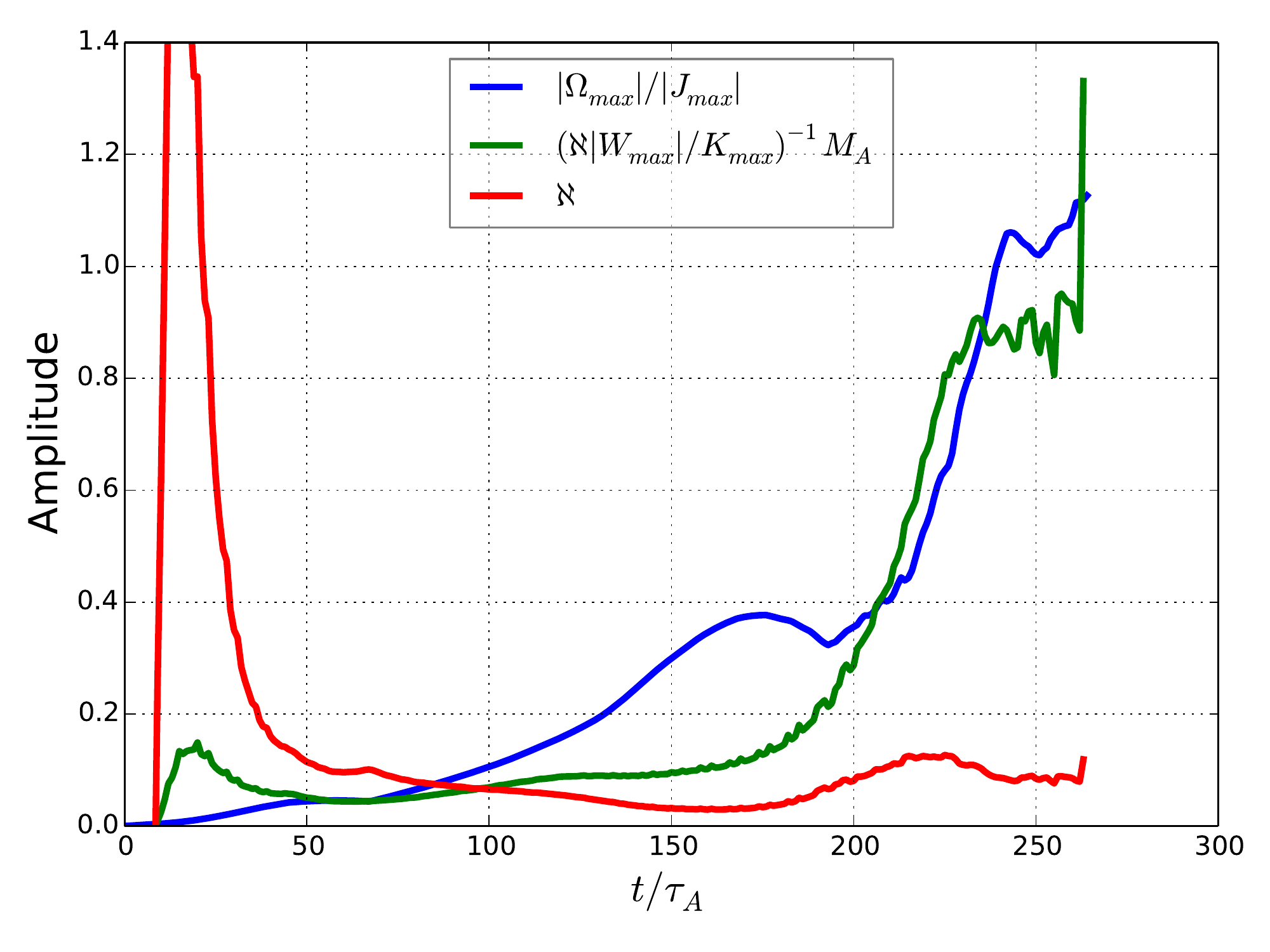}}\\ 
	  (a)  & (b)  & (c)  \\
  \end{tabular}
    \caption{ Relation between turbulence and reconnection. (a) Maximum current density $J$, vorticity $\Omega$ and their ratio. (b) Maximum turbulent energy $K$, cross-helicity $W$ and their ratio. (c) Relation between $|\boldsymbol{\Omega}|/\boldsymbol{J}$ and $|W|/K$. Solid line $\eta=10^{-3}$, dashed lines $\eta=10^{-6}$}
  \label{fig:CompJOKW}
\end{figure*}
The energy spectra discussed in section \ref{sec:Cascade} give an explanation to the earlier growth of the reconnection rate for smaller $\eta$ (larger Reynolds number). At the peak of reconnection more energy is transfered for $\eta=10^{-3}$. This limit explains why
the reconnection rate is slightly higher for $\eta=10^{-3}$ than for smaller values. It has to be noted that the initial turbulence level is in this case $\beta_0=1\cdot10^{-3}$ giving $\eta/\beta_0=1$ initially.
In the limit of $\beta\to\infty$ the system behaves like a perfect conductor. An increase of the Reynolds number by a decrease of the molecular resistivity will also
fasten dynamical processes of the simulation, explaining the earlier occurrence of reconnection in the case of large Reynolds number (but not an increase of its maximum value). These findings are in accordance with previous studies.\cite{Matthaeus_Lamkin_1985} This implies that a higher Reynolds number (smaller molecular resistivity) causes stronger turbulence leading to fast reconnection. The turbulence therefore determines the rate of fast reconnection.
\Fref{fig:CompJOKW} illustrates another interesting result: when the ratio $\boldsymbol{|\Omega|}/\boldsymbol{|J|}\cong 1$ saturates. This corresponds
to reaching a maximum value of the reconnection rate for fixed parameters $\eta$ and $\tau$. This situation also correspond to reaching a maximum value of the ratio $|W|/K$ which represents
the level of turbulence for the model. Since $\boldsymbol{J}^2$, $\boldsymbol{\Omega}^2$ and $\boldsymbol{\Omega}\cdot\boldsymbol{J}$ are
the main sources of the turbulent energy $K$ and the cross-helicity $W$, their saturation will result in a saturation of the turbulence. In the framework of the turbulence model, the equation for the additional magnetic induction $\delta \boldsymbol{B}$ created by the cross-helicity, can be expressed in steady state as\cite{Yokoi2}
\begin{equation}
\delta \boldsymbol{B} = \frac{\gamma_t}{\beta}\boldsymbol{U}=\aleph \frac{W}{K}\boldsymbol{U}
\label{eq:dB}
\end{equation}
For a given large scale (mean) magnetic field $\boldsymbol{B}_0$ and $\aleph$ a model constant of order $\mathcal{O}(10^{-1})$. Supposing that the mean current density $\boldsymbol{J}$ is the source for the additional magnetic induction $\delta\boldsymbol{B}$ across the current sheet, the ratio can be written as
\begin{equation}
\frac{|\boldsymbol{\Omega}|}{\boldsymbol{J}}\cong \frac{\boldsymbol{U}}{\delta\boldsymbol{B}}\frac{\Delta}{L}
\end{equation}  
where $\Delta$ is the width of the diffusion region and $L$ its length. Using expression (\ref{eq:dB}), one can write
\begin{equation}
\frac{|\boldsymbol{\Omega}|}{\boldsymbol{J}}\cong \left(\aleph \frac{W}{K}\right)^{-1}\frac{\Delta}{L}
\label{eq:OJKWMa}
\end{equation}
Since the reconnection rate for an Harris-type current sheet is given by the aspect ratio $\Delta/L$,
the ratio $|\boldsymbol{\Omega}|/\boldsymbol{J}$ together with the value of the reconnection rate gives
an information about the amount of turbulence present in the system in the framework of the RANS model. Using the values
at the peak of reconnection from \fref{fig:RecHarrisFixed} (b) and for the ratio $|W|/K$ in (\ref{eq:OJKWMa}) (\fref{fig:CompJOKW})
, for a molecular resistivity $\eta=10^{-3}$ and $\tau=1.3$, results in, approximatively, 1 which is the value of $|\boldsymbol{\Omega}|/\boldsymbol{J}$ seen in \fref{fig:CompJOKW} when the reconnection rate reaches a maximum. \Fref{fig:CompJOKW} (c) shows the time history of this
comparison for $\aleph=0.1$. The value of $\aleph$ determined by (\ref{eq:OJKWMa}) is also plotted in the figure, it has very large initial value since the mean vorticity is small during the first $t=50\tau_A$. It confirms that the model constant should be of the order $\mathcal{O}(10^{-1})$.

\subsection{Effect of the initial amount of turbulence}
\label{Sec:Tearing}
\begin{figure}[h]
  \centering
\begin{tabular}{c}
	{\includegraphics[width=5.5cm, keepaspectratio]{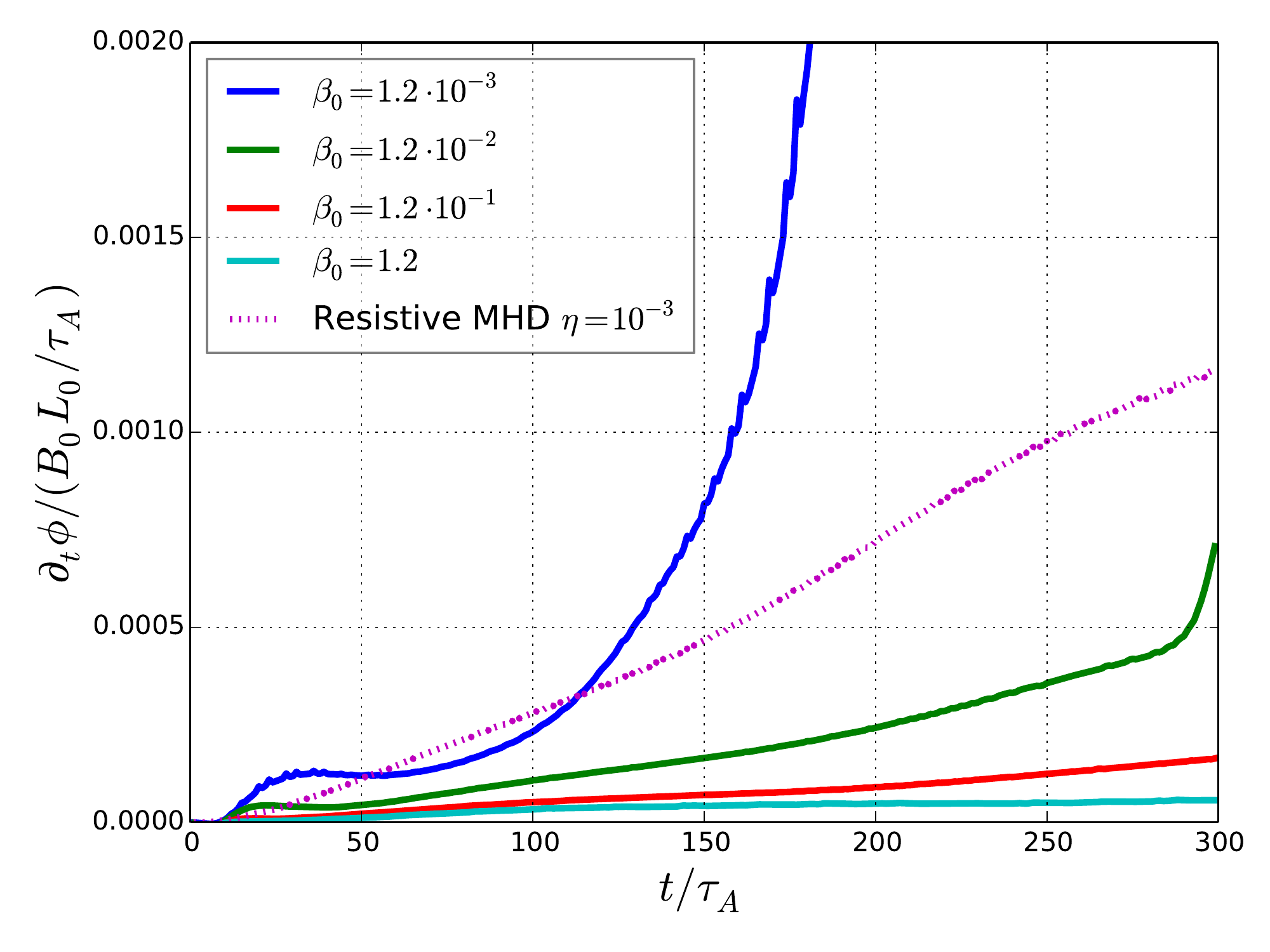}} \\ 
	(a) \small{Time dependence} \\
	{\includegraphics[width=5.5cm, keepaspectratio]{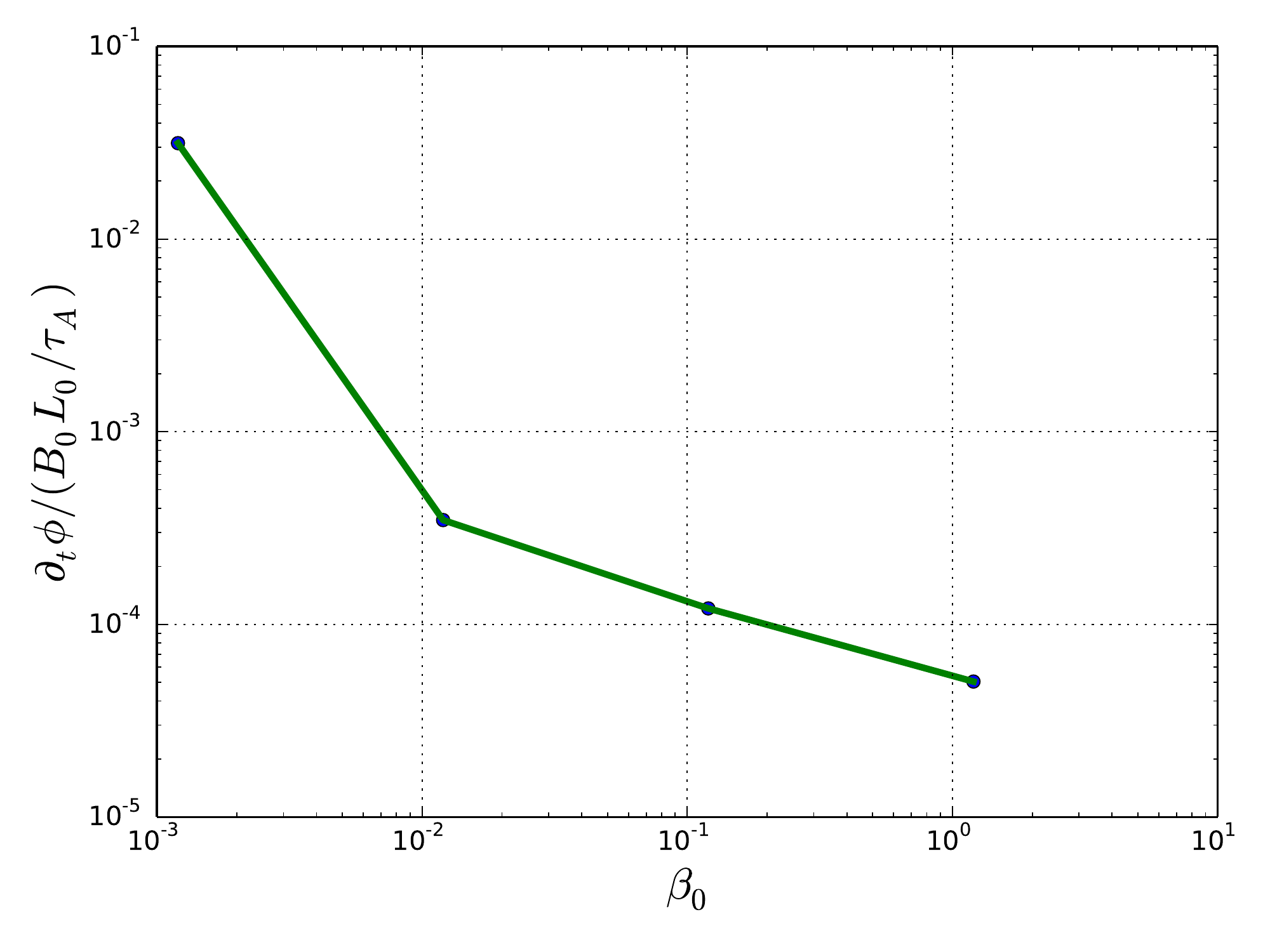}}\\
	(b) \small{Dependence at  $300\tau_A$}\\
\end{tabular}
\caption{Reconnection rates of the Harris equilibrium without guide magnetic field varying $\beta_0$ for $\tau=1.3$ }
\label{fig:RecHarrisBeta0}
\end{figure}

The rate of magnetic reconnection is highly dependent on the ratio $\eta/\beta$, it is increasing for low values
of the molecular resistivity. This can be understood by the energy spectrum at different times of the reconnection process as discussed in section \ref{sec:Cascade}.
\begin{figure}[h]
\centering 
 \includegraphics[width=0.4\textwidth,width=0.4\textwidth,keepaspectratio]{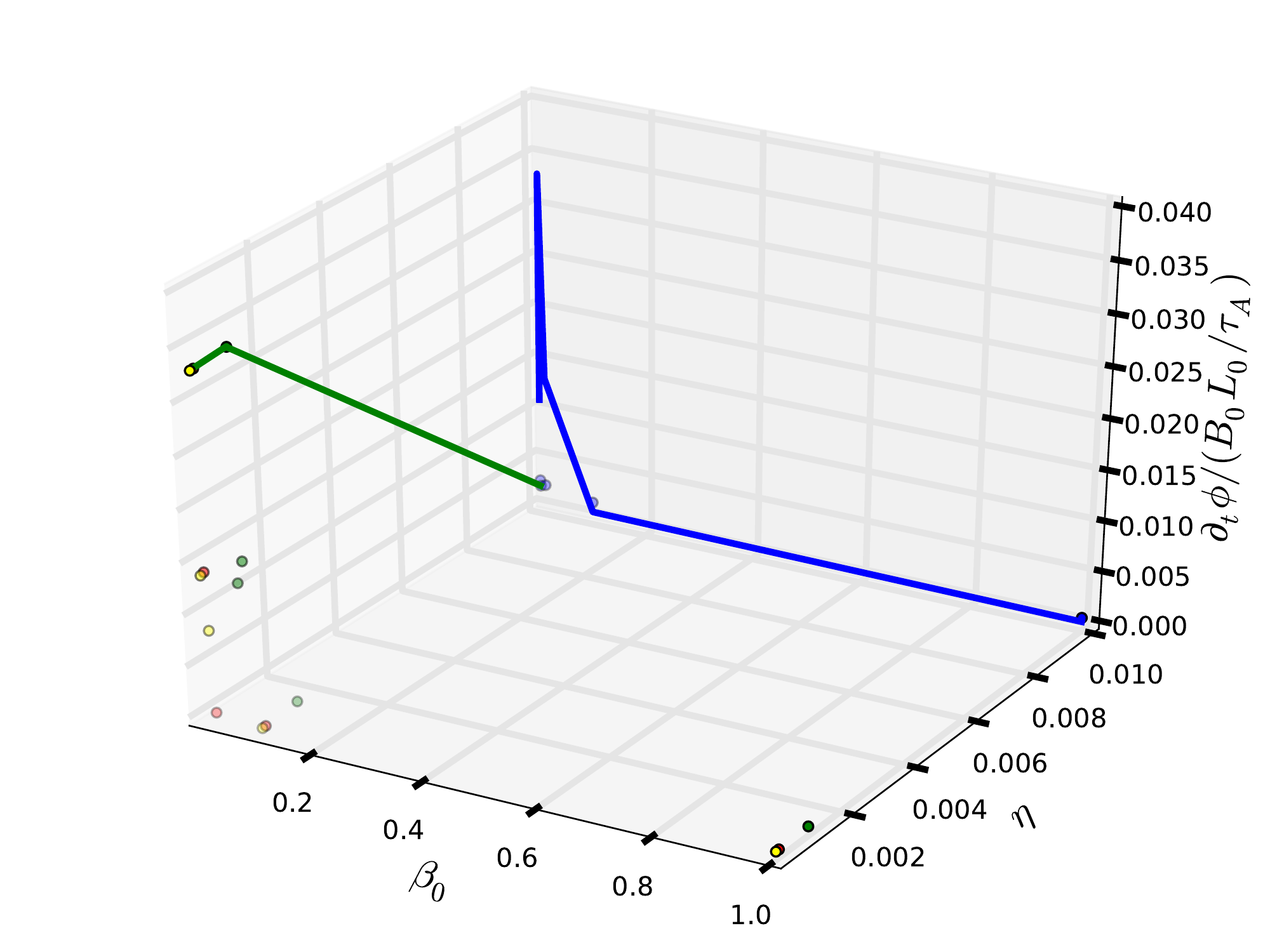}
 \caption{Maximum value of the reconnection rate as a function of $\beta_0$ and $\eta$ for the Harris equilibrium ($b_g=0$)}
  \label{fig:RRetabeta3d}
\end{figure}
\Fref{fig:CascadeEtasHarrisStage} show that before
the reconnection peak is reached for $\eta=10^{-3}$ not as much energy is transfered to smaller scales as in the case of smaller $\eta$. However, a large initial value of turbulence, represented by the quantity $\beta_0=C_\beta\tau K_0$, will delay or even suppress the onset of reconnection (\fref{fig:RRetabeta3d}). A large value of turbulence
in the system can be given either by a large $\tau$ or a large $K_0$. If at the initial state too much turbulence is imposed, then no reconnection takes place and magnetic diffusion dominates in similar situation as large $\tau$ (\fref{fig:RecHarrisBeta0}) while the reconnection rate reaches a maximum when $\eta=\beta_0$.
In the present work the constant turbulent timescale $\tau$ and turbulent energy $K$ are treated as separate quantities and can so independently produce fast reconnection or \textit{turbulent diffusion}. However, the algebraic relation between the turbulent timescale and the turbulent energy explain the similar behavior obtained by either a large $\tau$ or large $K_0$. The regimes obtained by variation of $K_0$ is only an artefact of the algebraic model for the turbulence timescale. A more realistic model thus requires that the timescale of turbulence is determined by turbulence itself, by solving, for example, evolution equations of quantities representing a turbulence timescale such as $K_0$ and $\epsilon$. In such a case, it is clear that only the turbulence timescale, determined by $K_0/\epsilon$, will distinguish if \textit{laminar reconnection}, \textit{turbulent reconnection} or \textit{turbulent diffusion} will take place.

\subsection{Case of Harris-type equilibrium with finite external guide field}
\begin{figure}[h]
  \centering
  \includegraphics[width=4.5cm, keepaspectratio]{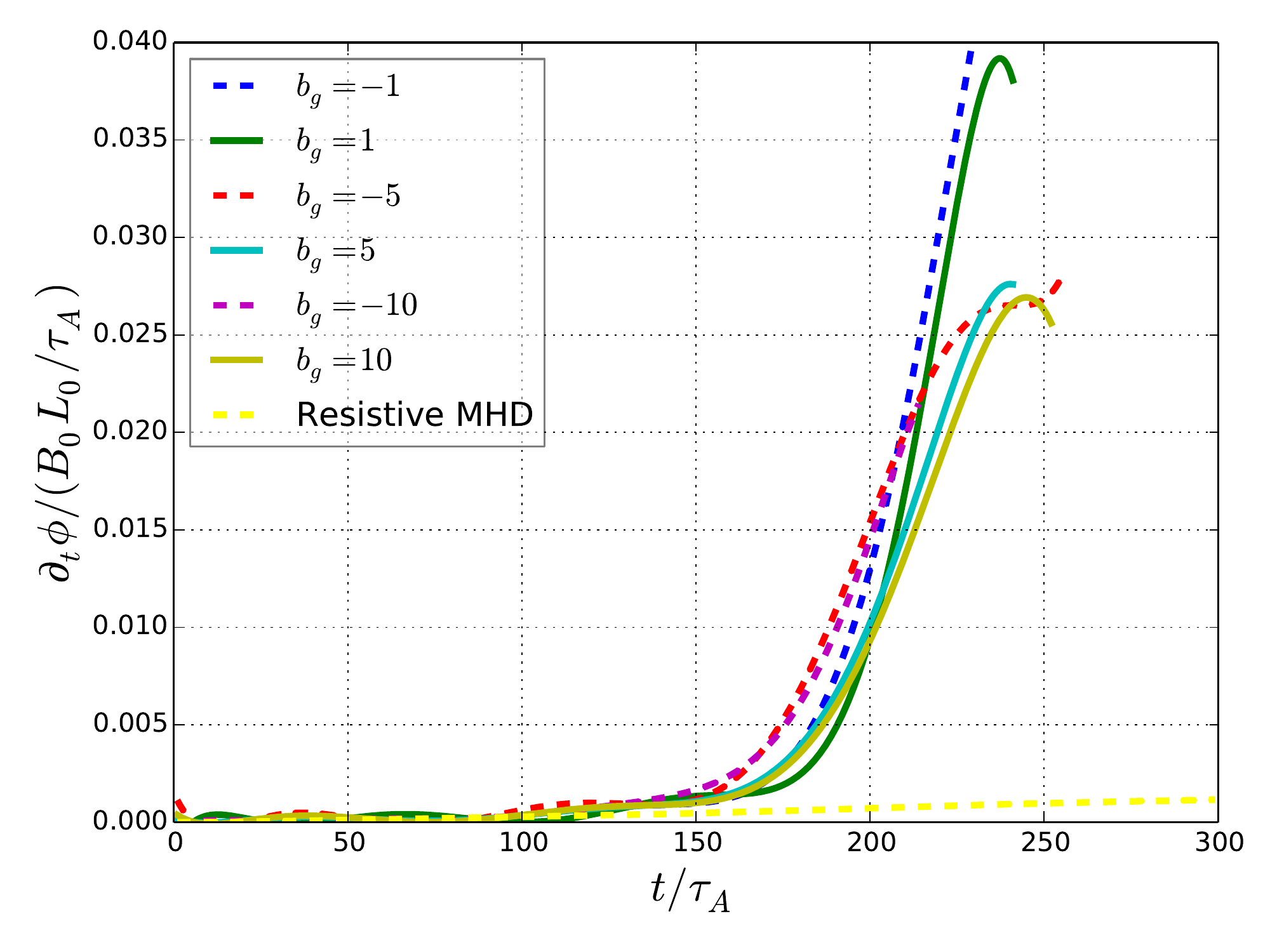}
  \caption{Reconnection rate for different value of the guide field}
  \label{fig:HarrisBGOPP}
\end{figure}
\begin{figure}
	\centering
	\begin{tabular}{cc}
		\hspace{-0.5cm}{\includegraphics[width=4.5cm, keepaspectratio]{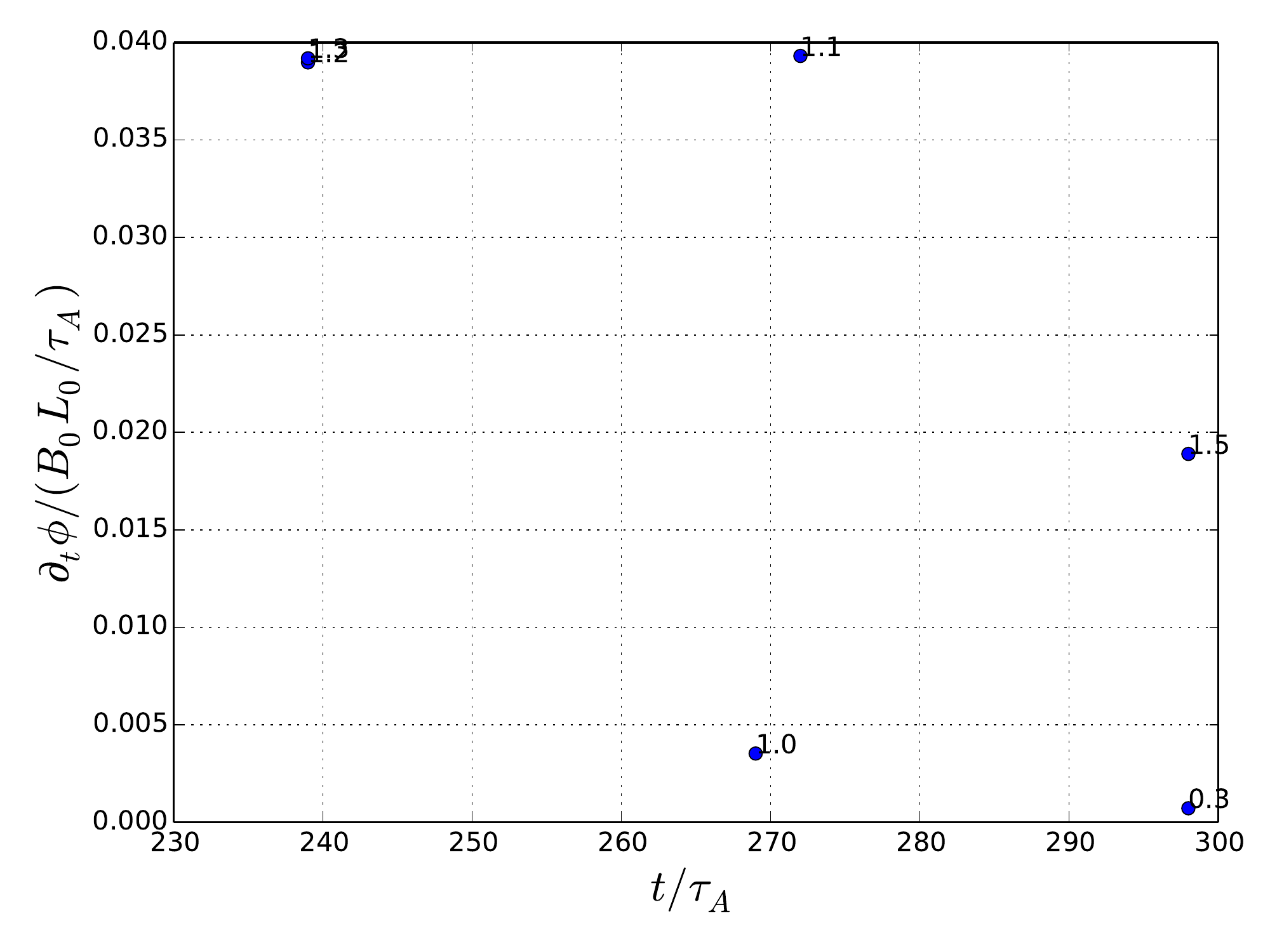}} & {\includegraphics[width=4.5cm, keepaspectratio]{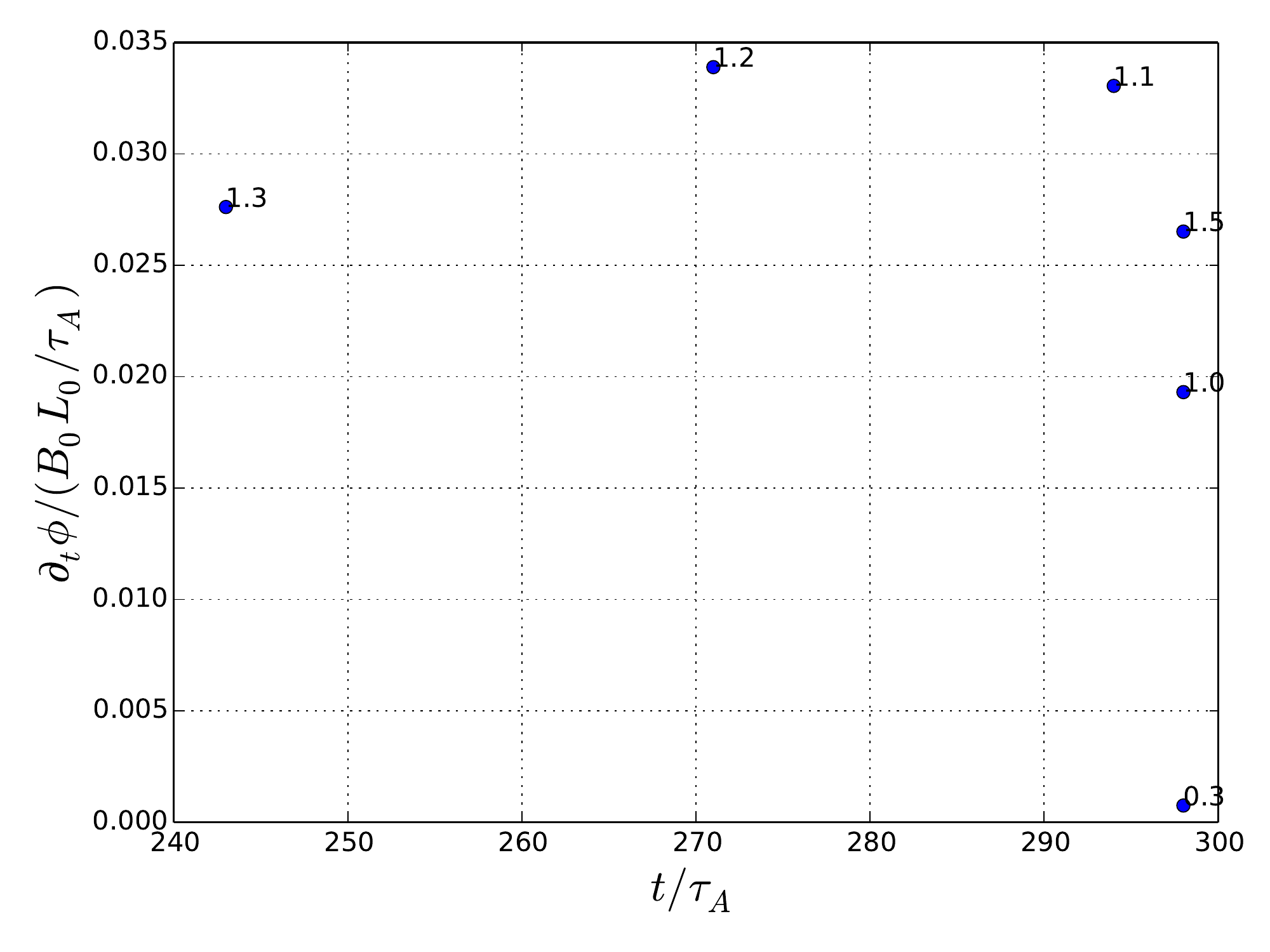}}\\
		(a) $b_g=1$ & (b) $b_g=5$ \\
	\hspace{-0.5cm}{\includegraphics[width=4.5cm, keepaspectratio]{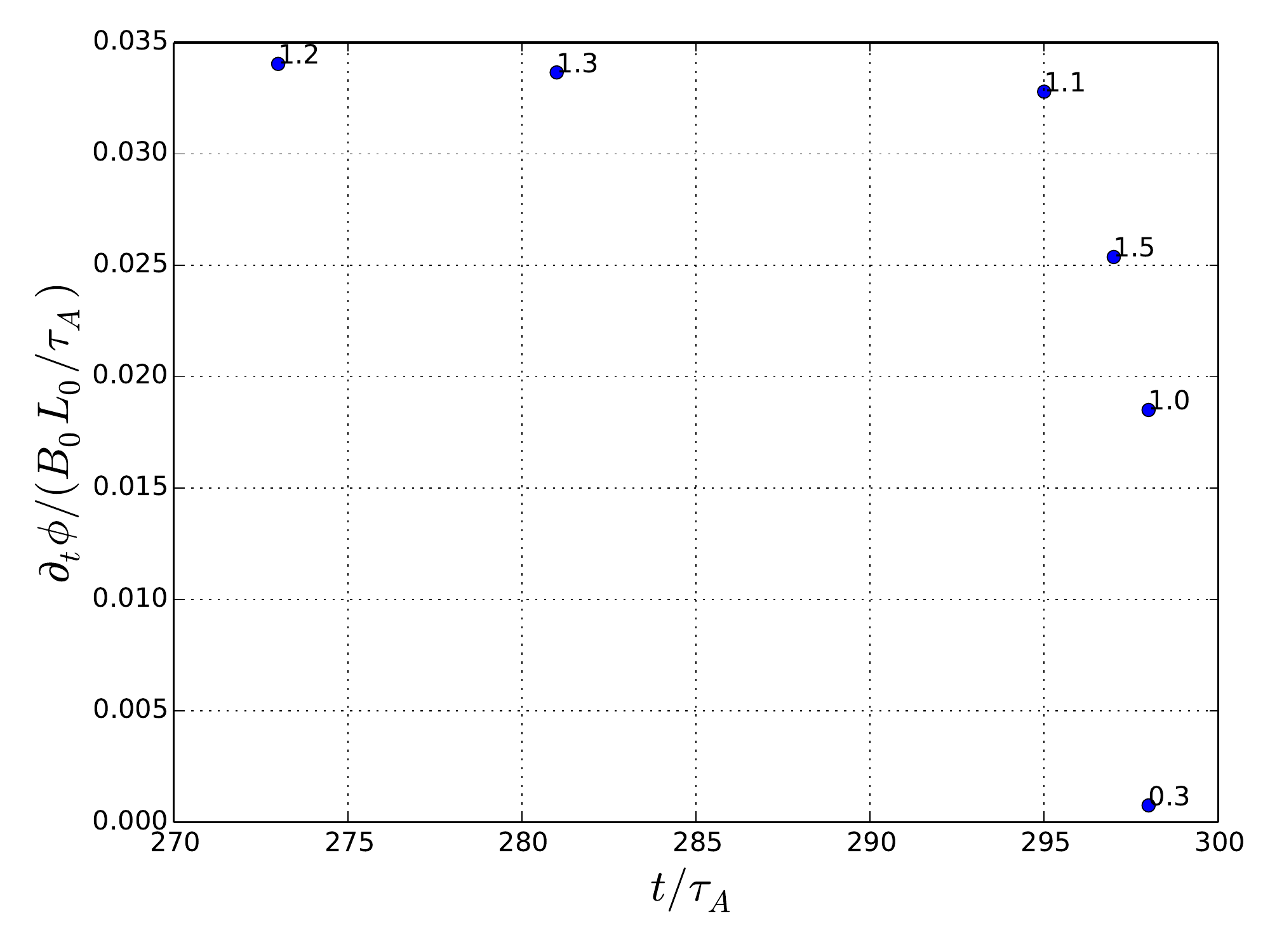}}&\\
		(c) $b_g=10$ \\
	\end{tabular}
\caption{Maximum reconnection rate in time for different guide magnetic field $b_g$ in Harris equilibrium }
\label{fig:HarrisBGAll}
\end{figure}

Since the addition of an external constant guide field $b_g=B_g/B_0$ does not change the equilibrium properties of an Harris-type current sheet, the effect of an out-of-plane guide magnetic field is tested on turbulence. Three different
values of the guide field are considered and their sign is also changed such that $b_g\cdot \boldsymbol{J}<0$ and $b_g\cdot \boldsymbol{J}>0$.
When the guide field is co-aligned to the mean current density, reconnection is affected in two ways. On one hand the maximum value reached is smaller and on the other hand the time needed for the system to reach the maximum value of the reconnection rate is increased (\fref{fig:HarrisBGAll}). As similar situation is obtained for an anti-aligned guide magnetic field even though it produces faster reconnection than an co-aligned one (\fref{fig:HarrisBGOPP}) 
Following the same procedure as for the derivation of equation~(\ref{eq:mach}), the reconnection rate (Alfv\'en Mach number) can be rewritten as
\begin{equation}
M_A^2=\eta+\beta\left(1+\frac{|\gamma|}{\beta}\eta^{3/2}\right) -\alpha\sqrt{\eta}
\label{eq:MachA}
\end{equation}
Equation~(\ref{eq:MachA}) shows how depending on the sign of $\alpha$, the reconnection rate is expected to be enhanced or decreased depending on the strength and the alignment on the guide magnetic field $b_g$ with the mean current density $\boldsymbol{J}$. Note that for a larger Reynolds number, the influence of the turbulent magnetic helicity on the reconnection rate would be decreased. The effect of a strong guide field in two dimensions will therefore,
as we showed using a RANS theory, play a important role in magnetic reconnection.

\subsection{Force free current sheets}
\label{SimFF}
\begin{figure}[h]
\centering
	\begin{tabular}{cc}
		\hspace{-0.5cm}{\includegraphics[width=4.5cm, keepaspectratio]{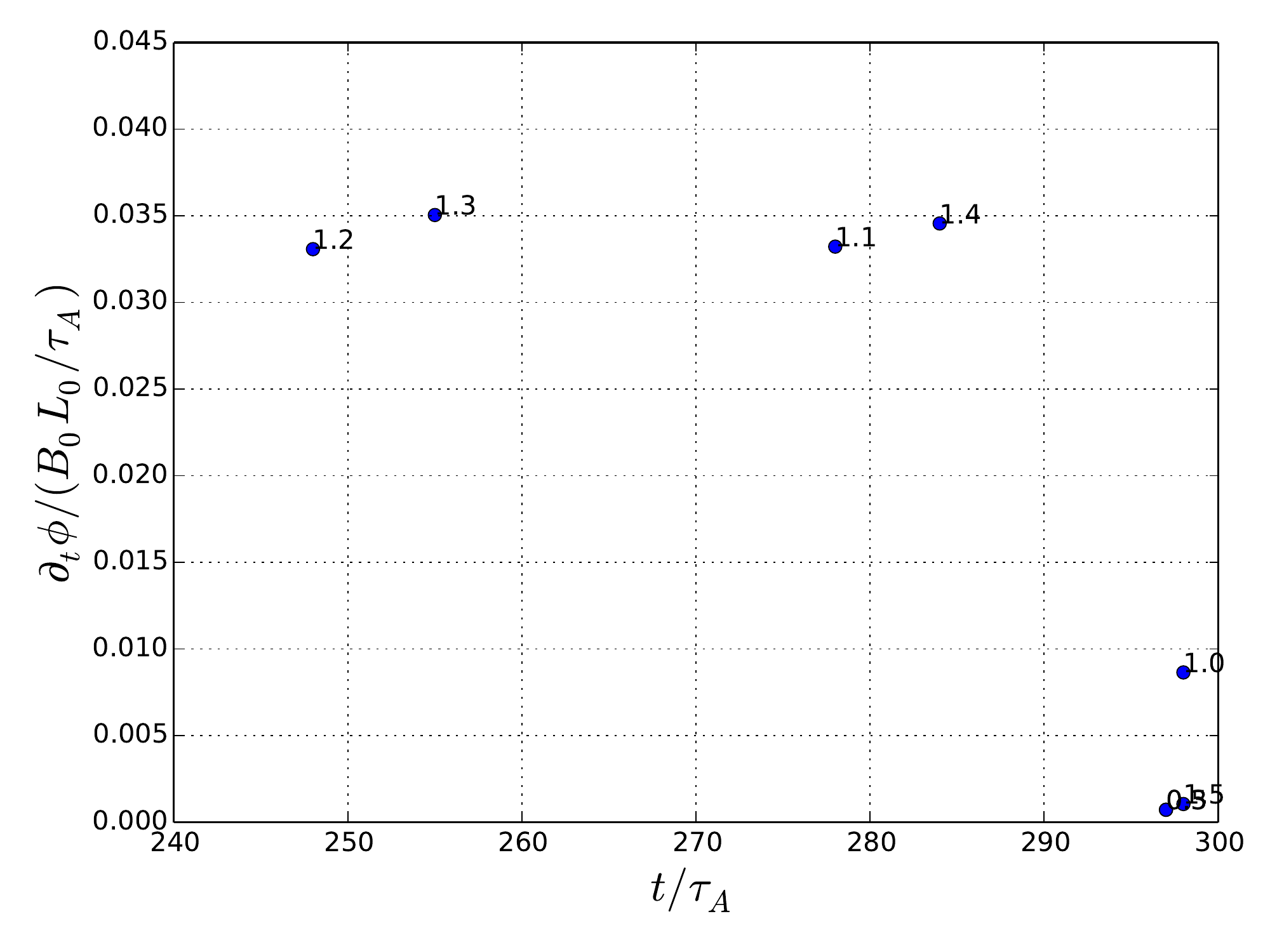}} & {\includegraphics[width=4.5cm, keepaspectratio]{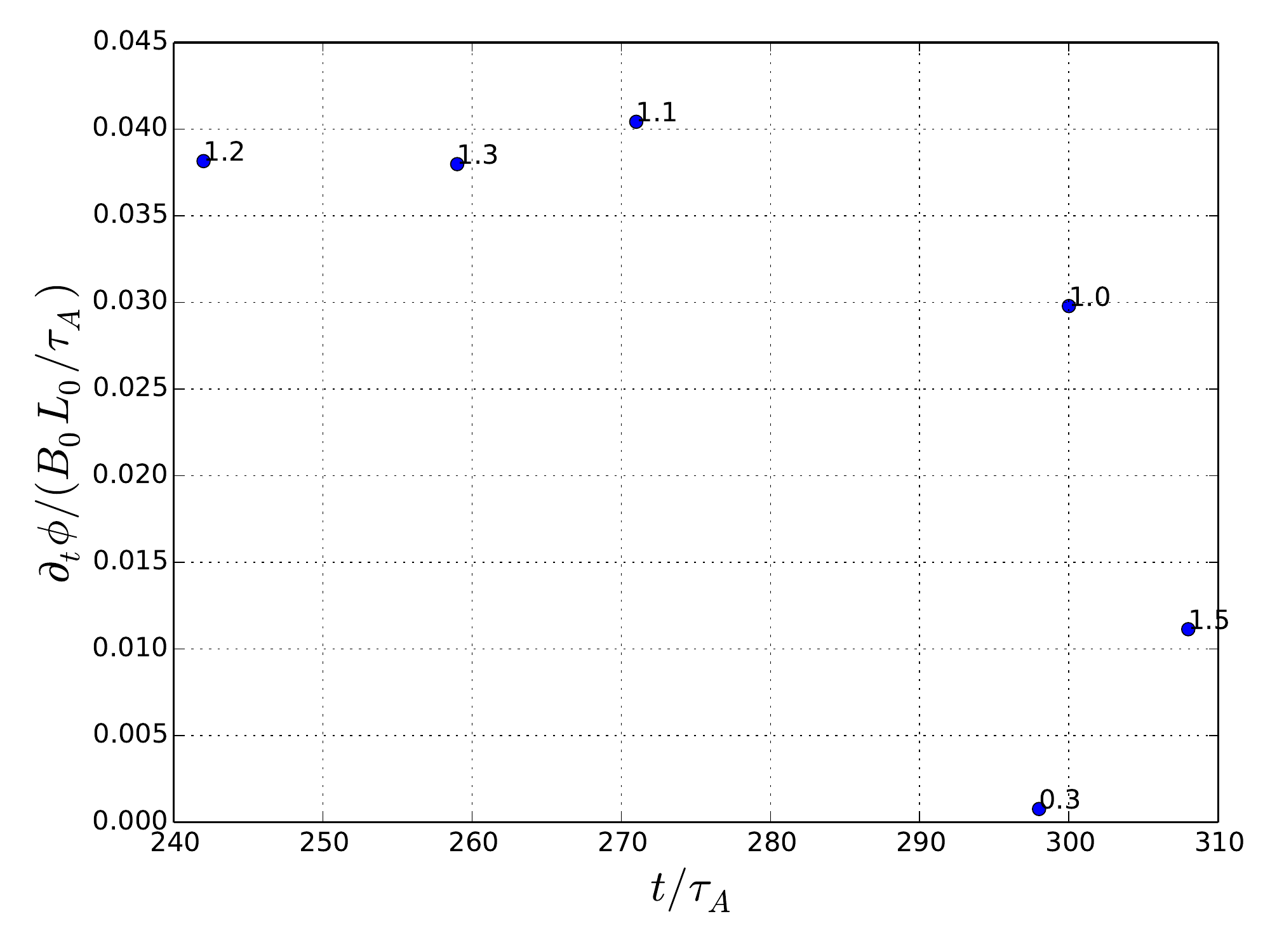}} \\
		(a) Harris & (b) Force free $b_g=1$ \\
		\hspace{-0.5cm}{\includegraphics[width=4.5cm, keepaspectratio]{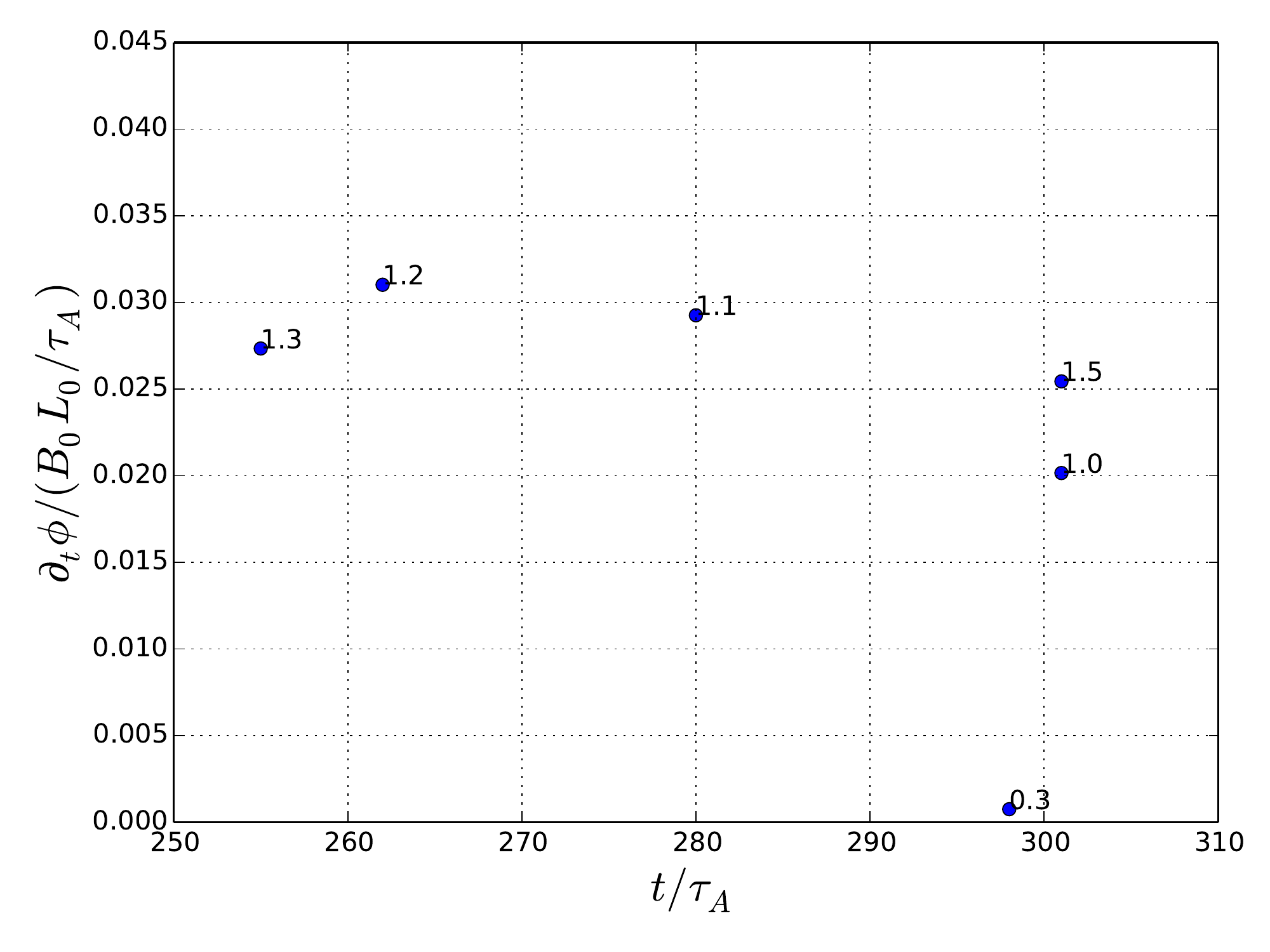}} & {\includegraphics[width=4.5cm, keepaspectratio]{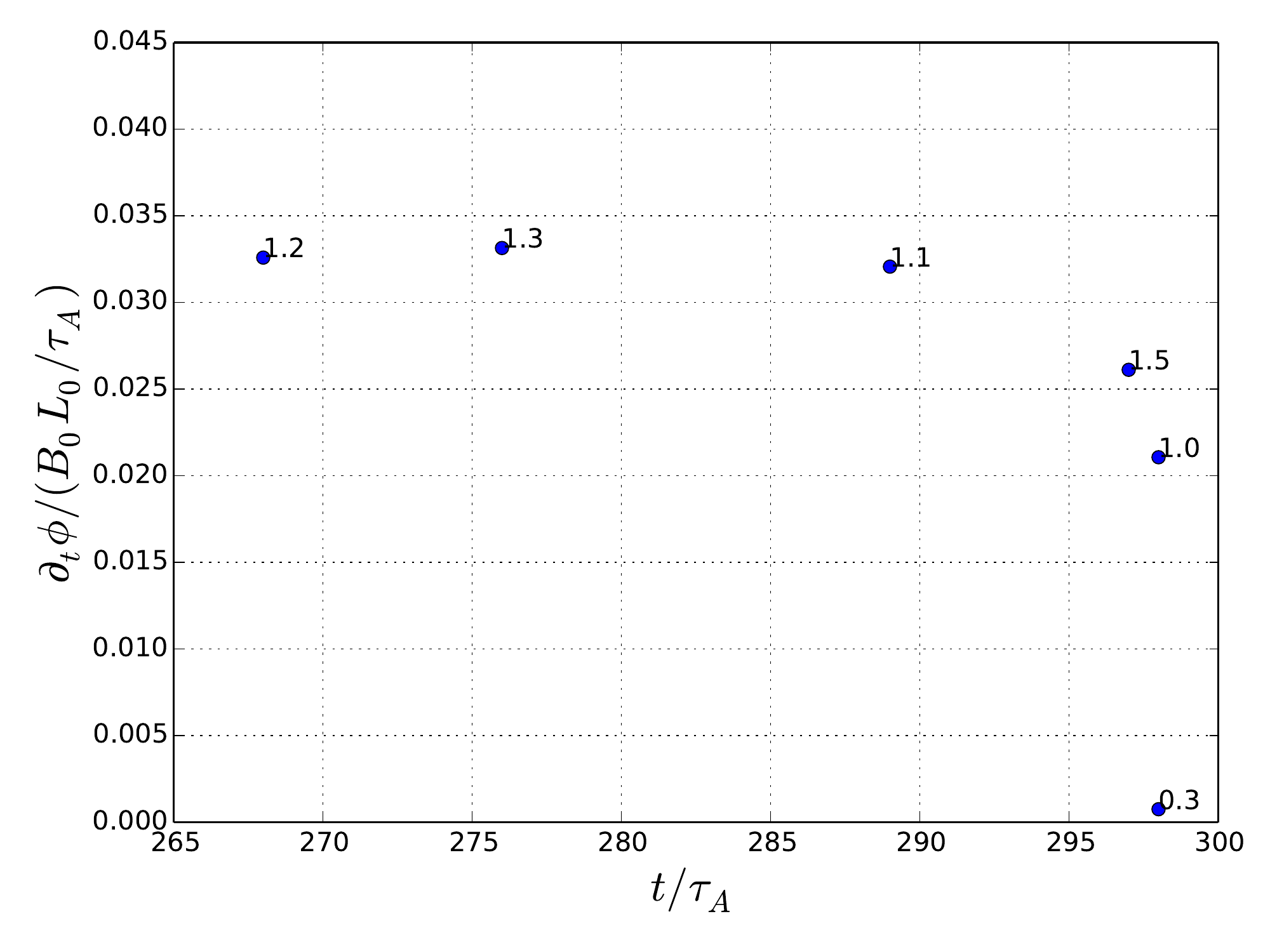}}\\
		(c) Force free $b_g=5$ & (b) Force free $b_g=10$ \\
	\end{tabular}
  \caption{Maximum reconnection rates in time, $\tau=1.3$ and $\eta=10^{-3}$}
  \label{fig:MAXBGs}
\end{figure}

For better applicability to the solar corona current sheets, we investigated the role of turbulence described by a RANS turbulence theory for force free current sheets with a guide field $b_g$ in the out-of-the-reconnection-plane direction.
The values tested for the guide magnetic field are the same as in the Harris equilibrium: 1, 5 and 10. The turbulence timescale $\tau$ controls the regime of the reconnection similar to the Harris-type equilibrium case (section \ref{SimHarris}) but for the fastest reconnection which is now obtained for $\tau=1.2$ (\fref{fig:MAXBGs}).
For $b_g=5$ and $10$, the maximum reconnection rate is lower than in the Harris-type and the force free $b_g=1$ case (as seen in \fref{fig:MAXBGs}). As shown when the turbulent magnetic helicity $H$ is taken into account,\cite{YokRub} a strong guide magnetic field reduces the reconnection rate because of the generation of a strong turbulent magnetic helicity $H$ acting against the production of turbulent energy $K$.
Even though the term related to $\alpha$ describing the turbulent magnetic helicity is not used in the present simulations, a similar reduction of the reconnection rate is obtained. The reason is that a strong guide field reduces the effect of turbulence on the magnetic reconnection rate as predicted by equation (\ref{eq:MachA}).
\begin{figure}
\centering
	\begin{tabular}{cc}
		\hspace{-0.5cm}{\includegraphics[width=4.5cm, keepaspectratio]{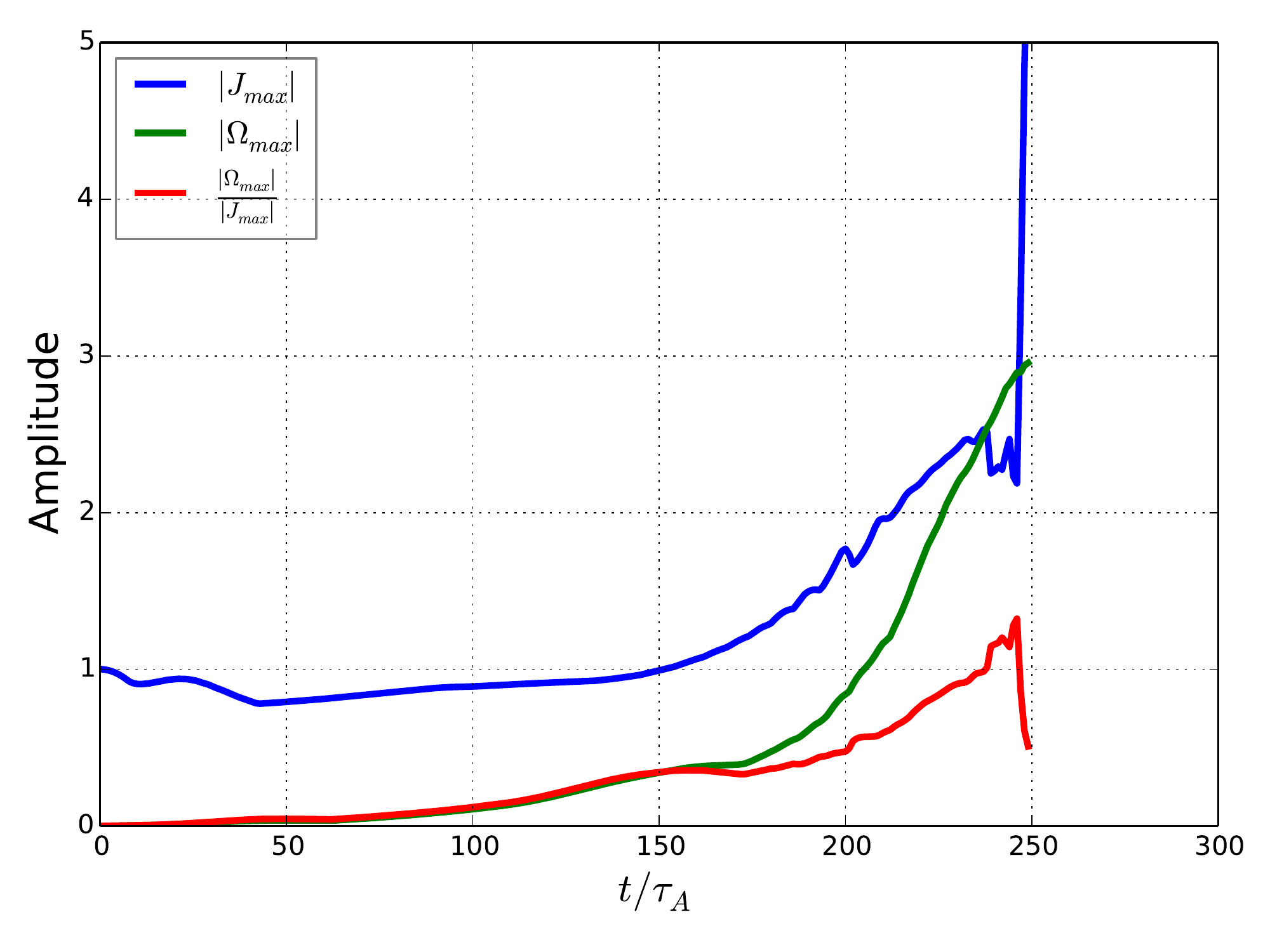}} & {\includegraphics[width=4.5cm, keepaspectratio]{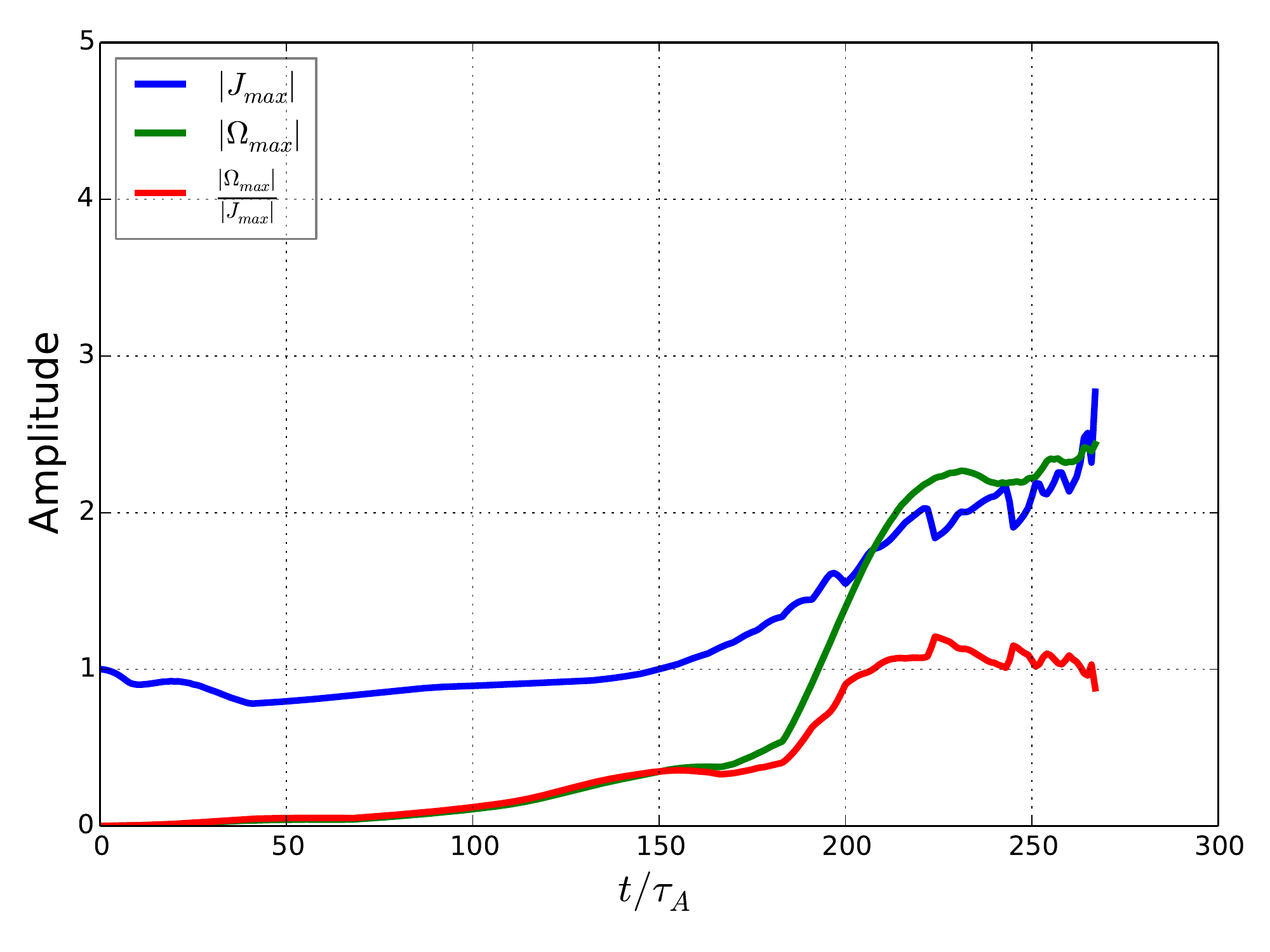}}\\
		(a) Guide field $b_g=1$ & (b) Guide field $b_g=5$ \\
	\end{tabular}
  \caption{Time history of the maximum current density $\boldsymbol{J}$ and vorticity $\boldsymbol\Omega$ in force free equilibrium for $\tau=1.2$ and $\eta=10^{-3}$}
  \label{fig:JOmaxBGs}
\end{figure}
A possible cause is that the mean current density and vorticity grow earlier in the force free equilibrium
but reach a smaller value than in the Harris-type equilibrium (see \fref{fig:JOmaxBGs}). In the model turbulence is driven by $\boldsymbol{J}^2$, $\boldsymbol{\Omega}^2$ and their product. Hence less turbulence is produced in the force free equilibrium than in Harris-type equilibrium. The reason of the decreased turbulence production is again that the main current density and vorticity are reduced by the presence of the guide field. As the Harris equilibrium, the reconnection rate is faster for larger Reynolds number. It also is dependent on the molecular resistivity but after decreasing it lower than $10^{-6}$ there is no significantly faster magnetic reconnection produced since the numerical resistivity level is reached. Hence, again, turbulence plays an important role for reconnection when the Reynolds number is large as is the case in most astrophysical plasma. 
\section{Guide field and helicity related term}
\begin{figure}[h]
  \centering
  \begin{tabular}{cc}
		\hspace{-0.5cm}{\includegraphics[width=4.5cm, keepaspectratio]{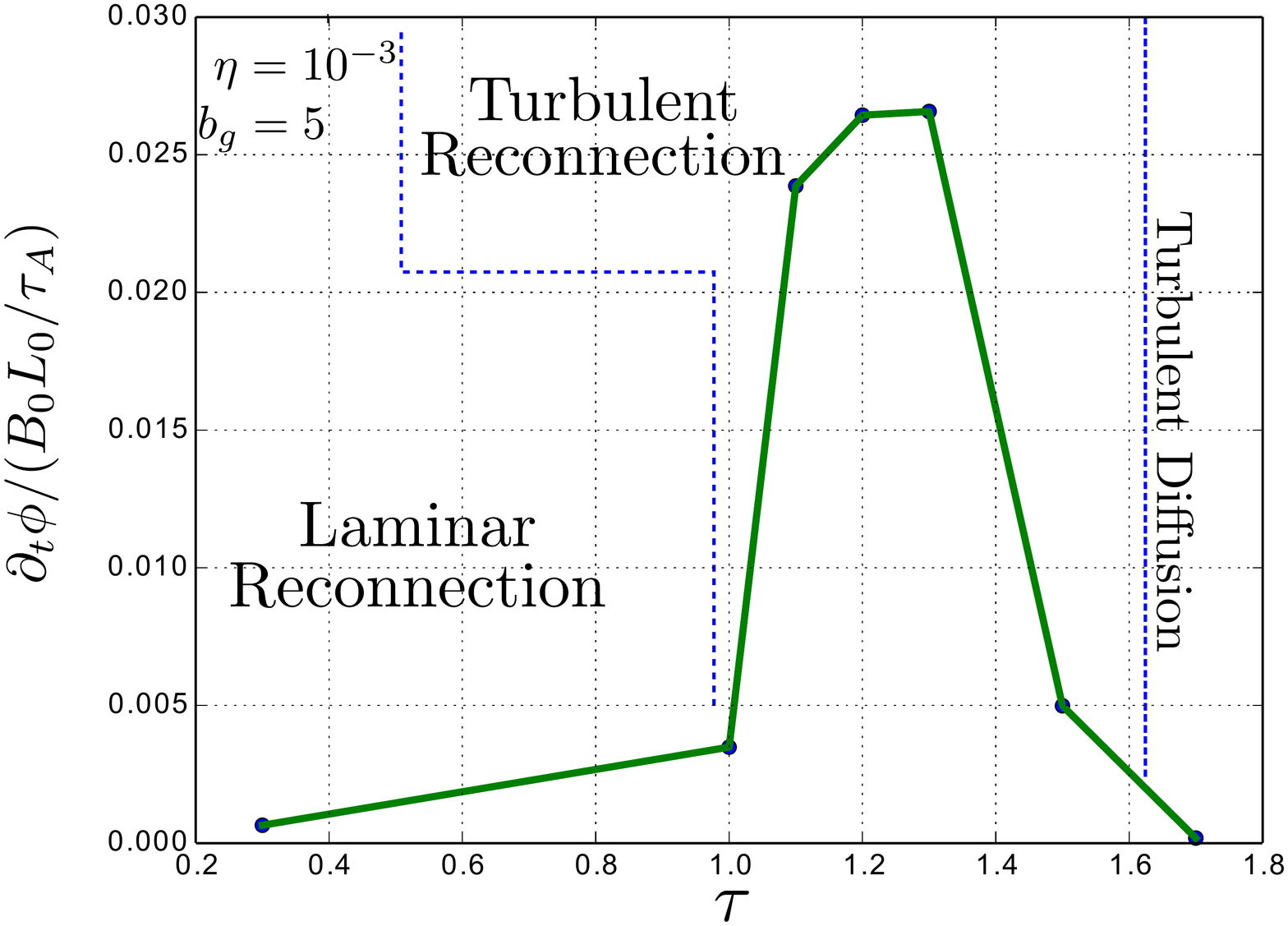}} & {\includegraphics[width=4.5cm, keepaspectratio]{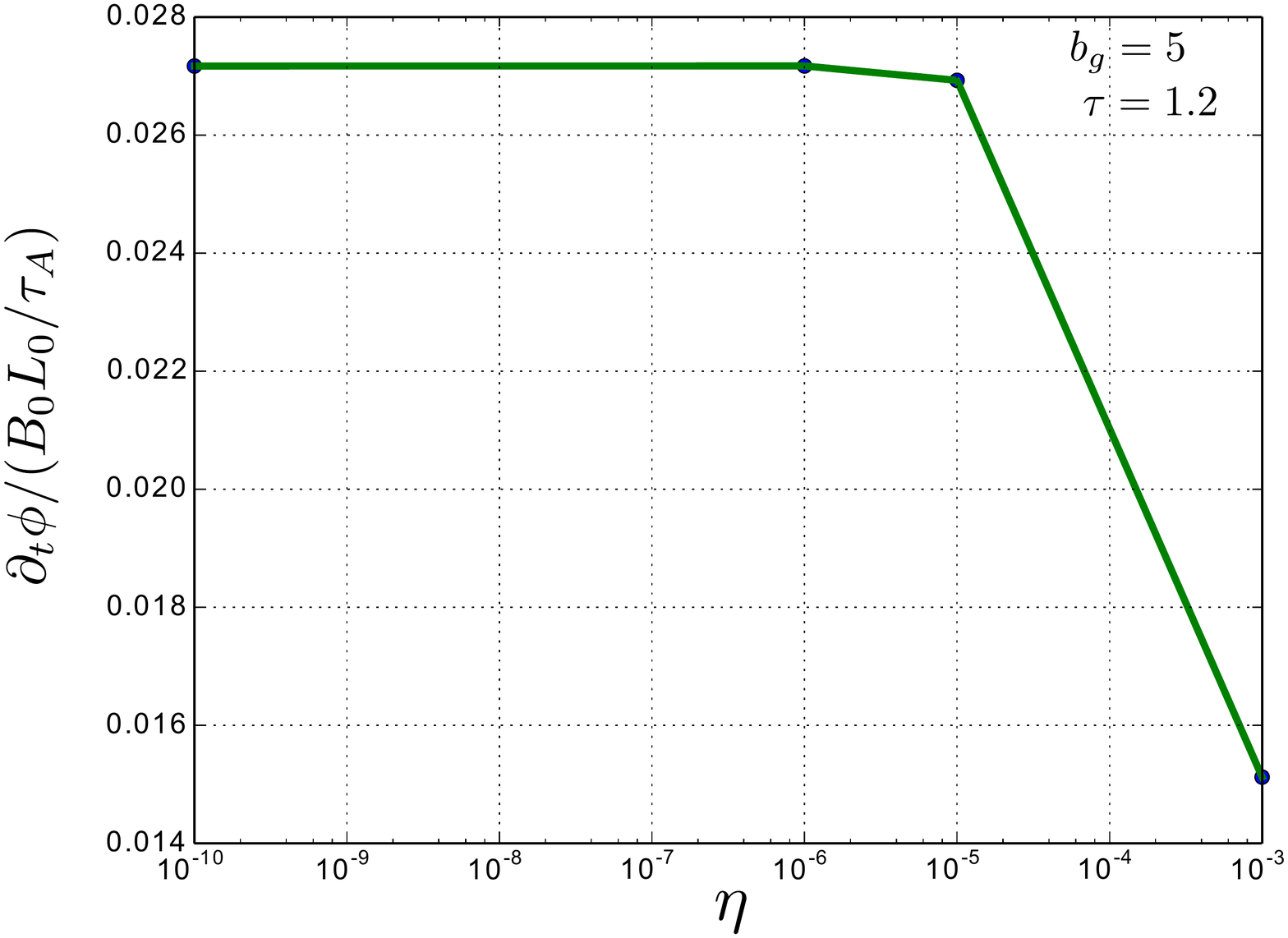}} \\
		(a) $t=245\tau_A$ & (b) $t=230\tau_A$ \\
		\hspace{-0.5cm}{\includegraphics[width=4.5cm, keepaspectratio]{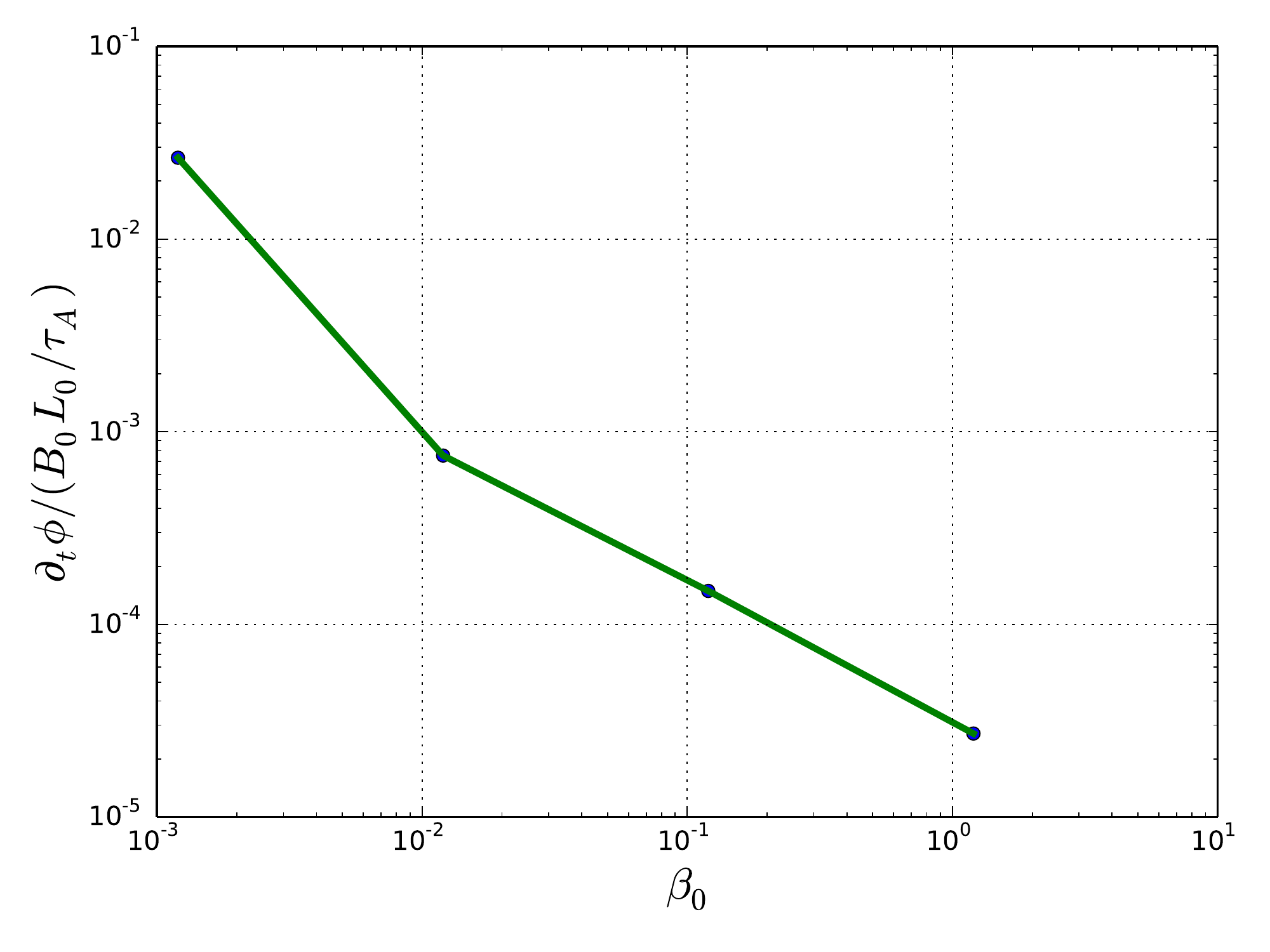}} &  \\
		(c) $t=300\tau_A$ & \\
	\end{tabular}
\caption{Reconnection rate for a guide field of 5 at fixed time for different $\tau$, $\eta$ and $\beta_0$}
\end{figure}
\begin{figure}
 \centering
	\begin{tabular}{cc}
		\hspace{-0.5cm}{\includegraphics[width=4.5cm, keepaspectratio]{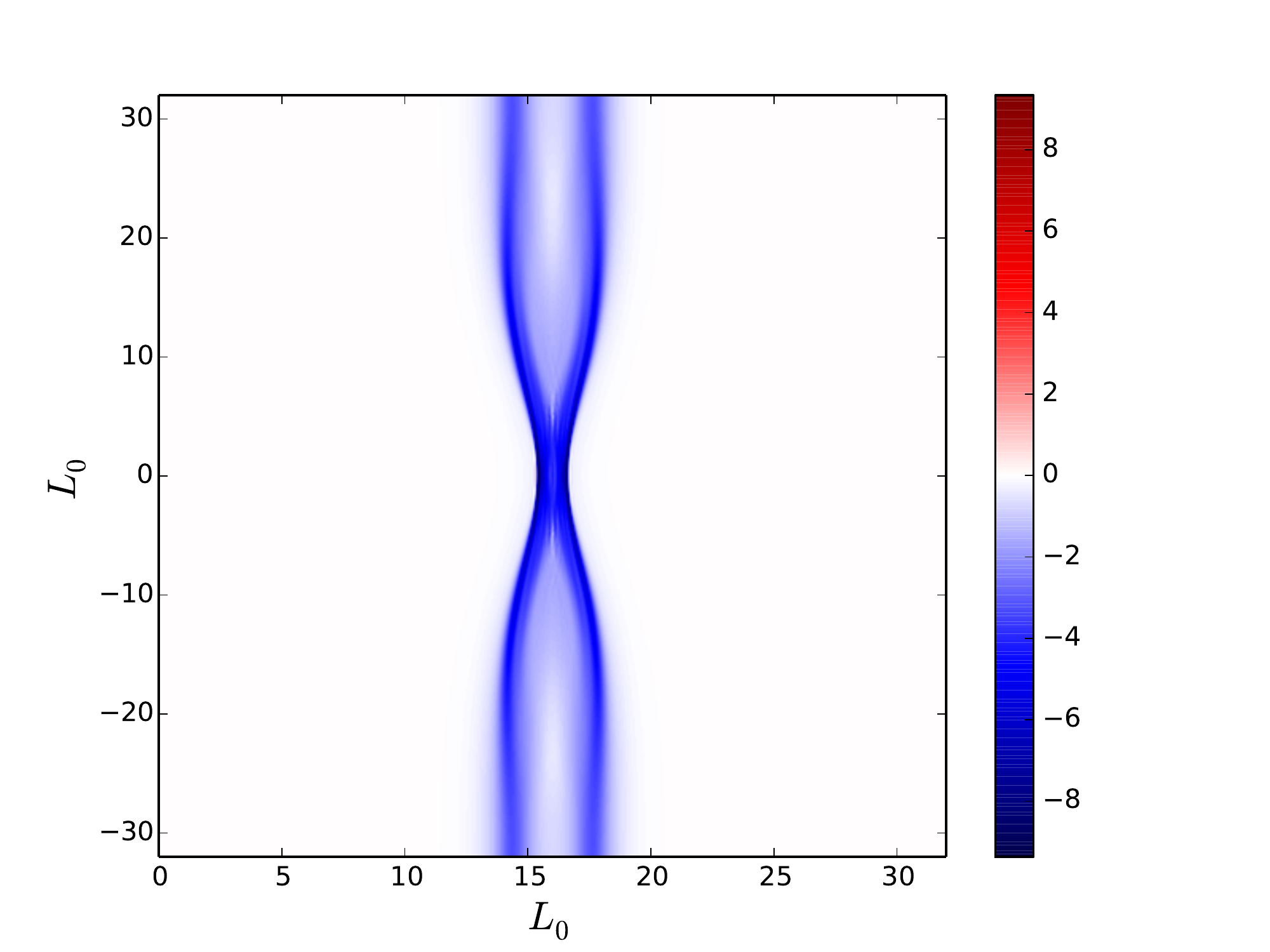}} & {\includegraphics[width=4.5cm, keepaspectratio]{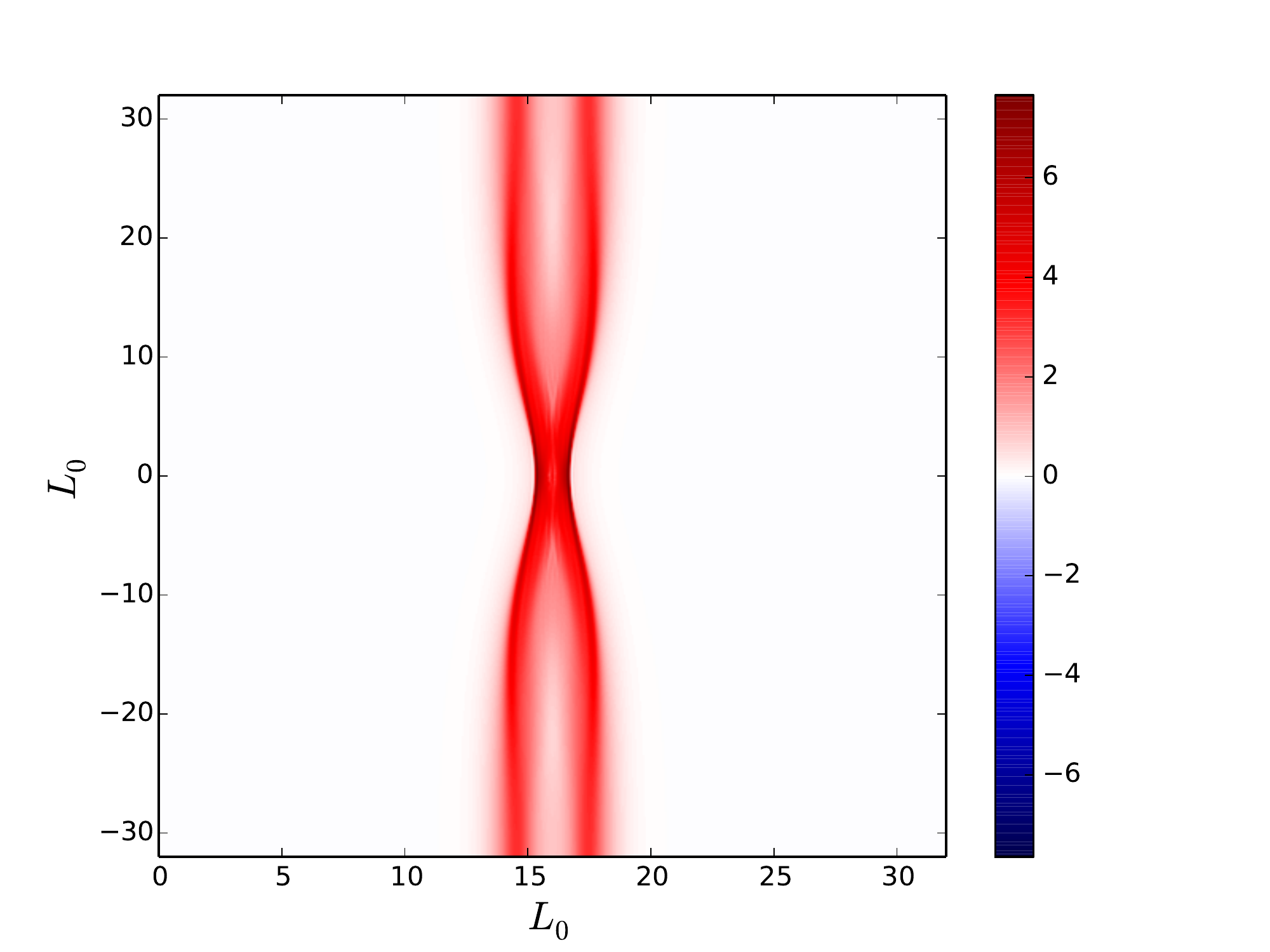}} \\
		(a) Harris, $b_g=-5$ & (b) Harris, $b_g=5$ \\
		\hspace{-0.5cm}{\includegraphics[width=4.5cm, keepaspectratio]{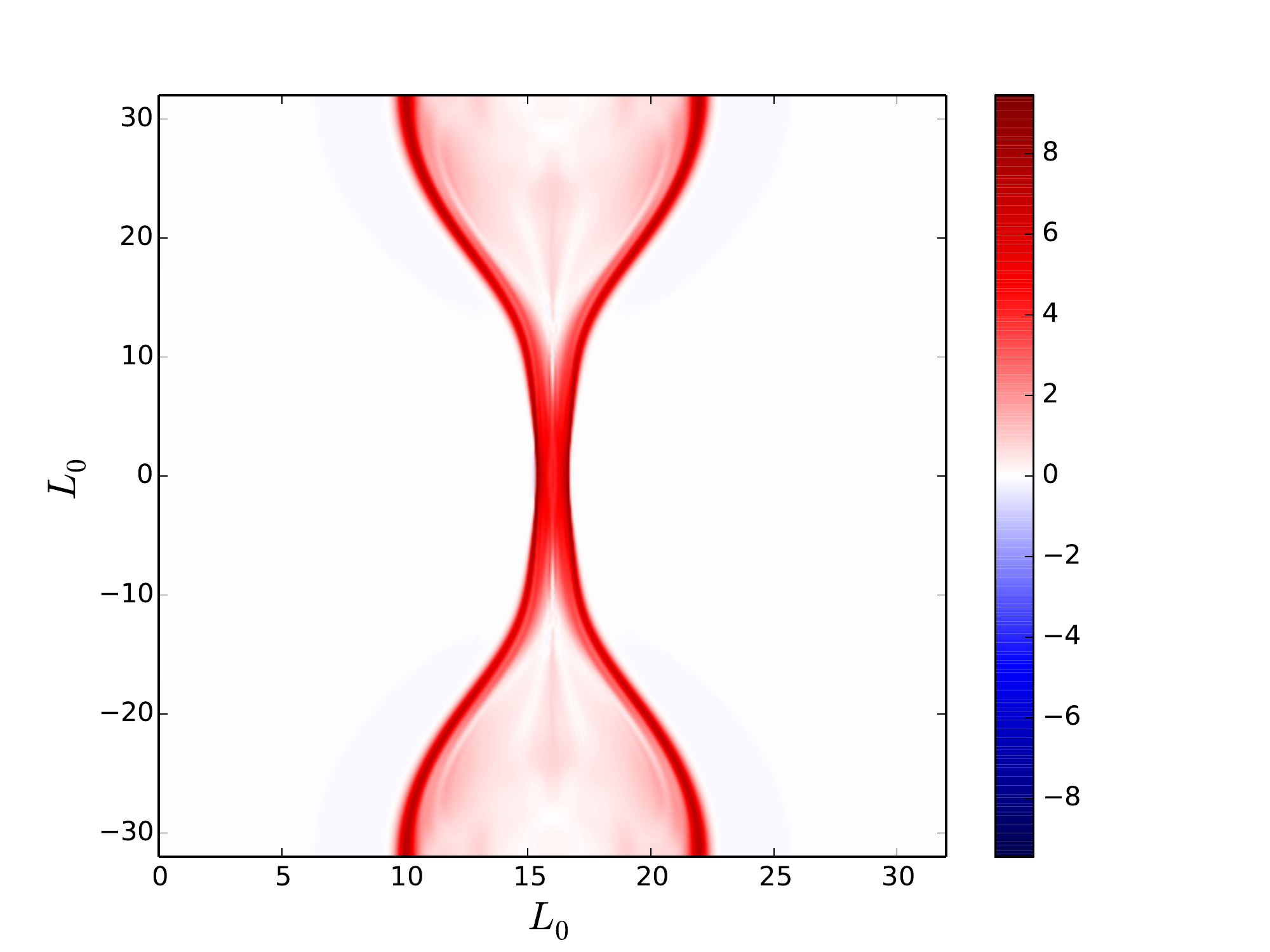}} & {\includegraphics[width=4.5cm, keepaspectratio]{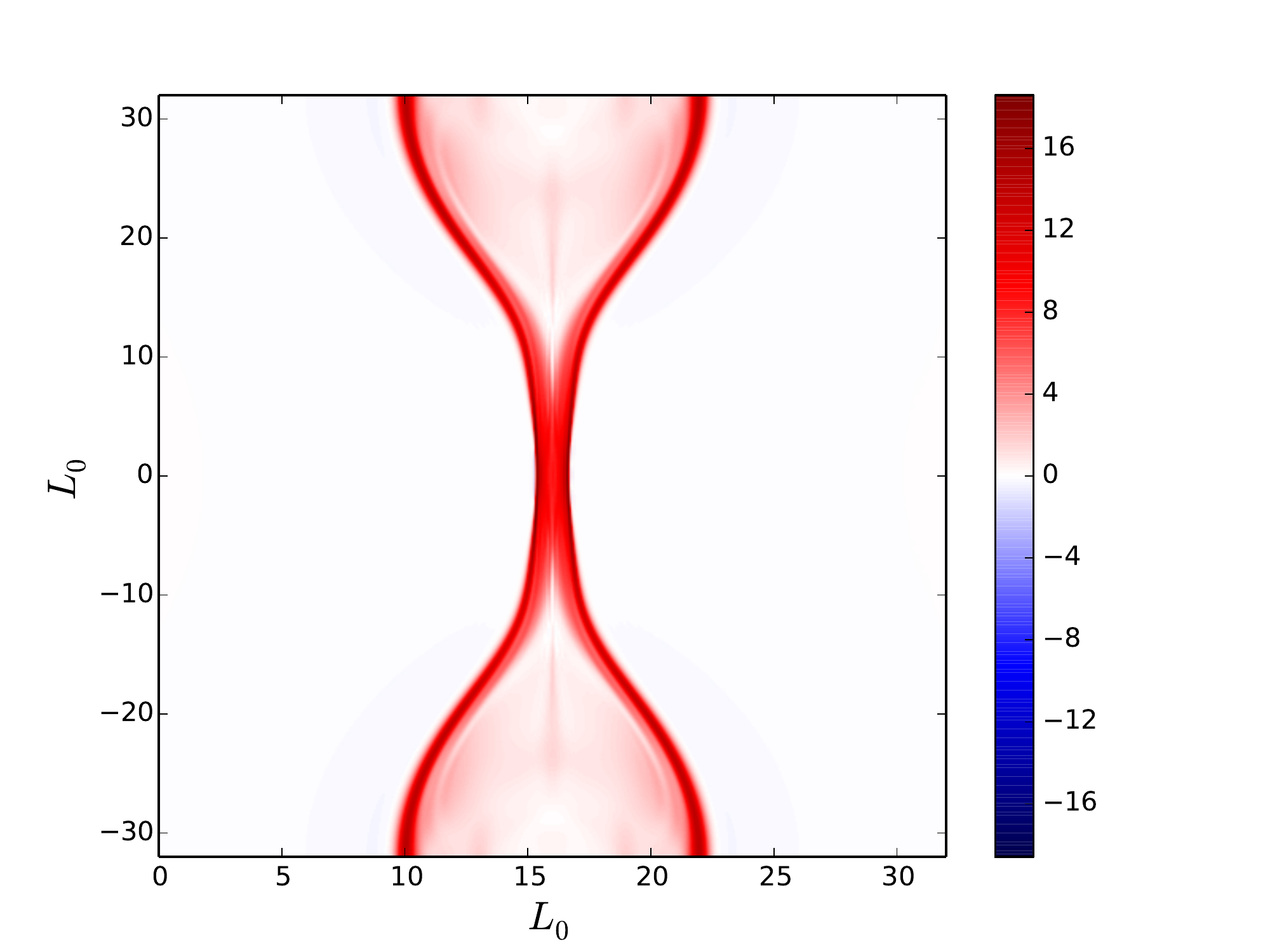}} \\
		(c) Force free, $b_g=5$ & (d) Force free, $b_g=10$ \\
	\end{tabular}
 \caption{Turbulent magnetic helicity related term for different guide field $b_g$}
 \label{fig:Alphas}
\end{figure}

In the framework of the current turbulence model, the turbulent magnetic helicity related term proportional to $\alpha$ may take the form
\begin{equation}
	\frac{D\alpha}{Dt}\simeq -\frac{1}{\beta}\langle\bf{u}'\times\bf{b}'\rangle\cdot\boldsymbol{B}\simeq\boldsymbol{J}\cdot\boldsymbol{B}
\label{Palpha}
\end{equation}
Hence, in the case of an out-of-plane finite guide magnetic field, whether in the parallel or anti-parallel direction with respect to the mean current density and electric field, the turbulent magnetic helicity may play a role even in two dimensions. 
In order to understand the role played by a guide field, the $\alpha$ term can be estimated using expression (\ref{Palpha}). In a Harris-type equilibrium, the sign of the guide field can be reversed to obtained an anti-alignment of the electromotive force with the out-of-plane guide magnetic field. \Fref{fig:Alphas} shows that the turbulent magnetic helicity related
term is, in two dimensions, a function of the guide field, increasing with it and following its sign.
The turbulent magnetic helicity related term is therefore a function of the guide field: $\alpha \mapsto \alpha(\boldsymbol{J},b_g)$. Considering it in the main contribution of the production of turbulent energy leads to
\begin{equation}
\frac{1}{\tau_t}\frac{D\beta}{Dt}\simeq -\langle\bf{u}'\times\bf{b}'\rangle\cdot\boldsymbol{J} = \beta\boldsymbol{J}^2-\alpha\boldsymbol{B}\cdot\boldsymbol{J} - \gamma\boldsymbol{\Omega}\cdot\boldsymbol{J} 
\end{equation}
Since the sign of $\alpha$ is the same as $\boldsymbol{B}\cdot\boldsymbol{J}$, the turbulent helicity will always contribute to decrease the production of turbulent energy together with the cross-helicity.
Rewriting the magnetic field in (\ref{eq:Emf}) as $\boldsymbol{B} = \boldsymbol{B_b}(1+ \boldsymbol{B_g}/\boldsymbol{B_b}) $, where $\boldsymbol{B}_b$ refers to the background magnetic field, shows that a large guide field will efficiently suppress the turbulent energy production and so decrease its effect as turbulent diffusivity affecting the reconnection rate.
\section{Energy Cascade}
\label{sec:Cascade}
\begin{figure}[h]
 \centering
 	\begin{tabular}{cc}
		\hspace{-0.5cm}{\includegraphics[width=4.5cm, keepaspectratio]{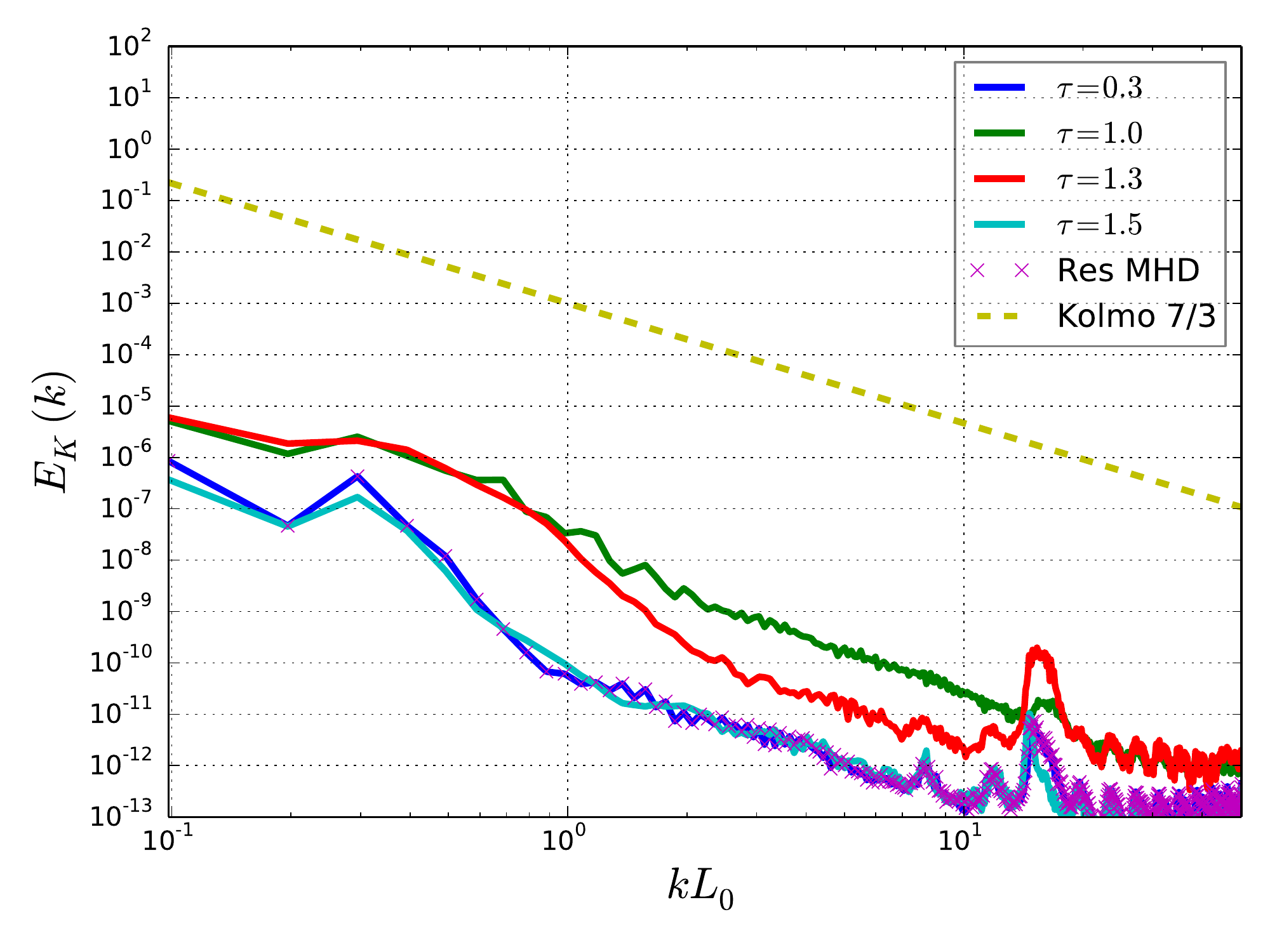}} & {\includegraphics[width=4.5cm, keepaspectratio]{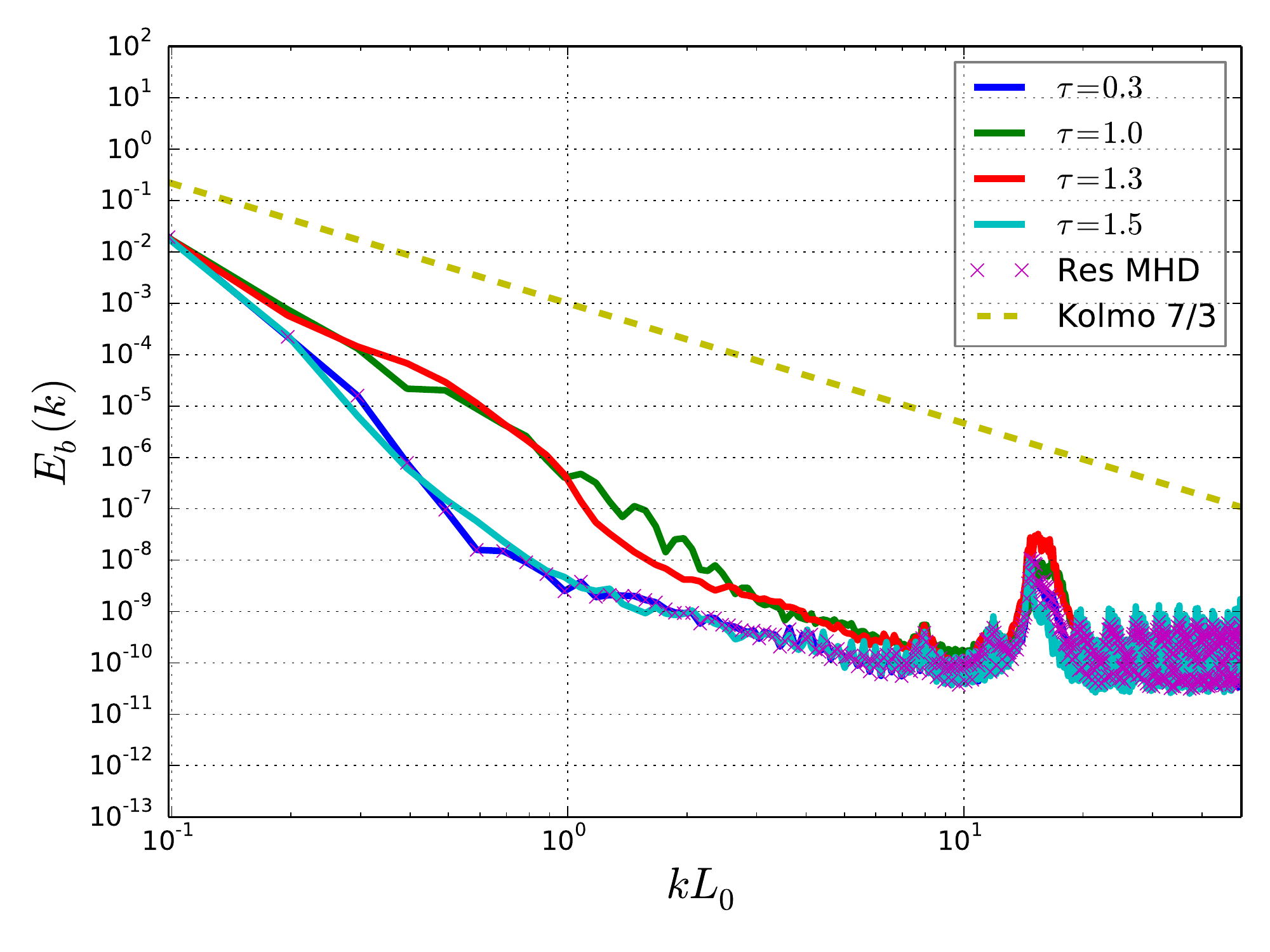}} \\
		(a) Kinetic energy, start & (b) Magnetic energy, start \\
		\hspace{-0.5cm}{\includegraphics[width=4.5cm, keepaspectratio]{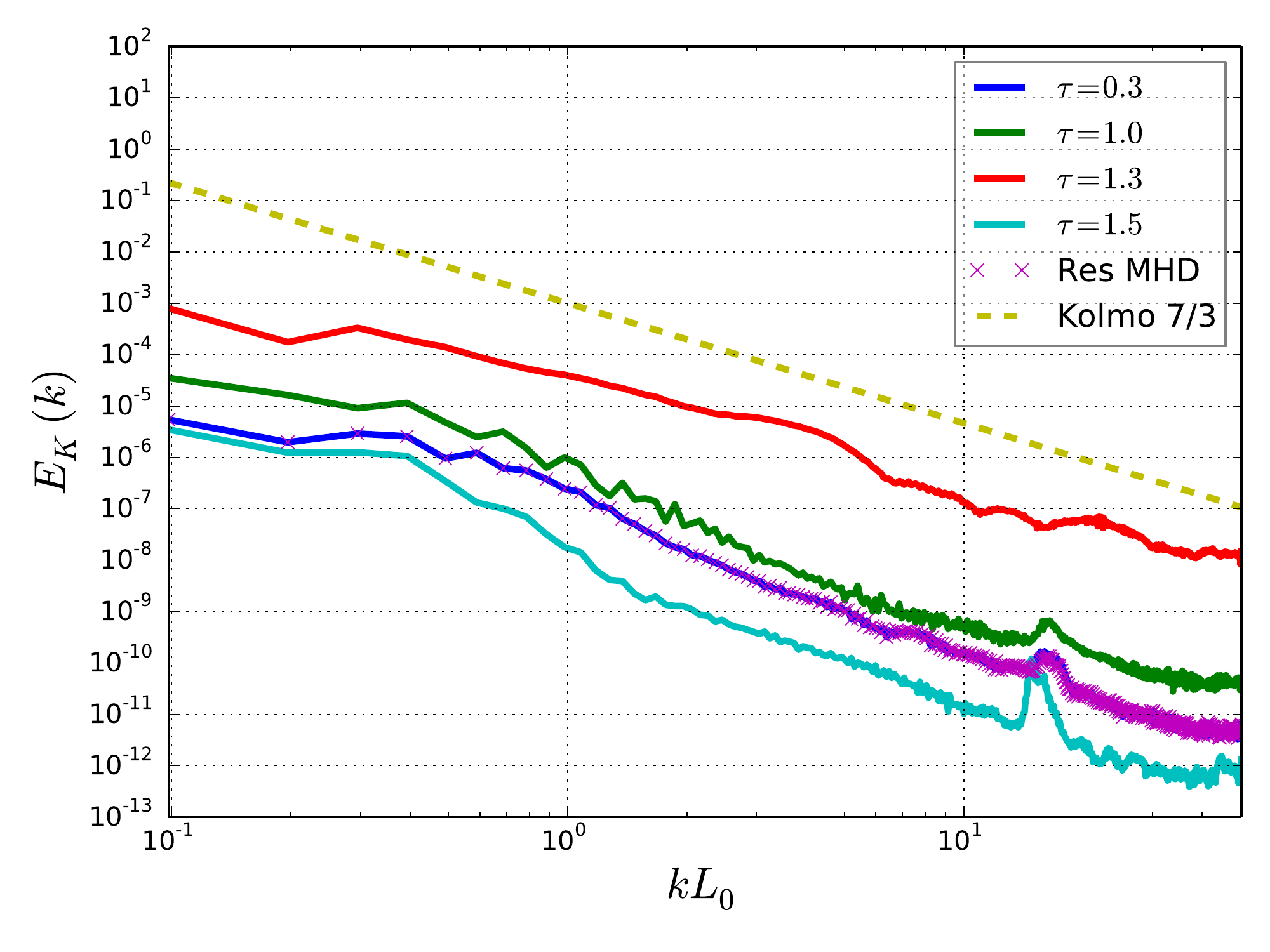}} & {\includegraphics[width=4.5cm, keepaspectratio]{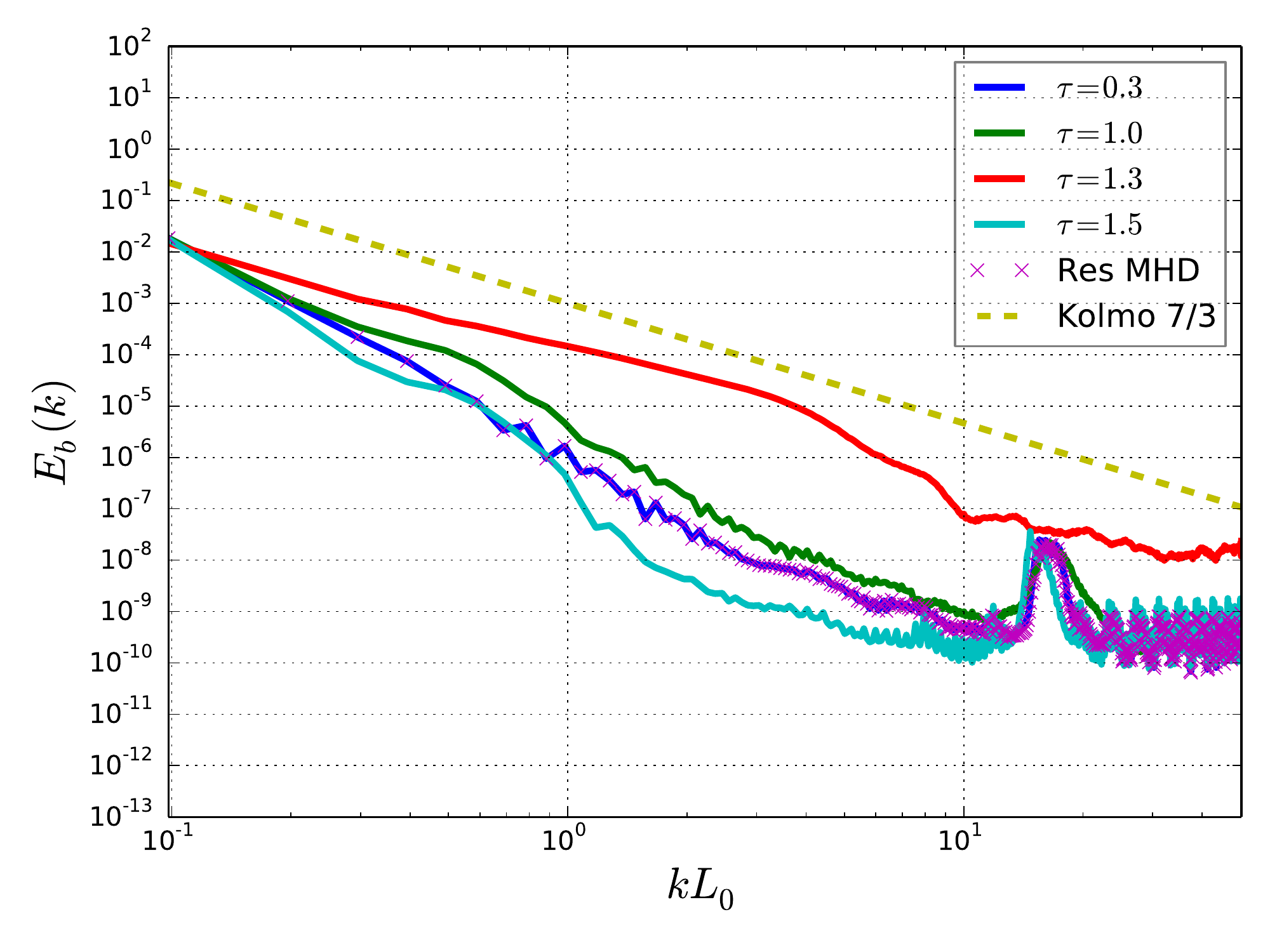}} \\
		(c) Kinetic energy, peak & (d) Magnetic energy, peak \\
	\end{tabular}
 \caption{Energy cascade for different $\tau$, Harris equilibrium}
 \label{fig:CascadeTausHarris}
\end{figure}

\begin{figure}
\centering
	\begin{tabular}{cc}
		\hspace{-0.5cm}{\includegraphics[width=4.5cm, keepaspectratio]{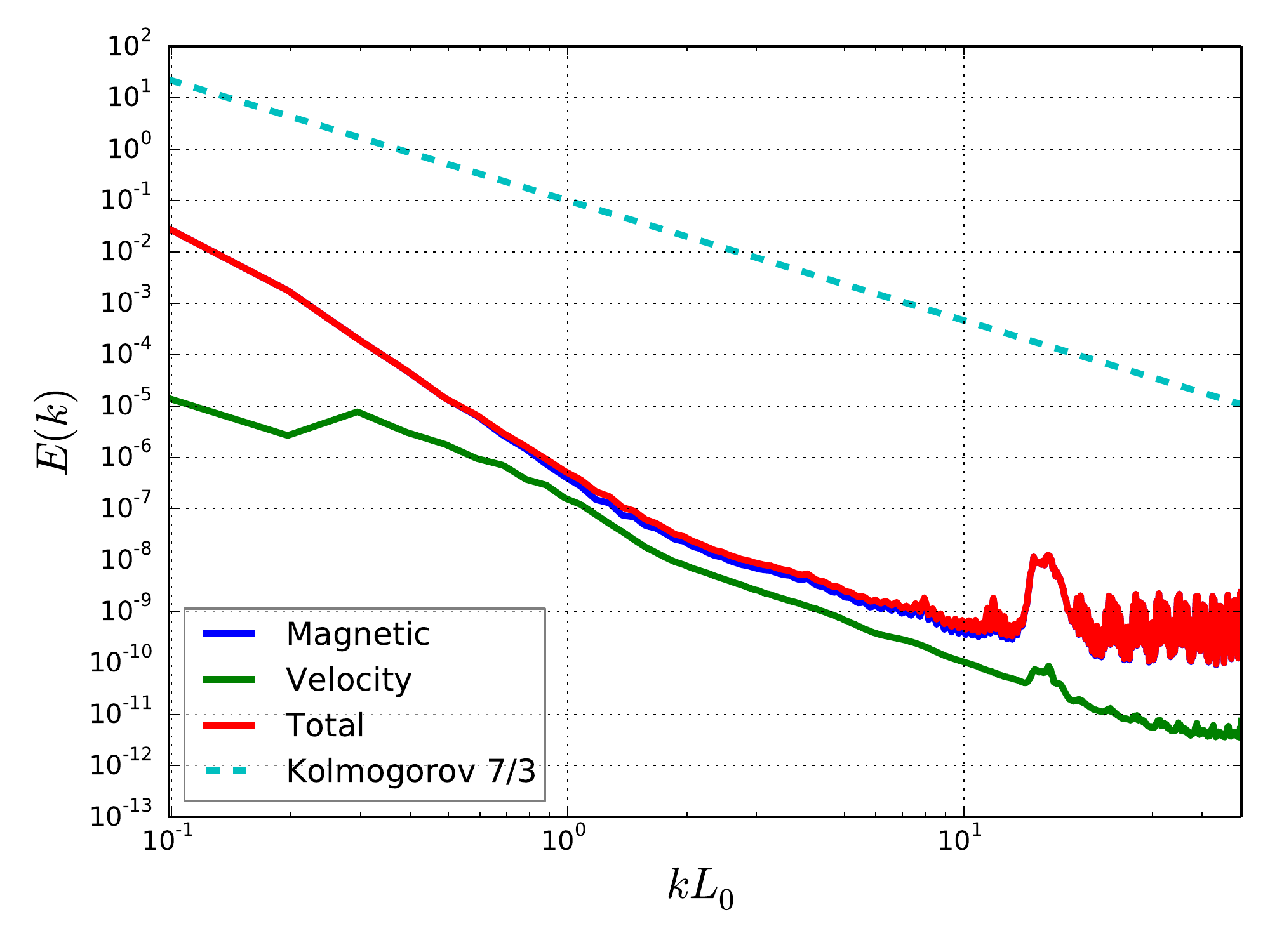}} & {\includegraphics[width=4.5cm, keepaspectratio]{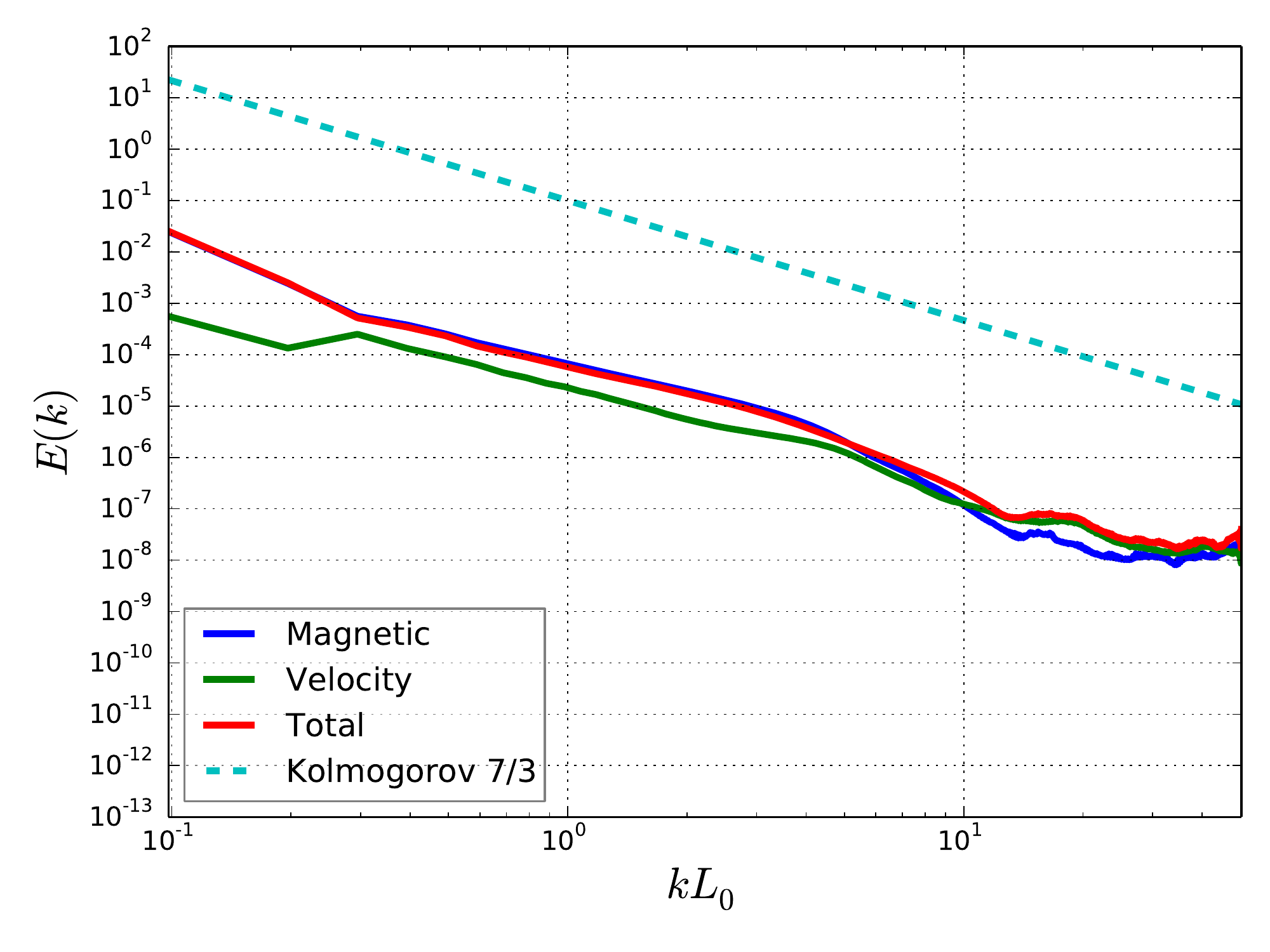}} \\
		(a) $\tau=0.3$ & (b) $\tau=1.3$ \\
		\hspace{-0.5cm}{\includegraphics[width=4.5cm, keepaspectratio]{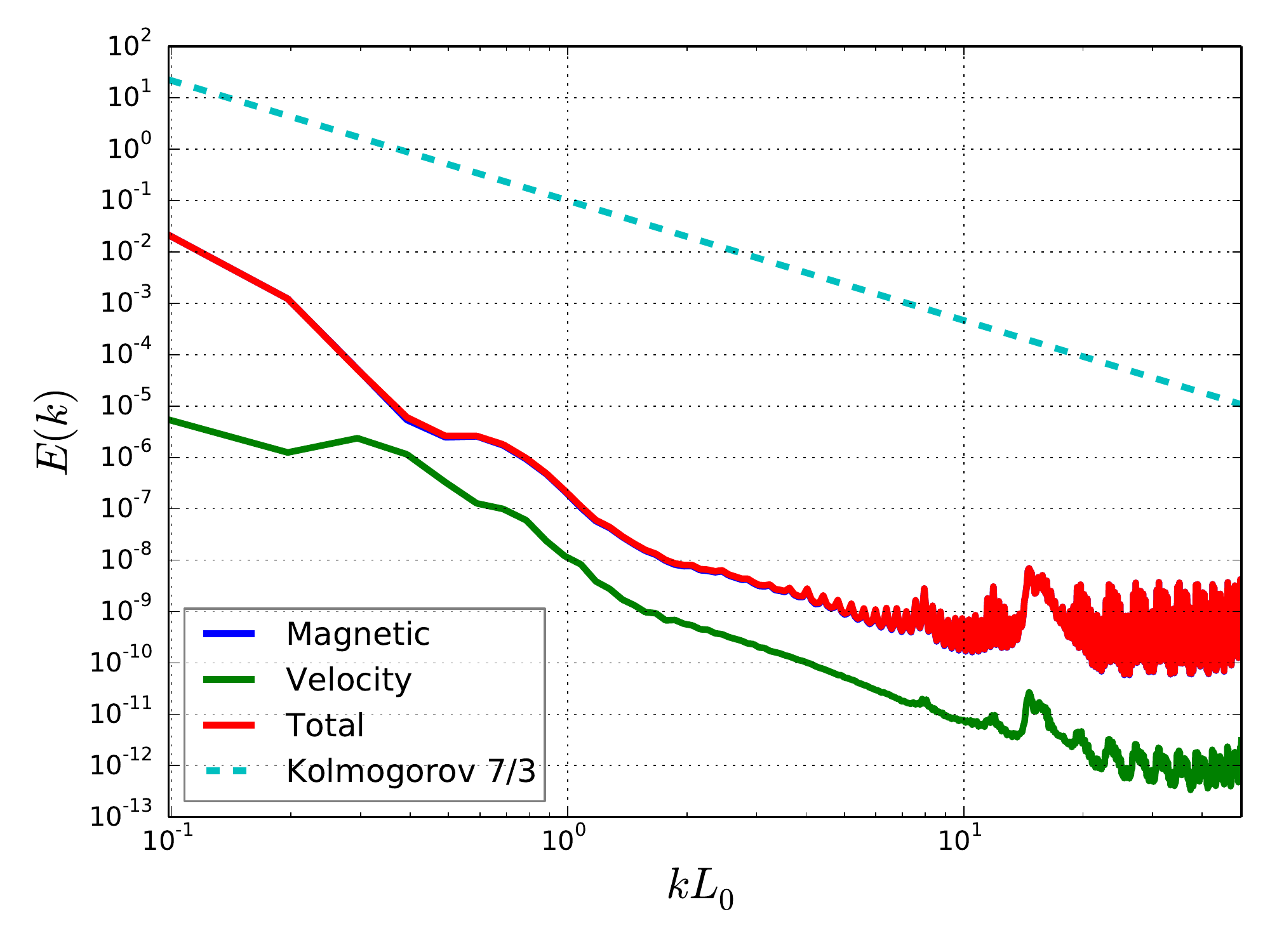}} & \\
		(c) $\tau=1.5$ & \\
	\end{tabular}
\caption{Time average of the energy cascade for different $\tau$, $\eta=10^{-3}$, Harris-type equilibrium}
\label{fig:CascadeTausTime}
\end{figure}

\begin{figure}[h]
 \centering

	\begin{tabular}{cc}
		\hspace{-0.5cm}{\includegraphics[width=4.5cm, keepaspectratio]{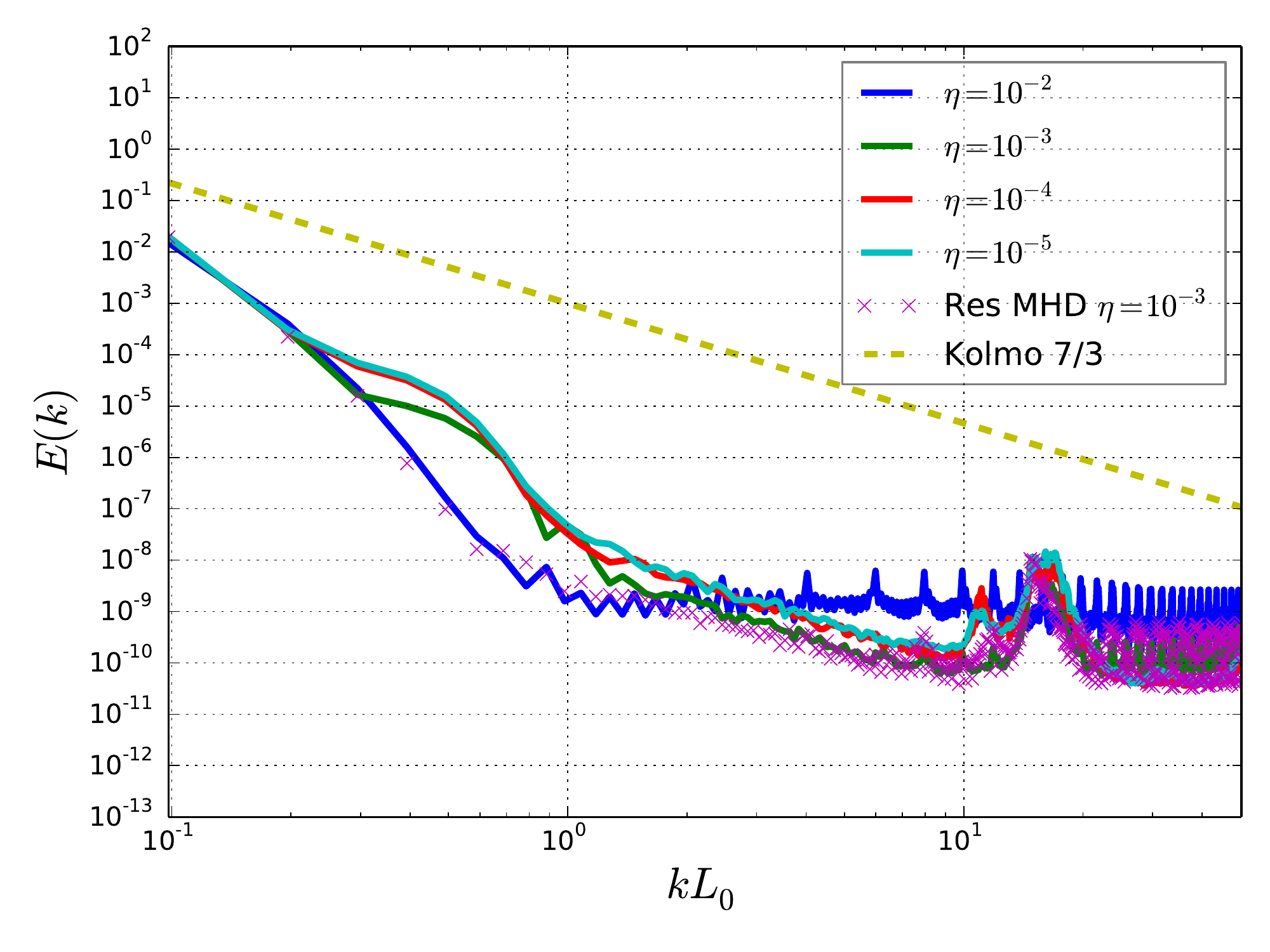}} & {\includegraphics[width=4.5cm, keepaspectratio]{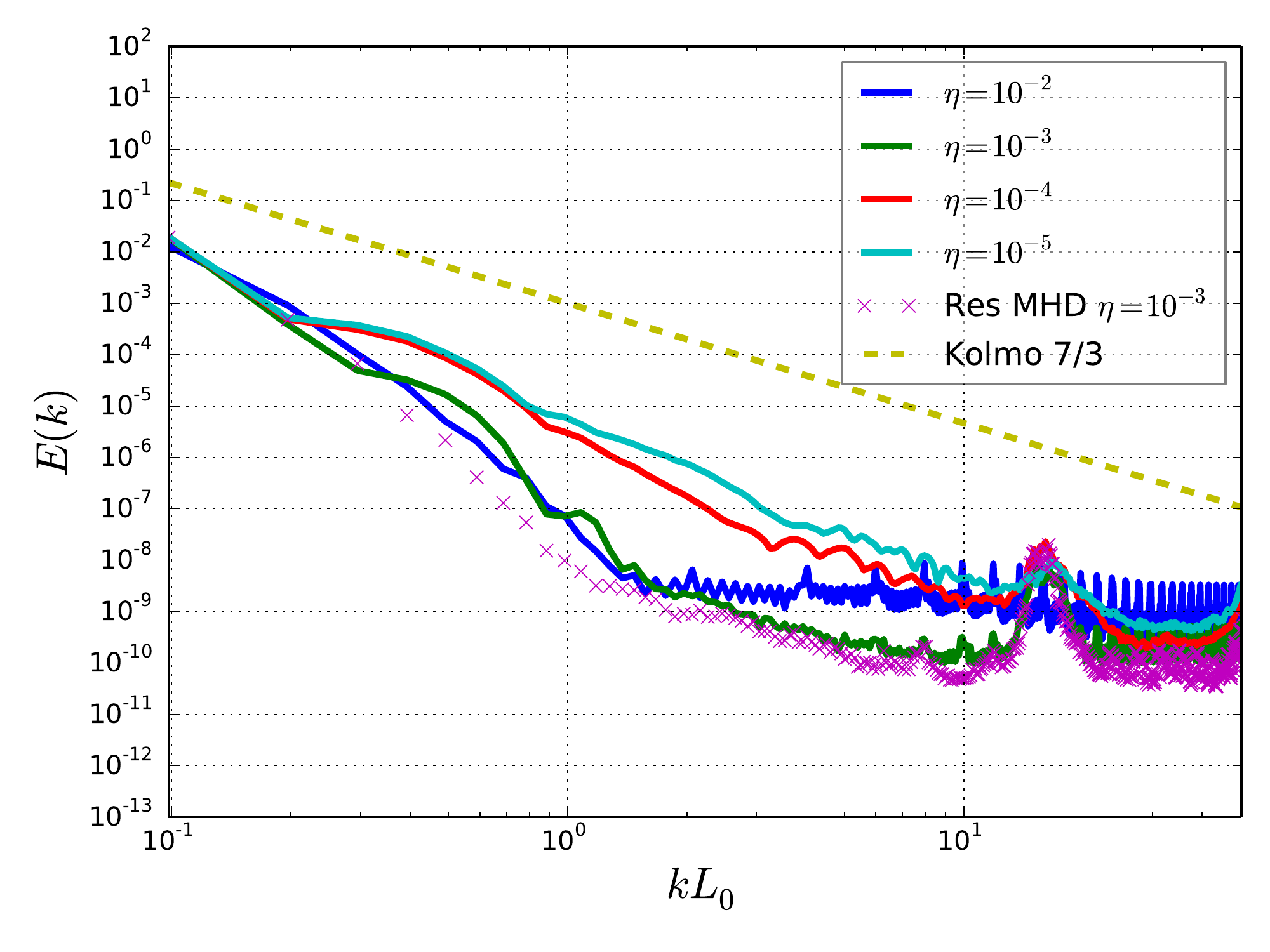}} \\
		(a) Start of reconnection & (b) Middle of reconnection \\
		\hspace{-0.5cm}{\includegraphics[width=4.5cm, keepaspectratio]{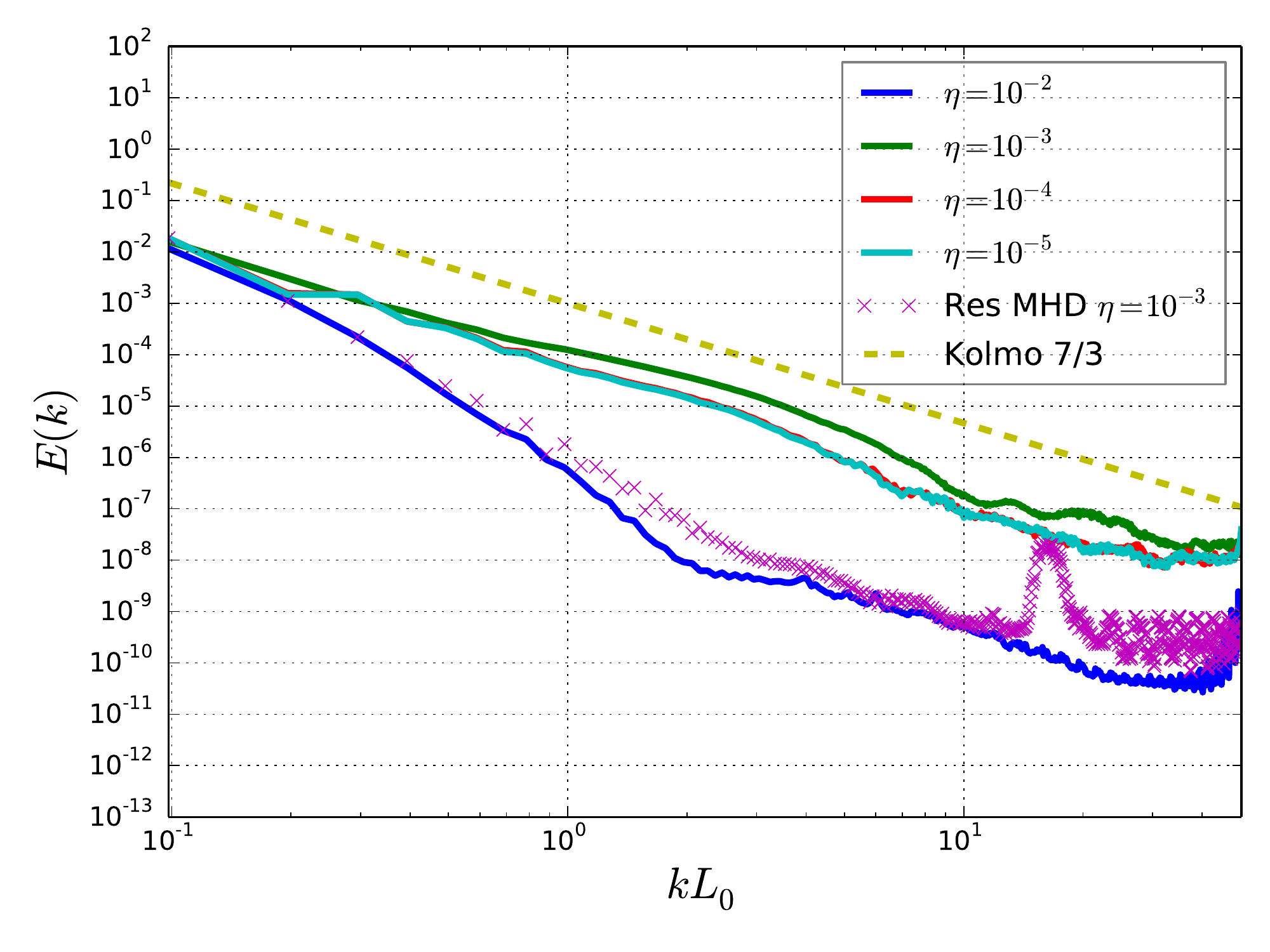}} &  \\
		(c) Peak of reconnection & \\
	\end{tabular}
 \caption{Total energy spectrum for different $\eta$, $\tau=1.3$, Harris-type equilibrium}
 \label{fig:CascadeEtasHarrisStage}
\end{figure}

\begin{figure}[h]
 \centering
\begin{tabular}{cc}
		\hspace{-0.5cm}{\includegraphics[width=4.5cm, keepaspectratio]{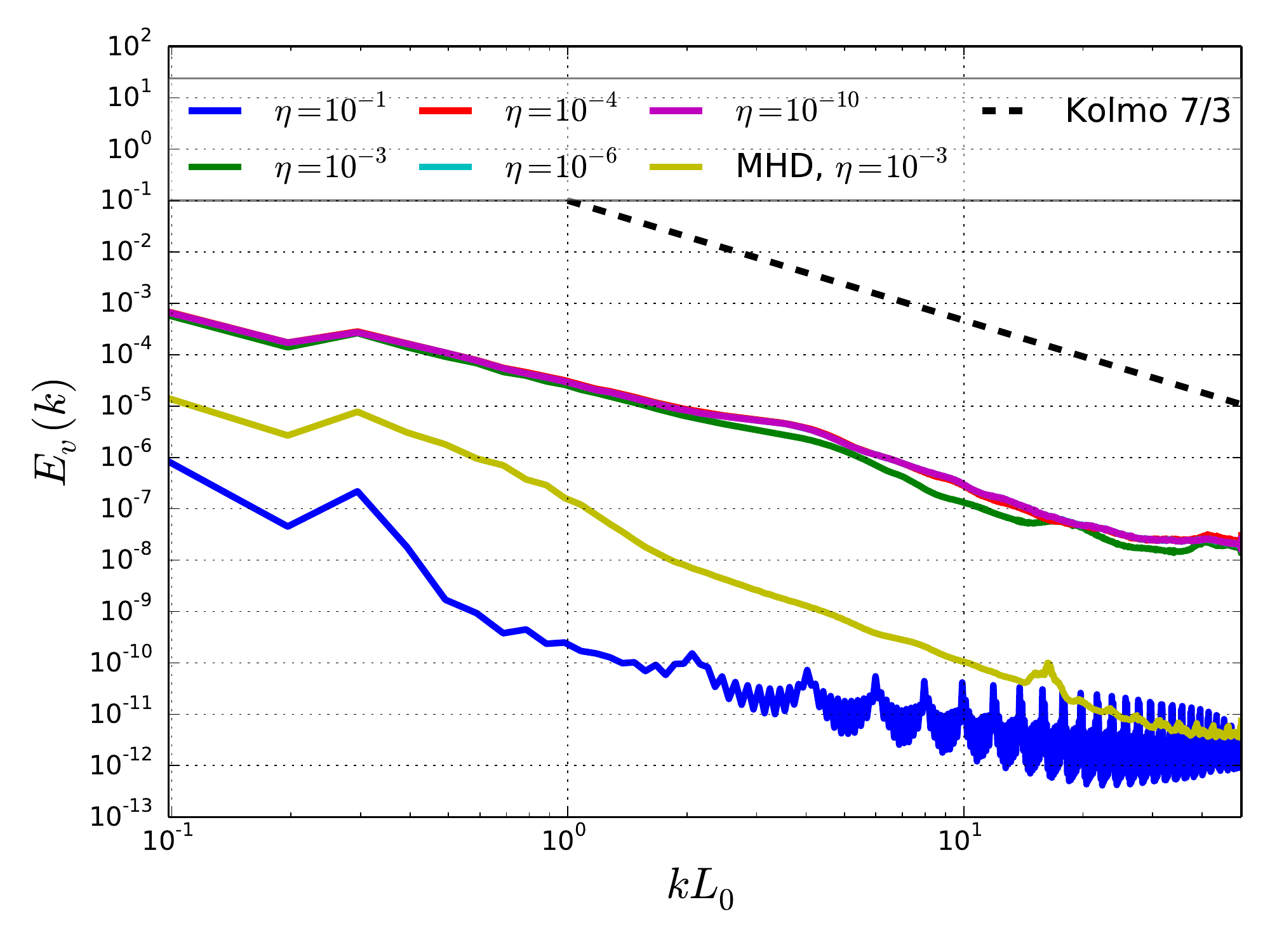}} & {\includegraphics[width=4.5cm, keepaspectratio]{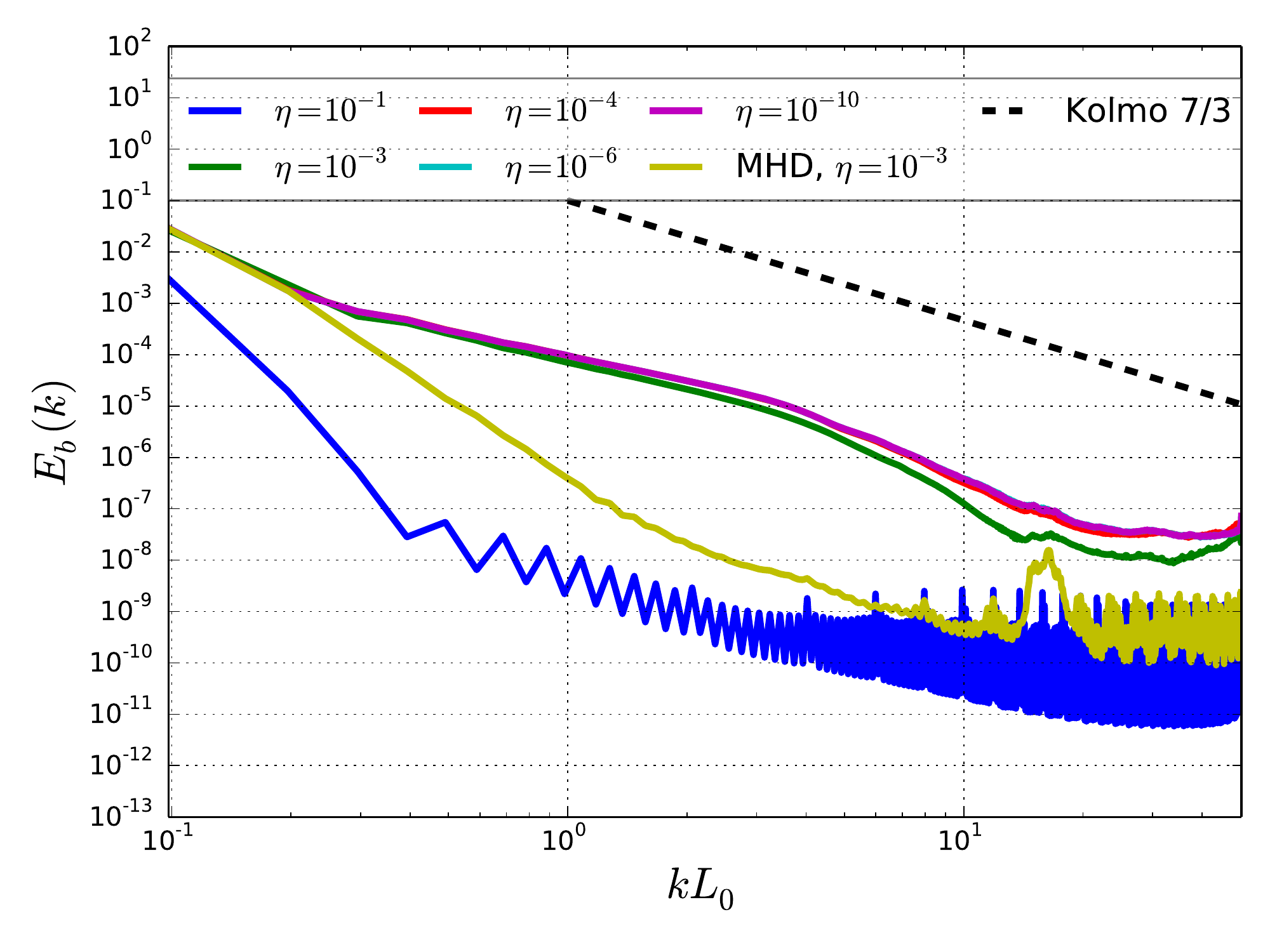}} \\
		(a) Kinetic energy & (b) Magnetic energy \\
	\end{tabular}
 \caption{Time average of the energy cascade for different $\eta$, $\tau=1.3$, Harris-type equilibrium}
 \label{fig:CascadeEtasHarris}
\end{figure}

In order to better understand how a specific value of the turbulence timescale parameter $\tau$ influences the regime of energy dissipation change from a \textit{laminar} to a
\textit{turbulent diffusive} state passing by a \textit{turbulent reconnection} regime, we plot the energy spectra for the kinetic and magnetic energy at two different times; When the reconnection rate increases and when the reconnection peaks. For this two times, the energy spetra for the different $\tau$ as well as for the resistive MHD (no turbulence) case are plotted in \fref{fig:CascadeTausHarris}. The energy bump present at small scales for all simulations are numerical artefacts due to the finite grid size.
Looking at the kinetic energy spectra, it can be seen that for values of $\tau=1.0$ and 1.3, the energy is larger than for 0.3 and 1.5. This timescale range corresponds to the \textit{turbulent reconnection} regime (\fref{fig:RecHarrisFixed}). The kinetic energy is already increased by turbulence in comparison to the resistive MHD case. When the turbulent reconnection starts, the kinetic energy for 1.0 and 1.3 is similar while the kinetic energy for $\tau=1.3$ is approximatively two orders of magnitude larger than for $\tau=1.0$ when the reconnection rate reaches its maximum. Also the width of the inertial range is much larger.
For $\tau=0.3$, on the other hand, the cascade in each situation perfectly matches the result for the resistive MHD reconnection explaining the similar reconnection rate for this specific $\tau$ at which no turbulence is produced. Only small amounts of energy are
transported from the large to small scales. Therefore the evolution of the current sheet is not driven by turbulence but is purely in a '\textit{laminar}'
non-turbulent regime. Finally,
for $\tau=1.5$, turbulence is so strong that energy is just diffused away on large scales in short time. Later the energy is lower than resistive
MHD case and the inertial range of the turbulence cascade is very short. Hence a too high intensity of turbulence does not lead to faster reconnection since practically no energy is transfered to the smaller scales. This is in accordance with the trend seen in \fref{fig:RecHarrisBeta0} and more explicitly in \fref{fig:RRetabeta3d}. The magnetic energy spectra (\fref{fig:CascadeTausHarris}), which represent the magnetic energy at the beginning and at the time of maximum reconnection, clearly show that for $\tau=1.3$ a large amount of energy is transfered from large to small scales at the maximum reconnection rate. This energy transfer results in a larger reconnection rate compared to the other time scales of turbulence. Especially this can be seen comparing with $\tau=1.5$ for which even less energy is transfered than in the case of resistive MHD reconnection.
The different energy transfer explain  the different regimes of reconnection observed in the different equilibria as first pointed out for the Harris equilibrium without guide magnetic field.\cite{Yokoi3}
A better understanding of \textit{turbulent} reconnection can be reached by looking at \fref{fig:CascadeTausTime}. The figure clearly show that, in average, a large amount of energy is transfered from the large to the small scales for $\tau=1.3$ while only a small amount is transfered for $\tau=1.5$.
A similar analysis can be done for the different values of the molecular resistivity. The total energy spectrum
(\fref{fig:CascadeEtasHarrisStage}) shows that at the time of maximum reconnection rate more energy is transfered to the small
scales for $\eta=10^{-3}$ than for smaller value of the molecular resistivity. 
\Fref{fig:CascadeEtasHarris} depicts the time average for different molecular resistivity.
Increasing the Reynolds number (decreasing the molecular resistivity $\eta$) to larger value increases, in average, the energy transfer from large to small scales. This explains the earlier start of reconnection for a smaller resistivity in comparison to larger ones. However, the same amount of energy is transfered to the small scales. This, in turn, explains why decreasing the resistivity does not
cause a larger reconnection rate. This also emphasises the importance of the turbulence for high Reynolds number. This is especially important for astrophysical plasmas whose Reynolds number may be of $10^{12}$. The energy spectra clearly show the role played by turbulence in the transport of energy from large to
small scales and its important role for fast reconnection.

\section{Discussion and conclusions}
\label{Conclusions}
In all considered Harris and force free equilibria, magnetic reconnection is influenced by MHD turbulence through the turbulence
timescale $\tau$ and the molecular resistivity $\eta$.
As predicted by a dimensional analysis (\eref{eq:mach}), the reconnection
rate is faster for smaller values of the resistivity. This is mainly due to the turbulent energy, related to the $\beta$ term, which
acts as a turbulent resistivity localised in the center of the diffusion
region. The effect of the turbulent cross-helicity, on the other hand, is important for the localisation of the turbulent energy around the diffusion region.
The production of turbulent energy and cross-helicity, $P_K$ (\ref{eq:PK}) and $P_W$ (\ref{eq:PW}), enhances the $\beta-$ and $\gamma-$terms proportional
to the mean current density $\boldsymbol{J}$ and the mean vorticity $\Omega$, respectively. By decreasing the
molecular resistivity, steeper magnetic and velocity gradients are generated, hence the mean current density and vorticity are increased (\fref{fig:CompJOKW} (a)). This will, therefore, fasten the production
of turbulence, as it can be seen in \fref{fig:CompJOKW} (b).
The latter leads to faster magnetic reconnection. The energy spectrum allows to understand the different regimes of reconnection in turbulence as they are controlled by the amount of turbulence and consequently by the energy transport. The different regimes, \textit{laminar reconnection}, \textit{turbulent reconnection}
and \textit{turbulent diffusion}, are directly related to the amount of energy transfered from large to small scales which largely depends on the initial amount of turbulence represented by $\beta_0=\tau K_0$. 
For a fixed $K_0$, it is $\tau$ which dictates the regimes while a variation of $K_0$ for a fixed $\tau$ is responsible for
the different regimes of reconnection which will be \textit{turbulent diffusive} if the initial level of turbulence is too large. Too high level of initial turbulence is either obtained for $\tau\gg1$ or for a large $K_0$ compared to the molecular resistivity. 
These limits correspond to the small amount of energy transfered to small scales as well as small rate of magnetic reconnection obtained either by large $\tau$ or large $K_0$. However, like discussed in section \ref{Sec:Tearing}, it is only an artefact of the algebraic modelling of the turbulent timescale. For instance, if the timescale of turbulence is represented by $\tau=K_0/\epsilon$, where $\epsilon$ is the dissipation rate, then $\tau\gg1 \rightarrow K_0\gg\epsilon$ and the transport of energy by $\epsilon$ through the energy cascade towards the smaller scales is limited. In this picture, $\beta_0=K_0^2/\epsilon$ and it is only when the turbulent energy is of the order of its dissipation rate that the reconnection rate can be enhanced by turbulence in situation of large magnetic Reynolds number. Therefore it is only the turbulence timescale which should control the reconnection regimes. A more realistic picture of mean field turbulent reconnection will be obtained by a closure for the turbulence timescale obtained from the non-linear dynamics of turbulence.\linebreak In situation of high Lundquist number, the Reynolds number due to turbulence is larger than the one based on molecular resistivity $\eta/\beta_0\equiv R_T/R >1$. In such situation a large amount of energy is transfered to smaller scales resulting in an efficient \textit{turbulent reconnection}. The energy spectrum average over time for different resistivity allows also to understand better the earlier start of reconnection for a smaller resistivity since larger amount of energy is transfered compared to the case of a larger resistivity. Decreasing the value of $\eta$ will enhance the reconnection rate since it does not result in a higher initial value of turbulence in the sense that the system still has $K_0\simeq \epsilon$. Oppositely, increasing the initial turbulence, by increasing $\tau$ or $K_0$, to a level causing $\beta_0$ to be greater than the molecular resistivity $\eta$ leads to smaller amount of reconnected flux since in such situation $K_0\gg\epsilon$. This situation is similar to the case of large $\tau$ as discussed previously. A larger Lundquist number (smaller molecular resistivity) allows steeper magnetic and velocity gradients which are the main source for the production of the turbulent energy and cross-helicity. Their production is fastened and turbulence can enhance the rate of magnetic reconnection earlier.
\linebreak Hence we confirmed the results stating that the turbulent RANS model produces three different regimes of reconnection,\cite{Yokoi3} \textit{laminar reconnection}, \textit{turbulent reconnection} and \textit{turbulent diffusion} regimes, controlled by the turbulence timescale $\tau$ and provided an explanation of the regimes through the energy spectrum.
Varying the molecular resistivity $\eta$, we could show that production $P_K$ and $P_W$ can be enhanced for a larger Reynolds number, based on the molecular resistivity, since it will increase the velocity and magnetic field gradients. The latter is the main source for the production
of turbulence which, in turn, results in fast magnetic reconnection for higher Reynolds number. The relation between turbulence, mean fields inhomogeneities and reconnection has been pointed out by the relation (\ref{eq:OJKWMa}). Finally, we estimated that the $\alpha$ term may also play an important role in the reconnection process even in two dimensions when an out-of-plane guide magnetic field is considered and that turbulent magnetic helicity related terms is depending on the guide field. Its counter-balancing effect to turbulent diffusivity decreases the turbulent transport enhancement. Hence it may decrease the amount of reconnected flux in two dimensional reconnection which may explain the suppression of the reconnection rate in the force free current sheets.
\begin{acknowledgments}
This work as been done under the SFB Project A15 of the CRC 963 at the Max-Planck-Institute for Solar System
Research, G\"ottingen, Germany and greatly enhanced during the extended programme of the \textit{Magnetic Reconnection in Plasmas} workshop at NORDITA, Stockholm. Thanks also to W.Schmidt, D. Vlaykov and P. Grete for the discussions and J. Skala for his great help in developing an appropriate numerical code.
\end{acknowledgments}

\bibliography{Bibliography}

\providecommand{\noopsort}[1]{}\providecommand{\singleletter}[1]{#1}%
\begin{thebibliography}{24}%
\makeatletter
\providecommand \@ifxundefined [1]{%
 \@ifx{#1\undefined}
}%
\providecommand \@ifnum [1]{%
 \ifnum #1\expandafter \@firstoftwo
 \else \expandafter \@secondoftwo
 \fi
}%
\providecommand \@ifx [1]{%
 \ifx #1\expandafter \@firstoftwo
 \else \expandafter \@secondoftwo
 \fi
}%
\providecommand \natexlab [1]{#1}%
\providecommand \enquote  [1]{``#1''}%
\providecommand \bibnamefont  [1]{#1}%
\providecommand \bibfnamefont [1]{#1}%
\providecommand \citenamefont [1]{#1}%
\providecommand \href@noop [0]{\@secondoftwo}%
\providecommand \href [0]{\begingroup \@sanitize@url \@href}%
\providecommand \@href[1]{\@@startlink{#1}\@@href}%
\providecommand \@@href[1]{\endgroup#1\@@endlink}%
\providecommand \@sanitize@url [0]{\catcode `\\12\catcode `\$12\catcode
  `\&12\catcode `\#12\catcode `\^12\catcode `\_12\catcode `\%12\relax}%
\providecommand \@@startlink[1]{}%
\providecommand \@@endlink[0]{}%
\providecommand \url  [0]{\begingroup\@sanitize@url \@url }%
\providecommand \@url [1]{\endgroup\@href {#1}{\urlprefix }}%
\providecommand \urlprefix  [0]{URL }%
\providecommand \Eprint [0]{\href }%
\providecommand \doibase [0]{http://dx.doi.org/}%
\providecommand \selectlanguage [0]{\@gobble}%
\providecommand \bibinfo  [0]{\@secondoftwo}%
\providecommand \bibfield  [0]{\@secondoftwo}%
\providecommand \translation [1]{[#1]}%
\providecommand \BibitemOpen [0]{}%
\providecommand \bibitemStop [0]{}%
\providecommand \bibitemNoStop [0]{.\EOS\space}%
\providecommand \EOS [0]{\spacefactor3000\relax}%
\providecommand \BibitemShut  [1]{\csname bibitem#1\endcsname}%
\let\auto@bib@innerbib\@empty
\bibitem [{\citenamefont {{Yokoi}}(2013)}]{Yokoi4}%
  \BibitemOpen
  \bibfield  {author} {\bibinfo {author} {\bibfnamefont {N.}~\bibnamefont
  {{Yokoi}}},\ }\href {\doibase 10.1080/03091929.2012.754022} {\bibfield
  {journal} {\bibinfo  {journal} {Geophysical and Astrophysical Fluid
  Dynamics}\ }\textbf {\bibinfo {volume} {107}},\ \bibinfo {pages} {114}
  (\bibinfo {year} {2013})},\ \Eprint {http://arxiv.org/abs/1306.6348}
  {arXiv:1306.6348 [astro-ph.SR]} \BibitemShut {NoStop}%
\bibitem [{\citenamefont {Higashimori}, \citenamefont {Yokoi},\ and\
  \citenamefont {Hoshino}(2013)}]{Yokoi3}%
  \BibitemOpen
  \bibfield  {author} {\bibinfo {author} {\bibfnamefont {K.}~\bibnamefont
  {Higashimori}}, \bibinfo {author} {\bibfnamefont {N.}~\bibnamefont {Yokoi}},
  \ and\ \bibinfo {author} {\bibfnamefont {M.}~\bibnamefont {Hoshino}},\ }\href
  {\doibase 10.1103/PhysRevLett.110.255001} {\bibfield  {journal} {\bibinfo
  {journal} {Physical Review Letters}\ }\textbf {\bibinfo {volume} {110}},\
  \bibinfo {pages} {255001} (\bibinfo {year} {2013})}\BibitemShut {NoStop}%
\bibitem [{\citenamefont {Parker}(1957)}]{JGR:JGR677}%
  \BibitemOpen
  \bibfield  {author} {\bibinfo {author} {\bibfnamefont {E.~N.}\ \bibnamefont
  {Parker}},\ }\href {\doibase 10.1029/JZ062i004p00509} {\bibfield  {journal}
  {\bibinfo  {journal} {Journal of Geophysical Research}\ }\textbf {\bibinfo
  {volume} {62}},\ \bibinfo {pages} {509} (\bibinfo {year} {1957})}\BibitemShut
  {NoStop}%
\bibitem [{\citenamefont {Yamada}, \citenamefont {Kulsrud},\ and\ \citenamefont
  {Ji}(2010)}]{RevModPhys.82.603}%
  \BibitemOpen
  \bibfield  {author} {\bibinfo {author} {\bibfnamefont {M.}~\bibnamefont
  {Yamada}}, \bibinfo {author} {\bibfnamefont {R.}~\bibnamefont {Kulsrud}}, \
  and\ \bibinfo {author} {\bibfnamefont {H.}~\bibnamefont {Ji}},\ }\href
  {\doibase 10.1103/RevModPhys.82.603} {\bibfield  {journal} {\bibinfo
  {journal} {Reviews of Modern Physics}\ }\textbf {\bibinfo {volume} {82}},\
  \bibinfo {pages} {603} (\bibinfo {year} {2010})}\BibitemShut {NoStop}%
\bibitem [{\citenamefont {Schindler}(2007)}]{Schindler}%
  \BibitemOpen
  \bibfield  {author} {\bibinfo {author} {\bibfnamefont {K.}~\bibnamefont
  {Schindler}},\ }\href@noop {} {\emph {\bibinfo {title} {Physics of Space
  Plasma Activity}}}\ (\bibinfo  {publisher} {Cambridge University Press},\
  \bibinfo {year} {2007})\BibitemShut {NoStop}%
\bibitem [{\citenamefont {Yokoyama}\ and\ \citenamefont
  {Shibata}(1994)}]{yokoyama1994condition}%
  \BibitemOpen
  \bibfield  {author} {\bibinfo {author} {\bibfnamefont {T.}~\bibnamefont
  {Yokoyama}}\ and\ \bibinfo {author} {\bibfnamefont {K.}~\bibnamefont
  {Shibata}},\ }\href@noop {} {\bibfield  {journal} {\bibinfo  {journal} {The
  Astrophysical Journal}\ }\textbf {\bibinfo {volume} {436}},\ \bibinfo {pages}
  {L197} (\bibinfo {year} {1994})}\BibitemShut {NoStop}%
\bibitem [{\citenamefont {Ugai}\ and\ \citenamefont
  {Tsuda}(1977)}]{ugai1977magnetic}%
  \BibitemOpen
  \bibfield  {author} {\bibinfo {author} {\bibfnamefont {M.}~\bibnamefont
  {Ugai}}\ and\ \bibinfo {author} {\bibfnamefont {T.}~\bibnamefont {Tsuda}},\
  }\href@noop {} {\bibfield  {journal} {\bibinfo  {journal} {Journal of Plasma
  Physics}\ }\textbf {\bibinfo {volume} {17}},\ \bibinfo {pages} {337}
  (\bibinfo {year} {1977})}\BibitemShut {NoStop}%
\bibitem [{\citenamefont {Terasawa}(1983)}]{terasawa1983hall}%
  \BibitemOpen
  \bibfield  {author} {\bibinfo {author} {\bibfnamefont {T.}~\bibnamefont
  {Terasawa}},\ }\href@noop {} {\bibfield  {journal} {\bibinfo  {journal}
  {Geophysical research letters}\ }\textbf {\bibinfo {volume} {10}},\ \bibinfo
  {pages} {475} (\bibinfo {year} {1983})}\BibitemShut {NoStop}%
\bibitem [{\citenamefont {Birn}\ \emph {et~al.}(2001)\citenamefont {Birn},
  \citenamefont {Drake}, \citenamefont {Shay}, \citenamefont {Rogers},
  \citenamefont {Denton}, \citenamefont {Hesse}, \citenamefont {Kuznetsova},
  \citenamefont {Ma}, \citenamefont {Bhattacharjee}, \citenamefont {Otto},\
  and\ \citenamefont {Pritchett}}]{JGRA:JGRA15381}%
  \BibitemOpen
  \bibfield  {author} {\bibinfo {author} {\bibfnamefont {J.}~\bibnamefont
  {Birn}}, \bibinfo {author} {\bibfnamefont {J.~F.}\ \bibnamefont {Drake}},
  \bibinfo {author} {\bibfnamefont {M.~A.}\ \bibnamefont {Shay}}, \bibinfo
  {author} {\bibfnamefont {B.~N.}\ \bibnamefont {Rogers}}, \bibinfo {author}
  {\bibfnamefont {R.~E.}\ \bibnamefont {Denton}}, \bibinfo {author}
  {\bibfnamefont {M.}~\bibnamefont {Hesse}}, \bibinfo {author} {\bibfnamefont
  {M.}~\bibnamefont {Kuznetsova}}, \bibinfo {author} {\bibfnamefont {Z.~W.}\
  \bibnamefont {Ma}}, \bibinfo {author} {\bibfnamefont {A.}~\bibnamefont
  {Bhattacharjee}}, \bibinfo {author} {\bibfnamefont {A.}~\bibnamefont {Otto}},
  \ and\ \bibinfo {author} {\bibfnamefont {P.~L.}\ \bibnamefont {Pritchett}},\
  }\href {\doibase 10.1029/1999JA900449} {\bibfield  {journal} {\bibinfo
  {journal} {Journal of Geophysical Research: Space Physics}\ }\textbf
  {\bibinfo {volume} {106}},\ \bibinfo {pages} {3715} (\bibinfo {year}
  {2001})}\BibitemShut {NoStop}%
\bibitem [{\citenamefont {Loureiro}, \citenamefont {Schekochihin},\ and\
  \citenamefont {Cowley}(2007)}]{Loureiro:2007gv}%
  \BibitemOpen
  \bibfield  {author} {\bibinfo {author} {\bibfnamefont {N.}~\bibnamefont
  {Loureiro}}, \bibinfo {author} {\bibfnamefont {A.~A.}\ \bibnamefont
  {Schekochihin}}, \ and\ \bibinfo {author} {\bibfnamefont {S.~C.}\
  \bibnamefont {Cowley}},\ }\href {\doibase 10.1063/1.2783986} {\bibfield
  {journal} {\bibinfo  {journal} {Physics of Plasmas}\ }\textbf {\bibinfo
  {volume} {14}},\ \bibinfo {pages} {100703} (\bibinfo {year} {2007})},\
  \Eprint {http://arxiv.org/abs/astro-ph/0703631} {arXiv:astro-ph/0703631
  [ASTRO-PH]} \BibitemShut {NoStop}%
\bibitem [{\citenamefont {{Bruno}}\ and\ \citenamefont
  {{Carbone}}(2013)}]{Carbone1}%
  \BibitemOpen
  \bibfield  {author} {\bibinfo {author} {\bibfnamefont {R.}~\bibnamefont
  {{Bruno}}}\ and\ \bibinfo {author} {\bibfnamefont {V.}~\bibnamefont
  {{Carbone}}},\ }\href {\doibase 10.12942/lrsp-2013-2} {\bibfield  {journal}
  {\bibinfo  {journal} {Living Reviews in Solar Physics}\ }\textbf {\bibinfo
  {volume} {10}},\ \bibinfo {pages} {2} (\bibinfo {year} {2013})}\BibitemShut
  {NoStop}%
\bibitem [{\citenamefont {{Osman}}\ \emph {et~al.}(2014)\citenamefont
  {{Osman}}, \citenamefont {{Matthaeus}}, \citenamefont {{Gosling}},
  \citenamefont {{Greco}}, \citenamefont {{Servidio}}, \citenamefont {{Hnat}},
  \citenamefont {{Chapman}},\ and\ \citenamefont {{Phan}}}]{Matthaus1}%
  \BibitemOpen
  \bibfield  {author} {\bibinfo {author} {\bibfnamefont {K.~T.}\ \bibnamefont
  {{Osman}}}, \bibinfo {author} {\bibfnamefont {W.~H.}\ \bibnamefont
  {{Matthaeus}}}, \bibinfo {author} {\bibfnamefont {J.~T.}\ \bibnamefont
  {{Gosling}}}, \bibinfo {author} {\bibfnamefont {A.}~\bibnamefont {{Greco}}},
  \bibinfo {author} {\bibfnamefont {S.}~\bibnamefont {{Servidio}}}, \bibinfo
  {author} {\bibfnamefont {B.}~\bibnamefont {{Hnat}}}, \bibinfo {author}
  {\bibfnamefont {S.~C.}\ \bibnamefont {{Chapman}}}, \ and\ \bibinfo {author}
  {\bibfnamefont {T.~D.}\ \bibnamefont {{Phan}}},\ }\href {\doibase
  10.1103/PhysRevLett.112.215002} {\bibfield  {journal} {\bibinfo  {journal}
  {Physical Review Letters}\ }\textbf {\bibinfo {volume} {112}},\ \bibinfo
  {eid} {215002} (\bibinfo {year} {2014})},\ \Eprint
  {http://arxiv.org/abs/1403.4590} {arXiv:1403.4590 [physics.space-ph]}
  \BibitemShut {NoStop}%
\bibitem [{\citenamefont {Matthaeus}\ and\ \citenamefont
  {Lamkin}(1985)}]{Matthaeus_Lamkin_1985}%
  \BibitemOpen
  \bibfield  {author} {\bibinfo {author} {\bibfnamefont {W.}~\bibnamefont
  {Matthaeus}}\ and\ \bibinfo {author} {\bibfnamefont {S.}~\bibnamefont
  {Lamkin}},\ }\href@noop {} {\bibfield  {journal} {\bibinfo  {journal}
  {Physics of Fluids}\ }\textbf {\bibinfo {volume} {28:1}} (\bibinfo {year}
  {1985})}\BibitemShut {NoStop}%
\bibitem [{\citenamefont {Lazarian}\ and\ \citenamefont
  {Vishniac}(1999)}]{lazarian1999reconnection}%
  \BibitemOpen
  \bibfield  {author} {\bibinfo {author} {\bibfnamefont {A.}~\bibnamefont
  {Lazarian}}\ and\ \bibinfo {author} {\bibfnamefont {E.~T.}\ \bibnamefont
  {Vishniac}},\ }\href@noop {} {\bibfield  {journal} {\bibinfo  {journal} {The
  Astrophysical Journal}\ }\textbf {\bibinfo {volume} {517}},\ \bibinfo {pages}
  {700} (\bibinfo {year} {1999})}\BibitemShut {NoStop}%
\bibitem [{\citenamefont {Kowal}\ \emph {et~al.}(2009)\citenamefont {Kowal},
  \citenamefont {Lazarian}, \citenamefont {Vishniac},\ and\ \citenamefont
  {Otmianowska-Mazur}}]{kowal2009numerical}%
  \BibitemOpen
  \bibfield  {author} {\bibinfo {author} {\bibfnamefont {G.}~\bibnamefont
  {Kowal}}, \bibinfo {author} {\bibfnamefont {A.}~\bibnamefont {Lazarian}},
  \bibinfo {author} {\bibfnamefont {E.}~\bibnamefont {Vishniac}}, \ and\
  \bibinfo {author} {\bibfnamefont {K.}~\bibnamefont {Otmianowska-Mazur}},\
  }\href@noop {} {\bibfield  {journal} {\bibinfo  {journal} {The Astrophysical
  Journal}\ }\textbf {\bibinfo {volume} {700}},\ \bibinfo {pages} {63}
  (\bibinfo {year} {2009})}\BibitemShut {NoStop}%
\bibitem [{\citenamefont {{Grete}}\ \emph {et~al.}(2015)\citenamefont
  {{Grete}}, \citenamefont {{Vlaykov}}, \citenamefont {{Schmidt}},
  \citenamefont {{Schleicher}},\ and\ \citenamefont
  {{Federrath}}}]{GreteDimitar}%
  \BibitemOpen
  \bibfield  {author} {\bibinfo {author} {\bibfnamefont {P.}~\bibnamefont
  {{Grete}}}, \bibinfo {author} {\bibfnamefont {D.~G.}\ \bibnamefont
  {{Vlaykov}}}, \bibinfo {author} {\bibfnamefont {W.}~\bibnamefont
  {{Schmidt}}}, \bibinfo {author} {\bibfnamefont {D.~R.~G.}\ \bibnamefont
  {{Schleicher}}}, \ and\ \bibinfo {author} {\bibfnamefont {C.}~\bibnamefont
  {{Federrath}}},\ }\href {\doibase 10.1088/1367-2630/17/2/023070} {\bibfield
  {journal} {\bibinfo  {journal} {New Journal of Physics}\ }\textbf {\bibinfo
  {volume} {17}},\ \bibinfo {eid} {023070} (\bibinfo {year} {2015})},\ \Eprint
  {http://arxiv.org/abs/1501.07170} {arXiv:1501.07170 [physics.flu-dyn]}
  \BibitemShut {NoStop}%
\bibitem [{\citenamefont {{Yokoi}}, \citenamefont {{Higashimori}},\ and\
  \citenamefont {{Hoshino}}(2013)}]{Yokoi1}%
  \BibitemOpen
  \bibfield  {author} {\bibinfo {author} {\bibfnamefont {N.}~\bibnamefont
  {{Yokoi}}}, \bibinfo {author} {\bibfnamefont {K.}~\bibnamefont
  {{Higashimori}}}, \ and\ \bibinfo {author} {\bibfnamefont {M.}~\bibnamefont
  {{Hoshino}}},\ }\href {\doibase 10.1063/1.4851976} {\bibfield  {journal}
  {\bibinfo  {journal} {Physics of Plasmas}\ }\textbf {\bibinfo {volume}
  {20}},\ \bibinfo {eid} {122310} (\bibinfo {year} {2013})},\ \Eprint
  {http://arxiv.org/abs/1401.1498} {arXiv:1401.1498 [physics.plasm-ph]}
  \BibitemShut {NoStop}%
\bibitem [{\citenamefont {{Yokoi}}\ and\ \citenamefont
  {{Hamba}}(2007)}]{2007PhPl...14k2904Y}%
  \BibitemOpen
  \bibfield  {author} {\bibinfo {author} {\bibfnamefont {N.}~\bibnamefont
  {{Yokoi}}}\ and\ \bibinfo {author} {\bibfnamefont {F.}~\bibnamefont
  {{Hamba}}},\ }\href {\doibase 10.1063/1.2792337} {\bibfield  {journal}
  {\bibinfo  {journal} {Physics of Plasmas}\ }\textbf {\bibinfo {volume}
  {14}},\ \bibinfo {pages} {112904} (\bibinfo {year} {2007})}\BibitemShut
  {NoStop}%
\bibitem [{\citenamefont {{Yokoi}}\ and\ \citenamefont
  {{Balarac}}(2011)}]{YokBal}%
  \BibitemOpen
  \bibfield  {author} {\bibinfo {author} {\bibfnamefont {N.}~\bibnamefont
  {{Yokoi}}}\ and\ \bibinfo {author} {\bibfnamefont {G.}~\bibnamefont
  {{Balarac}}},\ }\href {\doibase 10.1088/1742-6596/318/7/072039} {\bibfield
  {journal} {\bibinfo  {journal} {Journal of Physics Conference Series}\
  }\textbf {\bibinfo {volume} {318}},\ \bibinfo {eid} {072039} (\bibinfo {year}
  {2011})},\ \Eprint {http://arxiv.org/abs/1107.1154} {arXiv:1107.1154
  [astro-ph.SR]} \BibitemShut {NoStop}%
\bibitem [{\citenamefont {{Mu{\~n}oz}}\ \emph {et~al.}(2015)\citenamefont
  {{Mu{\~n}oz}}, \citenamefont {{Told}}, \citenamefont {{Kilian}},
  \citenamefont {{B{\"u}chner}},\ and\ \citenamefont {{Jenko}}}]{Pato2015}%
  \BibitemOpen
  \bibfield  {author} {\bibinfo {author} {\bibfnamefont {P.~A.}\ \bibnamefont
  {{Mu{\~n}oz}}}, \bibinfo {author} {\bibfnamefont {D.}~\bibnamefont {{Told}}},
  \bibinfo {author} {\bibfnamefont {P.}~\bibnamefont {{Kilian}}}, \bibinfo
  {author} {\bibfnamefont {J.}~\bibnamefont {{B{\"u}chner}}}, \ and\ \bibinfo
  {author} {\bibfnamefont {F.}~\bibnamefont {{Jenko}}},\ }\href@noop {}
  {\bibfield  {journal} {\bibinfo  {journal} {ArXiv e-prints}\ } (\bibinfo
  {year} {2015})},\ \Eprint {http://arxiv.org/abs/1504.01351} {arXiv:1504.01351
  [physics.plasm-ph]} \BibitemShut {NoStop}%
\bibitem [{\citenamefont {Ricci}, \citenamefont {Lapenta},\ and\ \citenamefont
  {Brackbill}(2004)}]{Ricci:2003yc}%
  \BibitemOpen
  \bibfield  {author} {\bibinfo {author} {\bibfnamefont {P.}~\bibnamefont
  {Ricci}}, \bibinfo {author} {\bibfnamefont {G.}~\bibnamefont {Lapenta}}, \
  and\ \bibinfo {author} {\bibfnamefont {J.}~\bibnamefont {Brackbill}},\ }\href
  {\doibase 10.1063/1.1768552} {\bibfield  {journal} {\bibinfo  {journal}
  {Physics of Plasmas}\ }\textbf {\bibinfo {volume} {11}},\ \bibinfo {pages}
  {4102} (\bibinfo {year} {2004})},\ \Eprint
  {http://arxiv.org/abs/astro-ph/0304224} {arXiv:astro-ph/0304224 [astro-ph]}
  \BibitemShut {NoStop}%
\bibitem [{\citenamefont {Skala}\ \emph {et~al.}(2014)\citenamefont {Skala},
  \citenamefont {Baruffa}, \citenamefont {Buechner},\ and\ \citenamefont
  {Rampp}}]{Skala:2014cwa}%
  \BibitemOpen
  \bibfield  {author} {\bibinfo {author} {\bibfnamefont {J.}~\bibnamefont
  {Skala}}, \bibinfo {author} {\bibfnamefont {F.}~\bibnamefont {Baruffa}},
  \bibinfo {author} {\bibfnamefont {J.}~\bibnamefont {Buechner}}, \ and\
  \bibinfo {author} {\bibfnamefont {M.}~\bibnamefont {Rampp}},\ }\href@noop {}
  {\  (\bibinfo {year} {2014})},\ \Eprint {http://arxiv.org/abs/1411.1289}
  {arXiv:1411.1289 [astro-ph.SR]} \BibitemShut {NoStop}%
\bibitem [{\citenamefont {{Yokoi}}\ and\ \citenamefont
  {{Hoshino}}(2011)}]{Yokoi2}%
  \BibitemOpen
  \bibfield  {author} {\bibinfo {author} {\bibfnamefont {N.}~\bibnamefont
  {{Yokoi}}}\ and\ \bibinfo {author} {\bibfnamefont {M.}~\bibnamefont
  {{Hoshino}}},\ }\href {\doibase 10.1063/1.3641968} {\bibfield  {journal}
  {\bibinfo  {journal} {Physics of Plasmas}\ }\textbf {\bibinfo {volume}
  {18}},\ \bibinfo {pages} {111208} (\bibinfo {year} {2011})},\ \Eprint
  {http://arxiv.org/abs/1105.6343} {arXiv:1105.6343 [astro-ph.SR]} \BibitemShut
  {NoStop}%
\bibitem [{\citenamefont {{Yokoi}}\ \emph {et~al.}(2008)\citenamefont
  {{Yokoi}}, \citenamefont {{Rubinstein}}, \citenamefont {{Yoshizawa}},\ and\
  \citenamefont {{Hamba}}}]{YokRub}%
  \BibitemOpen
  \bibfield  {author} {\bibinfo {author} {\bibfnamefont {N.}~\bibnamefont
  {{Yokoi}}}, \bibinfo {author} {\bibfnamefont {R.}~\bibnamefont
  {{Rubinstein}}}, \bibinfo {author} {\bibfnamefont {A.}~\bibnamefont
  {{Yoshizawa}}}, \ and\ \bibinfo {author} {\bibfnamefont {F.}~\bibnamefont
  {{Hamba}}},\ }\href {\doibase 10.1080/14685240802433057} {\bibfield
  {journal} {\bibinfo  {journal} {Journal of Turbulence}\ }\textbf {\bibinfo
  {volume} {9}},\ \bibinfo {eid} {N37} (\bibinfo {year} {2008})}\BibitemShut
  {NoStop}%
\end{thebibliography}%
\end{document}